\documentclass[fleqn,usenatbib,useAMS]{mnras}

\usepackage{newtxtext,newtxmath}
\usepackage[T1]{fontenc}
\DeclareRobustCommand{\VAN}[3]{#2}
\let\VANthebibliography\thebibliography
\def\thebibliography{\DeclareRobustCommand{\VAN}[3]{##3}\VANthebibliography}
\usepackage{pdflscape}	
\usepackage{ae,aecompl}

\usepackage{graphicx} 
\graphicspath{{./}{figures/}}
\usepackage{subcaption}

\newcommand{\clfd}{\mbox{\textsc{clfd}}}
\newcommand{\dspsr}{\mbox{\textsc{dspsr}}}
\newcommand{\presto}{\mbox{\textsc{presto}}}

\newcommand{\psrchive}{\mbox{\textsc{psrchive}}}
\newcommand{\pulsarX}{\mbox{\textsc{pulsarX}}}
\newcommand{\peasoup}{\mbox{\textsc{peasoup}}}
\newcommand{\pics}{\mbox{\textsc{pics}}}
\newcommand{\iqrm}{\mbox{\textsc{iqrm}}}
\newcommand{\dedisp}{\mbox{\textsc{dedisp}}}
\newcommand{\mosaic}{\mbox{\textsc{mosaic}}}
\newcommand{\transientX}{\mbox{\textsc{transientX}}}

\newcommand{\psrfoldfil}{\mbox{\texttt{psrfold\_fil}}}
\newcommand{\aplpy}{\mbox{\textsc{aplpy}}}
\newcommand{\katpoint}{\mbox{\textsc{katpoint}}}
\newcommand{\candyjar}{\mbox{\textsc{CandyJar}}}
\newcommand{\seeKAT}{\mbox{\textsc{seeKAT}}}
\newcommand{\pygsm}{\mbox{\textsc{pyGSM}}}
\newcommand{\pulsarminer}{\mbox{\textsc{pulsar miner}}}

\newcommand{\accelunits}{\,m\,s$^{-2}$}
\newcommand{\dmunits}{\,pc\,cm$^{-3}$}

\newcommand{\psrfortyeight}{PSR\,J0048$-$7317}

\newcommand{\psrfortyfour}{PSR\,J0044$-$7314}
\newcommand{\fortyfour}{J0044$-$7314}
\newcommand{\psrthirtyseven}{PSR\,J0040$-$7337}
\newcommand{\thirtyseven}{J0040$-$7337}
\newcommand{\psrthirtyfive}{PSR\,J0040$-$7335}
\newcommand{\thirtyfive}{J0040$-$7335}
\newcommand{\psrfiftyfour}{PSR\,J0054$-$7228}

\newcommand{\psrtwentysix}{PSR\,J0040$-$7326}

\newcommand{\psronezerofive}{PSR\,J0105$-$7208}
\newcommand{\onezerofive}{J0105$-$7208}

\newcommand{\psrfortyfive}{PSR\,J0045$-$7319}
\newcommand{\fortyfive}{J0045$-$7319}
\newcommand{\psrfortythree}{PSR\,J0043$-$73}
\newcommand{\fortythree}{J0043$-$73}
\newcommand{\psrfiftytwo}{PSR\,J0052$-$72}
\newcommand{\fiftytwo}{J0052$-$72}
\newcommand{\psrfiftyeight}{PSR\,J0058$-$7218}
\newcommand{\fiftyeight}{J0058$-$7218}
\newcommand{\smcmagnetar}{CXOU\,J010043.1$-$721134}
\newcommand{\crabtwin}{PSR\,J0540$-$6919}

\newcommand{\h}{$^{\rm h}$} 
\newcommand{\m}{$^{\rm m}$} 
\newcommand{\s}{$^{\rm s}$}

\title[TRAPUM SMC pulsar survey I]{The TRAPUM Small Magellanic Cloud pulsar survey with MeerKAT:\\I. Discovery of seven new pulsars and two Pulsar Wind Nebula associations}

\author[E. Carli et al.]{\parbox{\textwidth}{
E. Carli,$^{1}$\thanks{E-mail: \href{mailto:emma.carli@outlook.com}{emma.carli@outlook.com}}
L. Levin,$^{1}$
B.~W.~Stappers,$^{1}$
E.~D.~Barr,$^{2}$
R.~P.~Breton,$^{1}$
S. Buchner,$^{3}$
M. Burgay,$^{4}$
M. Geyer,$^{9,3}$
M. Kramer,$^{2}$
P.~V.~Padmanabh,$^{5,6,2}$
A. Possenti,$^{4}$
V. Venkatraman Krishnan,$^{2}$
W. Becker,$^{8,2}$
M. D. Filipovi\' c,$^{7}$
C. Maitra,$^{8}$
J. Behrend,$^{2}$
D.~J.~Champion,$^{2}$
W. Chen,$^{2}$
Y.~P.~Men,$^{2}$
A. Ridolfi$^{4,2}$
}
\\ \\ \\
$^{1}$Jodrell Bank Centre for Astrophysics, Department of Physics and Astronomy, The University of Manchester, Manchester M13 9PL, UK \\
$^{2}$Max-Planck-Institut f\"{u}r Radioastronomie, Auf dem H\"{u}gel 69, D-53121 Bonn, Germany \\
$^{3}$South African Radio Astronomy Observatory (SARAO), 2 Fir Street, Black River Park, Observatory, Cape Town, 7925 \\
$^{4}$INAF-Osservatorio Astronomico di Cagliari, via della Scienza 5, 09047, Selargius, Italy \\
$^{5}$Max-Planck-Institut f\"{u}r Gravitationsphysik (Albert-Einstein-Institut), D-30167 Hannover, Germany\\
$^{6}$Leibniz Universit\"{a}t Hannover, D-30167 Hannover, Germany\\
$^{7}$Western Sydney University, Locked Bag 1797, Penrith South DC, NSW 2751, Australia \\
$^{8}$Max-Planck-Institut f\"{u}r extraterrestrische Physik, Gie\ss{}enbachstra\ss{}e 1, D-85748 Garching bei M\"{u}nchen, Germany\\
$^{9}$Department of Mathematics and Applied Mathematics, University of Cape Town,  Rondebosch 7701, South Africa\\
}

\date{Accepted XXX. Received YYY; in original form ZZZ}
\pubyear{2024}

\begin{document}
\label{firstpage}
\pagerange{\pageref{firstpage}--\pageref{lastpage}}
\maketitle

\begin{abstract}
The sensitivity of the MeerKAT radio interferometer is an opportunity to probe deeper into the population of rare and faint extragalactic pulsars.
The TRAPUM (TRAnsients and PUlsars with MeerKAT) collaboration has conducted a radio-domain search for accelerated pulsars and transients in the Small Magellanic Cloud (SMC).
This partially targeted survey, performed at L-band (856--1712\,MHz) with the core array of the MeerKAT telescope in 2-h integrations, is twice as sensitive as the latest SMC radio pulsar survey.
We report the discovery of seven new SMC pulsars, doubling this galaxy's radio pulsar population and increasing the total extragalactic population by nearly a quarter.
We also carried out a search for accelerated millisecond pulsars in the SMC Globular Cluster NGC\,121 using the full array of MeerKAT. This improved the previous upper limit on pulsed radio emission from this cluster by a factor of six. 
Our discoveries reveal the first radio pulsar-PWN systems in the SMC, with only one such system previously known outside our galaxy (the `Crab pulsar twin' in the Large Magellanic Cloud, PSR\,J0540$-$6919).
We associate the 59\,ms pulsar discovery PSR\,J0040$-$7337, now the fastest spinning radio pulsar in the SMC, with the bow-shock Pulsar Wind Nebula (PWN) of Supernova Remnant DEM\,S5. We also present a new young pulsar with a 79\,ms period, PSR\,J0048$-$7317, in a PWN recently discovered in a MeerKAT radio continuum image.
Using the multi-beam capability of MeerKAT, we localised our pulsar discoveries, and two previous Murriyang discoveries, to a positional uncertainty of a few arcseconds.  
\end{abstract}


\section{Introduction}
\label{introduction}
The Small and Large Magellanic Clouds (SMC and LMC) are the only galaxies outside our own in which radio pulsars have been discovered to date. They are nearby galaxies that are unobstructed by the Milky Way's (MW) galactic plane and therefore they are good targets for extragalactic pulsar searches. Indeed, the Small Magellanic Cloud is just 60\,kpc away \citep{Karachentsev2004}, and the expected Milky Way Dispersion Measure (DM) contribution in its direction is low: about $30$\dmunits{}  according to the YMW16 electron density model \citep{YMW2016}, and $42$\dmunits{} according to the NE2001 model \citep{NE2001}.

The SMC has a lower metallicity than our galaxy ($Z_{\text{SMC}} \simeq 0.2\,Z_{\text{MW}}$, \citealt{Luck1998}), which can lead to an increased Neutron Star (NS) birth rate \citep{Heger2003}. 
Indeed, the latest surveys confirm that the SMC has a much higher density of High-Mass X-ray Binaries (HMXBs) than the Milky Way \citep{Haberl2016}. The SMC has also undergone recent episodes of star formation \citep{Harris2004}, which should allow for the detection of a higher proportion of short-lived and young objects than in the Milky Way's older population. Together, these galactic properties make the SMC an excellent target to search for pulsars, particularly in rare young systems.

Of the nearly 3400 radio pulsars discovered, only 31 are extragalactic (see the \href{https://www.atnf.csiro.au/research/pulsar/psrcat/proc_form.php?version=1.68&Name=Name&JName=JName&RaJ=RaJ&DecJ=DecJ&P0=P0&P1=P1&DM=DM&S1400=S1400&Binary=Binary&Dist=Dist&Assoc=Assoc&Date=Date&Age=Age&Bsurf=Bsurf&Edot=Edot&startUserDefined=true&c1_val=&c2_val=&c3_val=&c4_val=&sort_attr=jname&sort_order=asc&condition=Dist%3E40&pulsar_names=&ephemeris=short&coords_unit=raj%2Fdecj&radius=&coords_1=&coords_2=&style=Long+with+last+digit+error&no_value=*&fsize=3&x_axis=&x_scale=linear&y_axis=&y_scale=linear&state=query&table_bottom.x=83&table_bottom.y=24}{ATNF pulsar catalogue} for a list, \citealt{ATNF}). Their radio emissions were discovered using Murriyang, the Parkes 64\,m radio telescope. Of these discoveries, seven  are in the Small Magellanic Cloud, discovered by \citealt{McConnell1991} (1 pulsar), \citealt{Crawford2001} (1), \citealt{Manchester2006} (3), and  \citealt{Titus2019} (2). Their spin periods range from 191 to 937\,ms, their characteristic ages span 1-4\,Myr and their median DM is approximately 115\dmunits{}.
    
In addition, two young (not accretion-powered) pulsars have been discovered, via their X-ray pulsations only, in the SMC: the magnetar \smcmagnetar{} by \citealt{Lamb2002} (recently associated with a new supernova remnant by \citealt{Cotton2024}) and the rotation-powered pulsar \fiftyeight{} by \citealt{Maitra2021} (see also \citealt{Carli2022}). The latter is embedded in a Pulsar Wind Nebula (PWN) \citep{Maitra2015} in the supernova remnant (SNR) IKT\,16 \citep{Inoue1983,Mathewson1984,Owen2011}. A second PWN in the SMC was discovered by \cite{Alsaberi2019}  in the SNR\,DEM\,S\,5 \citep{Haberl2000,Payne2007} but no pulsar was detected. There are no other SMC pulsars known to be associated with  supernova remnants so far.
None of the SMC's known radio pulsars is young: their characteristic ages are all larger than 1\,Myr. There are no known extragalactic millisecond pulsars (old pulsars which have been spun up to millisecond rotation periods by accretion from a companion). The SMC hosts the only discovered extragalactic binary pulsar: \psrfortyfive{}, an ordinary pulsar with a massive main sequence B star companion \citep{Kaspi1994}.  
        
The larger population of the LMC shows more extensively the wealth of exotic neutron star systems that can be found outside our galaxy. It includes the `Crab pulsar twin' \crabtwin{} in the PWN of SNR 0540$-$69.3 \citep{Manchester1993b, Gotthelf2000, Brantseg2013}, the only known extragalactic multiwavelength pulsar, discovered in the X-ray band by \cite{Seward1984}, then via optical pulses by \cite{Middleditch1985}, then via radio pulsations by \cite{Manchester1993a}, later seen to emit giant radio pulses \citep{Johnston2003}, and finally in gamma-rays by \cite{Marshall2016}.
Along with PSR\,J0534$-$6703 \citep{Manchester2006}, a long period ($\simeq$1.8\,s) radio pulsar with no SNR or PWN association and no high-energy emission, these are the only two known young (<100\,kyr characteristic age) radio pulsars in the LMC, and thus outside our galaxy. There is one additional young protation-powered ulsar in the LMC, PSR\,J0537$-$6910, in a PWN in the SNR\,N157B \citep{Wang2001, wang1998}. It is only detected via its X-ray pulsations and is the  fastest spinning (16.1\,ms) young pulsar known \citep{Marshall1998}. The LMC also hosts one magnetar, SGR\,0526$-$66  \citep{Helfand1979,kulkarni2003}, possibly associated with the SNR\,N\,49 \citep[e.g.][]{Gaensler2001}. Finally, PSR\,J0529$-$6652  is also known to emit giant pulses \citep{Crawford2012}. There are a further three known PWNe in the LMC with no pulsar associations \citep{gaensler2003,williams2005,bamba2006,haberl2012a}.

Thus, despite only a very small sample of pulsars outside our galaxy being known, many systems are showing characteristics that are relatively rare in the Milky Way population, and are prized for a variety of astrophysical studies. TRAPUM\footnote{\href{http://trapum.org/}{trapum.org}} (TRAnsients and PUlsars with MeerKAT) is a Large Survey Project of the MeerKAT telescope \citep{Stappers2016}, and one of its core science goals is to increase the known population of the rare extragalactic neutron stars. The collaboration has already discovered  over 200 pulsars\footnote{\href{http://trapum.org/discoveries/}{http://trapum.org/discoveries/}} in the Milky Way's galactic plane \citep{Padmanabh2023,Bernadich2023}, gamma-ray sources \citep{Clark2023a, Dodge2024}, and Globular Clusters \citep{Ridolfi2021,Abbate2023,Vleeschower2022,Zhang2022,Ridolfi2022,Douglas2022, barr2024}. We have conducted the first survey of the SMC with the MeerKAT telescope. This TRAPUM survey searches for ordinary and young pulsars, single pulses, Fast Radio Bursts (FRBs), binary systems and millisecond pulsars in a variety of environments.

We begin this paper by presenting the novel observing  (\autoref{observations_general}) and processing (\autoref{processing}) techniques of this survey. The resulting survey sensitivity is compared to other surveys in \autoref{sensitivity}. The different observing and processing methods used for the SMC Globular Cluster NGC\,121 are reported in \autoref{NGC121}. Our discoveries are described in \autoref{results}. A summary of this work, \autoref{summary}, concludes this paper.
The timing solutions and peculiarities of our discoveries will be shown in Paper II of this series. The discoveries will enable an analysis of  the impact of the low-metallicity, recent star formation environment  of the SMC on its neutron star population and comparisons with predictions  \citep[e.g.][]{Titus2020}. We will present this population study in Paper III of this series.

\section{Survey}
\label{survey}
\subsection{Observations}
\label{observations_general}
\subsubsection{Introduction}
\label{observations_introduction}
    
Historically, pulsar searching has been dominated by large single-dish surveys\footnote{See \href{www.jb.man.ac.uk/pulsar/surveys.html}{here} for a survey database from \cite{Lyon2016}.}, and in the Southern Hemisphere  by Murriyang, the Parkes 64-m radio-telescope in Australia (hereafter Murriyang). Multi-beam techniques were introduced for Murriyang by \cite{Staveley-Smith1996} and \cite{Crawford2001} conducted the first survey of the Magellanic Clouds that benefited from this increased field of view. This allowed for greater integration times per pointing, improving the sensitivity of surveys. To further increase the observing sensitivity and find more of the faint extragalactic pulsars, one can either turn to very large unsteerable aperture dishes (e.g. the FAST telescope, \citealt{Nan2012}), or to a combined array of smaller dishes: an interferometer.

The MeerKAT telescope \citep{MeerKAT,Camilo2018}, inaugurated in 2018, is of the latter type. It is a precursor to the mid-frequency component of the Square Kilometer Array, situated in the Karoo desert in South Africa. It consists of 64 13.5-m Gregorian antennas, 44 of which are arranged in a core of 1\,km diameter. The maximum baseline is 8\,km. The full coherent array of MeerKAT has a gain about four times that of Murriyang. This is a major sensitivity boost for finding faint or distant pulsars in the Southern Hemisphere, notably extragalactic pulsars. MeerKAT operates in three observing bands. Our observations were performed at L-band (see \autoref{tab:MeerKAT_parameters} for the specifications). The UHF band of MeerKAT is also available but was not used due to data rate and processing time considerations.

\begin{table}
\centering
\caption{MeerKAT L-band specifications \protect\citep{MKAT_L_band,Bailes2020}.}
\label{tab:MeerKAT_parameters}
\begin{tabular}{ll}
\hline
\textbf{Parameter}                            & \textbf{Value} \\ \hline
Bandwidth (MHz)      $\Delta\nu$                & 856   \\ 
Centre frequency $\nu$ (MHz)               & 1284  \\ 
System temperature $T_{\text{sys}}$ (K)               & 18    \\ 
Gain $G$ (64 coherent antennas or `full array', K\,Jy$^{-1}$)         & 2.8 \\ 
Gain $G$ (44 coherent antennas or `core array', K\,Jy$^{-1}$)         & 1.925 \\ 
Gain $G$ (62 incoherent antennas, K\,Jy$^{-1}$)       & 0.344  \\ 
Number of polarisations recorded $n_{\text{pol}}$     & 2     \\ \hline
\end{tabular}
\end{table}
        
This survey uses the Filterbanking Beamformer User Supplied Equipment (FBFUSE), a MeerKAT backend specific to pulsar and transient search users designed by the Max Planck Institute for Radio Astronomy (MPIfR). The beamformer can generate a chosen number of coherent `tied-array' beams on the sky by varying delays calculated by \katpoint{}\footnote{\href{https://developer.skao.int/projects/katpoint}{https://developer.skao.int/projects/katpoint}} between antennas before coherently combining their signals. The Point Spread Function (PSF), orientation and optimal tiling of the coherent beams (so that they overlap at a chosen fractional sensitivity  level of the PSF) is simulated by the \mosaic{} software suite \citep{Chen2021}. To increase the coherent beam size and therefore the field of view covered by the coherent beam tiling in each pointing of this survey, we use the core array configuration of MeerKAT (40 or 44\footnote{FBFUSE forms beams with a multiple of four antennas.} antennas out of 64) with a maximum baseline length of about 1\,km. This reduction from the full array's 8\,km maximum baseline strikes a good balance between beam area and sensitivity \citep{Chen2021}. The gain is reduced to just under three times that of Murriyang. We also observe the circumpolar SMC around the time it is furthest from the observer's Meridian, in the intermediate elevation range, to maximise beam size \citep{Chen2021}. With this setup, the area of a single coherent beam inside its 50 per cent sensitivity contour at the central frequency of L-band is around 1 square arcminute.
        
This is about 150 times smaller than Murriyang's single beam area. This means that a much larger number of beams must be recorded than in the Murriyang surveys to cover the galaxy's area. However, when a pulsar is detected in a MeerKAT coherent beam, its precise position is immediately pinpointed  to a beam size error of the order of an arcminute. It is even possible to constrain the pulsar location  to a few arcseconds by combining detections in multiple MeerKAT beams \citep{SeeKAT}, which is described in \autoref{localisation}. In comparison, the Murriyang telescope has a position error on discovered pulsars of 14\arcmin{} diameter, the Full Width at Half Maximum (FWHM) of its beam. The Parkes Multi Beam \citep[PMB,][]{Staveley-Smith1996} receiver of Murriyang can observe 13 simultaneous beams, covering 154 square arcminutes each with a sensitivity above 50 per cent of the maximum, or around 0.56 square degrees per pointing. The area covered with above 50 per cent of the maximum sensitivity by a 768 MeerKAT core array coherent beam tiling is only about 768 square arcminutes or 2.6 times less than a PMB pointing.

There is also an `incoherent' or `primary' beam obtained from summing all the antenna signals  with no delay corrections. It has a FWHM of 1.1\,degrees at the centre frequency of L-band \citep{Asad2021}. At the lowest frequency of the band, the incoherent beam has a FWHM of 1.7\,degrees, and at the highest, a FHWM of 0.9\,degrees.
The incoherent beam's gain is half that of Murriyang and covers 0.95 square degrees with over 50 per cent sensitivity at the central frequency of L-band, nearly double the area of the PMB. Coherent beams can be placed anywhere within  the MeerKAT incoherent beam, but their sensitivity is governed by the incoherent beam response. The gains of the different configurations of MeerKAT are provided in 
\autoref{tab:MeerKAT_parameters}. Beam tilings for each of our observations are shown in Appendix \ref{beam_maps}.

The filterbank data generated by FBFUSE are processed locally (\autoref{processing}), in the MPIfR APSUSE computing cluster (Accelerated Pulsar Searching User-Supplied Equipment)  located in the Karoo Array Processor Building. The cluster contains 60 nodes, each with 24 CPUs, two GPUs and 97\,GB of RAM. APSUSE pipeline processing is performed using MPIfR pipelines\footnote{ \href{https://github.com/MPIfR-BDG/trapum-pipeline-wrapper/tree/mongo_consumer}{https://github.com/MPIfR-BDG/trapum-pipeline-wrapper/tree/mongo\_consumer}} running on 56 nodes, while the remaining four are used for manual processing tasks and testing. Observations, processing and data products are tracked using a MPIfR database\footnote{\href{https://github.com/MPIfR-BDG/trapum-db}{https://github.com/MPIfR-BDG/trapum-db}}. The pulsar searching backends of MeerKAT are presented in further detail in \cite{Barr2017}, \cite{prajwals_thesis} and \cite{Padmanabh2023}.

\subsubsection{Survey pointings and targets}
\label{observations}
To maximise the number of beams available and thus the field of view observed in each pointing, we must reduce the data rate per beam. This decreases the amount of data that needs to be stored and processed as well as the APSUSE filterbank recording rate. We have thus opted to use 2048 frequency channels (a higher resolution 4096-channel mode is available), and a sampling time of 153\,$\upmu$s, rather than the minimum available 76\,$\upmu$s. This enabled us to record 768 simultaneous coherent beams\footnote{480 beams for the first pass of pointings 1 and 2, see \autoref{tab:observations}. We opted to reduce resolution to record more simultaneous beams.} and one incoherent beam, with a total data rate of 10 gigabytes per second.  We observed each pointing for 2 hours with 8-bit sampling. One pointing in the survey needs 74\,terabytes of APSUSE disk space.  The characteristics of the observations of each pointing of the survey are given in \autoref{tab:observations}. We observed each pointing twice to allow for candidate pulsar confirmations, except for pointings 5 and 7, which second passes we replaced by an additional pointing to the survey (\mbox{SMCPOINTING9}, observed once) and an observation of the SMC Globular Cluster NGC\,121 (see \autoref{NGC121}).

Pointing centres  were picked to loosely follow the High Mass X-ray Binary distribution of the SMC \citep{Haberl2016} to favour areas with high neutron star density. Furthermore, the pointing positions were optimised to include as many targets that could host pulsars as possible within the incoherent beam (IB) area with over 50 per cent of the maximum sensitivity. For each pointing, coherent beams were tiled on relevant targets, overlapping at a chosen sensitivity level. Any remaining beams were tiled at boresight, where sensitivity is highest.

\begin{landscape}
\begin{table}

	\caption{The SMC survey observation parameters. A high resolution coherent beam PSF was simulated with the latest version of \mosaic{} at the centre coordinates of the pointing (Right Ascension and Declination), at the central time of the observation, with the antennas used during the observation and at the central frequency of L-band. The coherent beam ellipse semi-axes are given in arcseconds in the Right Ascension and Declination directions, with the ellipse position angle. The observation parameters of the observation of the SMC Globular Cluster NGC\,121 are given in \autoref{tab:observations_NGC121}.}
	\label{tab:observations}

\resizebox{1.35\textwidth}{!}{
\

\begin{tabular}{ccccccc}
\hline
\begin{tabular}[c]{@{}c@{}}\textbf{Pointing name}\\\textbf{Date}\end{tabular}                                                & \textbf{Observation length (s)} & \begin{tabular}[c]{@{}c@{}}\textbf{Pointing position}\\\textbf{(J2000)}\end{tabular}                                                   & \begin{tabular}[c]{@{}c@{}}\textbf{Number of antennas}\\(coherent, incoherent)\end{tabular} & \textbf{Number of coherent beams} & \begin{tabular}[c]{@{}c@{}}\textbf{Coherent beam size}\\(50 per cent sensitivity level)\end{tabular}                  & \textbf{Notes}                                                                         \\ \hline
\begin{tabular}[c]{@{}c@{}}SMCPOINTING1\\14 Jan 2021\end{tabular}                & 6860                     & \begin{tabular}[c]{@{}c@{}}00\h49\m08\fs01\\ -73\textdegree{}12\arcmin53\farcs6\end{tabular} & \begin{tabular}[c]{@{}c@{}}44, 57\end{tabular}     & 480                               & 50\farcs8, 21\farcs2, $-$15\fdg6& \begin{tabular}[c]{@{}c@{}}76\,$\upmu$s sampling time and 4096 channels recorded: less beams.\\Segments searched in full resolution.\\Full observation downsampled in time by a factor of 2.\\No transient search.\end{tabular}                             \\[0.6cm] 
\begin{tabular}[c]{@{}c@{}}SMCPOINTING2\\14 Jan 2021\end{tabular}                & 7157                     & \begin{tabular}[c]{@{}c@{}}00\h55\m14\fs69\\ -72\textdegree{}26\arcmin06\farcs9\end{tabular} & \begin{tabular}[c]{@{}c@{}}44, 57\end{tabular}     & 480                               & 39\farcs7, 21\farcs3, 10\fdg0& \begin{tabular}[c]{@{}c@{}}76\,$\upmu$s sampling time and 4096 channels recorded: less beams.\\Segments searched in full resolution.\\Full observation downsampled in time by a factor of 2.\\No transient search.\end{tabular}                           \\[0.6cm] 
\begin{tabular}[c]{@{}c@{}}SMCPOINTING1\\Second pass\\24 April 2021\end{tabular} & 6902                     & \begin{tabular}[c]{@{}c@{}}00\h49\m08\fs01\\ -73\textdegree{}12\arcmin53\farcs6\end{tabular} & \begin{tabular}[c]{@{}c@{}}40, 62\end{tabular}     & 768                               &                                                                        63\farcs3, 28\farcs0, $-$18\fdg9& No transient search                                                                    \\[0.6cm] 
\begin{tabular}[c]{@{}c@{}}SMCPOINTING2\\Second pass\\24 April 2021\end{tabular} & 7140                     & \begin{tabular}[c]{@{}c@{}}00\h55\m14\fs69\\ -72\textdegree{}26\arcmin06\farcs9\end{tabular} & \begin{tabular}[c]{@{}c@{}}40, 62\end{tabular}     & 768                               &                                                                        49\farcs2, 28\farcs0, 7\fdg6& No transient search                                                                    \\[0.6cm] 
\begin{tabular}[c]{@{}c@{}}SMCPOINTING3\\27 Oct 2021\end{tabular}                & 7152                     & \begin{tabular}[c]{@{}c@{}}00\h41\m29\fs44\\ -73\textdegree{}33\arcmin34\farcs2\end{tabular} & \begin{tabular}[c]{@{}c@{}}44, 58\end{tabular}     & 768                               &                                                                        40\farcs6, 24\farcs8, $-$14\fdg9&                                                                                        \\[0.6cm] 
\begin{tabular}[c]{@{}c@{}}SMCPOINTING4\\27 Oct 2021\end{tabular}                & 7153                     & \begin{tabular}[c]{@{}c@{}}00\h57\m21\fs24\\ -71\textdegree{}52\arcmin35\farcs2\end{tabular} & \begin{tabular}[c]{@{}c@{}}44, 58\end{tabular}     & 762                               &                                                                        51\farcs1, 25\farcs0, 13\fdg4&                                                                                        \\[0.6cm] 
\begin{tabular}[c]{@{}c@{}}SMCPOINTING3\\Second pass\\4 March 2022\end{tabular}  & 6288                     & \begin{tabular}[c]{@{}c@{}}00\h41\m29\fs44\\ -73\textdegree{}33\arcmin34\farcs2\end{tabular} & \begin{tabular}[c]{@{}c@{}}60, 62\end{tabular}     & 287                               &                                                                        24\farcs7, 8\farcs4, $-$18\fdg1& \begin{tabular}[c]{@{}c@{}}Failed observation setup.\\Only part of central beam tiling observed.\\Full coherent array\end{tabular} \\[0.6cm] 
\begin{tabular}[c]{@{}c@{}}SMCPOINTING4\\Second pass\\4 March 2022\end{tabular}  & 7156                     & \begin{tabular}[c]{@{}c@{}}00\h57\m21\fs24\\ -71\textdegree{}52\arcmin35\farcs2\end{tabular} & \begin{tabular}[c]{@{}c@{}}44, 62\end{tabular}     & 285                               &                                                                        59\farcs4, 25\farcs6, 16\fdg3& \begin{tabular}[c]{@{}c@{}}Failed observation setup.\\Only part of central beam tiling observed.\end{tabular}                     \\[0.6cm] 
\begin{tabular}[c]{@{}c@{}}SMCPOINTING5\\14 June 2022\end{tabular}               & 7001                     & \begin{tabular}[c]{@{}c@{}}01\h04\m40\fs88\\ -72\textdegree{}14\arcmin50\farcs5\end{tabular} & \begin{tabular}[c]{@{}c@{}}44, 59\end{tabular}     & 765                               &                                                                        41\farcs9, 22\farcs9, $-$7.9&                                                                                        \\[0.6cm] 
\begin{tabular}[c]{@{}c@{}}SMCPOINTING6\\14 June 2022\end{tabular}               & 7029                     & \begin{tabular}[c]{@{}c@{}}01\h07\m15\fs54\\ -72\textdegree{}57\arcmin05\farcs6\end{tabular} & \begin{tabular}[c]{@{}c@{}}44, 59\end{tabular}     & 761                               &                                                                        53\farcs8, 23\farcs1 ,18\fdg7&                                                                                        \\[0.6cm] 
\begin{tabular}[c]{@{}c@{}}SMCPOINTING7\\22 Aug 2022\end{tabular}                & 7180                     & \begin{tabular}[c]{@{}c@{}}01\h16\m39\fs28\\ -73\textdegree{}24\arcmin35\farcs4\end{tabular} & \begin{tabular}[c]{@{}c@{}}44, 59\end{tabular}     & 764                               &                                                                        56\farcs6, 20\farcs0, $-$26.3&                                                                                        \\[0.6cm] 
\begin{tabular}[c]{@{}c@{}}SMCPOINTING8\\22 Aug 2022\end{tabular}                & 7181                     & \begin{tabular}[c]{@{}c@{}}01\h29\m11\fs16\\ -73\textdegree{}25\arcmin03\farcs3\end{tabular} & \begin{tabular}[c]{@{}c@{}}44, 59\end{tabular}     & 763                               &                                                                        44\farcs5, 20\farcs0, $-$1\fdg75&                                                                                        \\[0.6cm] 
\begin{tabular}[c]{@{}c@{}}SMCPOINTING6\\Second pass\\6 Dec 2022\end{tabular}    & 7168                     & \begin{tabular}[c]{@{}c@{}}01\h07\m15\fs54\\ -72\textdegree{}57\arcmin05\farcs6\end{tabular} & \begin{tabular}[c]{@{}c@{}}44, 59\end{tabular}     & 760                               &                                                                        43\farcs0, 24\farcs5, $-$4\fdg8&                                                                                        \\[0.6cm] 
\begin{tabular}[c]{@{}c@{}}SMCPOINTING8\\Second pass\\7 Dec 2022\end{tabular}    & 7182                     & \begin{tabular}[c]{@{}c@{}}01\h29\m11\fs16\\ -73\textdegree{}25\arcmin03\farcs3\end{tabular} & \begin{tabular}[c]{@{}c@{}}44, 59\end{tabular}     & 763                               &                                                                        53\farcs6, 24\farcs7, 18\fdg4&                                                                                        \\[0.6cm] 
\begin{tabular}[c]{@{}c@{}}SMCPOINTING9\\19 Mar 2023\end{tabular}                & 7169                     & \begin{tabular}[c]{@{}c@{}}00\h51\m26\fs80\\ -72\textdegree{}51\arcmin56\farcs0\end{tabular} & \begin{tabular}[c]{@{}c@{}}44, 62\end{tabular}     & 767                               &                                                                        43\farcs6, 23\farcs2, 5\fdg7&                                                                                        \\ \hline
\end{tabular}

}
\end{table}
\end{landscape}

We opted to tile most of our targeted beams on and around\footnote{Pulsars can be kicked out of their birth site due to a supernova kick \cite[e.g.][]{Lyne1994}.} the supernova remnants of the Small Magellanic Cloud to find associated young pulsars. We observed all the SNRs and SNR candidates in the multiwavelength census by \cite{Maggi2019}. 
A few additional SNR candidates were also obtained from \cite{Titus2019} and their observation proposal. A complete list of the observed SNRs is given in \autoref{upperlimits}. The new SNRs identified in the SMC by \cite{Cotton2024} were not published at the time of observing. For the SNRs IKT\,16 \citep{Maitra2015,Carli2022} and DEM\,S5 \citep{Alsaberi2019}, a single beam was placed on the associated PWN. 
Our target list also comprises five out of the seven known radio pulsars in the SMC, for localisation (see \autoref{titus_localisations}) or processing validation purposes, and the two known SMC X-ray pulsars that are not accretion-powered. There are no confirmed Central Compact Objects\footnote{CCOs  are  soft thermal X-ray sources  near the centre of a supernova remnant, with no detected hard X-ray emission, radio emission, or PWN environment \citep{DeLuca2017}.} (CCOs) in the Magellanic Clouds \citep[e.g.][]{Long2020}, so none was targeted by our tilings.  
In later pointings of the survey, we placed a beam on HMXBs with periods under 10\,s (\cite{Haberl2016},\cite{Bartlett2017},\cite{Vasilopoulos2020}, see \autoref{tab:fast_HMXBs}), as it is our longest period searched (see \autoref{peasoup}). We also targeted most steep spectrum continuum point sources (from the  \cite{Joseph2019} ASKAP EMU SMC image, see \autoref{tab:steep_spectrum_sources}) that fell within the IB FWHM and were not within planned tilings in those later pointings. Earlier in the survey, some beams were overlapping with the position of more of these fast-spinning HMXBs and steep spectrum point sources (see Appendix \ref{beam_maps}). Finally, we tiled star clusters with ages over 5\,Gyr near our pointings \citep{Bica2020}: Lindsay\,110 in both passes of pointing 8 (though it was outside of the IB FWHM), and NGC\,416 in the second pass of pointing 5. NGC\,361 was omitted due to being in a lower sensitivity area of a pointing.
An observation of the SMC's only Globular Cluster, NGC\,121, was conducted and processed with a different setup detailed in \autoref{NGC121}. The beams recorded for each observation are shown in Appendix \ref{beam_maps}. Sensitivity limits for various  targets are given in \autoref{upperlimits}.

\begin{table}

\caption{The accretion-powered X-ray pulsars with a period under 10\,s observed in this survey. These were taken from the \protect\cite{Haberl2016} census (unless specified) with \textit{Gaia} DR3 positions \protect\citep{Vallenari2023} to place a single beam on these sources  from \mbox{SMCPOINTING6} onwards.  Four fast-spinning HMXB sources were observed with beams overlapping their position while targeted at different sources (see Appendix \ref{beam_maps}).}
\label{tab:fast_HMXBs}
\begin{tabular}{cc}
\hline
\textbf{Name of accretion-powered X-ray pulsar} & \textbf{Pointing}      \\ \hline
SXP\,6.85                                         & \begin{tabular}[c]{@{}c@{}}SMCPOINTING6\\2 passes\end{tabular}\\[0.4cm] 
SXP\,2.16                                         &\begin{tabular}[c]{@{}c@{}}SMCPOINTING7\end{tabular}\\[0.4cm] 
SXP\,8.5                                          &\begin{tabular}[c]{@{}c@{}}SMCPOINTING9\end{tabular}\\[0.4cm] 
SXP\,7.78                                         &\begin{tabular}[c]{@{}c@{}}SMCPOINTING9\end{tabular}\\[0.4cm] 
SXP\,9.13 (untargeted)                                      &\begin{tabular}[c]{@{}c@{}}SMCPOINTING1\\2 passes\end{tabular}\\[0.4cm] 
SXP\,6.88 (untargeted)                                        &\begin{tabular}[c]{@{}c@{}}SMCPOINTING2 \\2 passes\end{tabular}\\[0.4cm] 
SXP\,7.59 \citep[untargeted,][]{Hong2017}                                      &\begin{tabular}[c]{@{}c@{}}SMCPOINTING3\end{tabular}\\[0.4cm] 
SXP\,0.72 (untargeted, SMC\,X-1)                                      &\begin{tabular}[c]{@{}c@{}}SMCPOINTING7\end{tabular}\\ \hline

\end{tabular}
\end{table}

\subsection{Data reduction}

\label{processing}
\subsubsection{Periodicity searches}
\label{peasoup}
We first processed the survey data with a periodicity search pipeline.
Radio Frequency Interference (RFI) cleaning was performed in the early stages of the survey with \iqrm{} \citep{IQRM}, \presto{}'s \texttt{rfifind} \citep{PRESTO}, and a multi-beam \texttt{rfifind} wrapper aiming to remove signals detected in several beams with sufficient spatial separation\footnote{\href{https://github.com/mcbernadich/multiTRAPUM}{https://github.com/mcbernadich/multiTRAPUM} by Miquel Colom i Bernadich}. These methods were eventually replaced by a new, performant RFI removal tool, \pulsarX{}'s \texttt{filtool} \citep{Men2023}.  We also  masked frequency channels that are known to be significantly RFI-corrupted.

We calculated the dispersion smearing for our observations and the corresponding DM trial step sizes with \presto{}'s \texttt{DDplan.py} script. The SMC survey de-dispersion plan spans 50 to 350\dmunits{}. We selected this maximum as the median DM of published SMC pulsars is 115\dmunits{} and the maximum is 205\dmunits{} \citep{Manchester2006}. Conversely, the minimum published DM for a SMC pulsar is 71\dmunits{} \citep{Manchester2006} and the Milky Way DM contribution in the direction of the SMC is 30 or 42\dmunits{} according to the YMW2016 or NE2001 models respectively \citep{YMW2016, NE2001}. Halving the frequency channels to 2048 after the first observation of the survey (to maximise the number of coherent beams, see \autoref{observations}) doubled dispersion smearing within channels, which prohibits detection of the fastest spinning MSPs at the highest DMs. Indeed, at the central frequency of L-band, a maximum total smearing (including de-dispersion step size, sample time, and intra-channel dispersion)  of 0.75\,ms  is reached at 350\dmunits{}. At the lowest frequency of L-band, the total smearing is greater than 1\,ms above a DM of 160\dmunits{}. We used the \dedisp{} library \citep{levin2012,Barsdell2012} for de-dispersion\footnote{As part of the \peasoup{} suite.}.

We performed periodicity searches using \peasoup{} \citep{PEASOUP}, a GPU-based time-domain linear acceleration pulsar searching software suite (described in \citealt{Morello2019} and \citealt{Sengar2022}). The 2-hour data were not corrected for acceleration before performing a Fast Fourier Transform (FFT), while 20-minute segments were resampled for constant accelerations from a binary companion up to $|50|$\accelunits{}. This segmenting reduces the portion of the orbit for which a potential binary is observed, and thus the range over which the Doppler-shifted pulsar spin frequency is smeared. While our potential to detect highly accelerated systems is thus increased, the shorter integration time reduces our sensitivity by a factor of $\sqrt{6}$ compared to the full observation search.
We chose an acceleration tolerance parameter of 10 per cent, resulting in 153 acceleration trials\footnote{From SMC pointing 7 onwards, a \href{https://github.com/ewanbarr/peasoup/commit/cf169abbd167713f66b2349110fe004067095f43}{fix} was rolled out in \peasoup{} which reduced the number of acceleration trials. The prior acceleration searches were oversampled in acceleration with 393 acceleration trials at 0\dmunits{}.} at 0\dmunits{}.
This means the acceleration broadening from one acceleration step to the next cannot exceed 10 per cent of the combined pulse smearing due to sampling time, intra-channel frequency dispersion, and de-dispersion step size (\citealt{levin2012}, used in \citealt{Morello2019}). 
The \peasoup{} searches were limited to consider candidates with a minimum FFT spectrum Signal to Noise Ratio (S/N) of 8 and to a maximum period of 10\,s. The 2-hour and 20-min timeseries were zero-padded to the next nearest power of 2 ($2^{23}$ and $2^{26}$ respectively), resulting in a better resolved and more quickly computed Fourier spectrum. Candidates were harmonically summed by \peasoup{} up to the eighth harmonic \citep{Taylor1969}.  A list of common local RFI fluctuation frequencies was removed from the searches.

The FFT candidates were sifted by \peasoup{}, i.e. clustered in DM, period, acceleration and harmonics. 
The full observation periodicity search processing time was de-dispersion dominated. These searches resulted in about 9$\times 10^{5}$ candidates per pointing, i.e. around 1200 candidates per full periodicity search of one beam. 
Due to an oversight, pointings 1 and 2 (first passes only) were limited to 1000 candidates per beam, resulting in a spectral S/N cut of about 8.4. They were also expected to retrieve around 1200 candidates per beam.

The segmented acceleration search processing time was FFT-search dominated and resulted in around 12 million candidates per pointing, i.e. about 2600 candidates per segment of a single beam. Again, survey pointings 1, 2 and 3 (first passes only) were erroneously limited to 1000 candidates per beam segment in the acceleration search, increasing the spectral S/N cut to about 8.5.
When acceleration trials were reduced by more than half from \mbox{SMCPOINTING7} due to oversampling, the acceleration search resulted in approximately 7 million candidates per pointing or 1500  per beam segment.

The candidates were then input into a multi-beam spatial sifting programme\footnote{\href{https://github.com/prajwalvp/candidate_filter}{https://github.com/prajwalvp/candidate\_filter} by Lars K\"{u}nkel, described in \cite{Padmanabh2023} and \cite{prajwals_thesis}.} which uses the expected spatial S/N distribution of real sources to identify RFI and performs clustering of candidates across beams, after which about 100 candidates per beam remained\footnote{The candidate counts here do not take into account the first pass of pointings 1 and 2, because of the applied candidate number cutoff and a 8.5 S/N cut applied at the multi-beam filter stage.} for the full periodicity search and 600 (200 after the reduction in acceleration trials) per segment for the acceleration search.  This reduced the total number of candidates per pointing, both searches combined, from around 13 million to 3 million (8 million to 1 million from \mbox{SMCPOINTING7}). 
We note that a recently discovered bug in the multi-beam filtering code over-removed candidates from an in-built list of RFI fluctuation frequencies. It is possible to retroactively check if a specific candidate has been removed in a list kept in the data products of this pipeline.

\subsubsection{Candidate folding}
\label{folding}
Due to the high total number of spectral candidates to fold, we applied the following cuts: we folded candidates with periods  longer than 8 time samples (1.216\,ms, 1\,ms for the segment search of the first pass of pointings 1 and 2), and a spectral S/N greater than 9. This is on the high end of S/N cuts compared to other surveys, while not being the highest \citep{Lyon2016}. 
This resulted in about 30 candidates to fold for each full integration periodicity search of a beam and 65 (32 after the reduction of acceleration trials) for each beam segment acceleration search. 
We folded the raw data of all 3.5$\times 10^{5}$ (2$\times 10^{5}$ from \mbox{SMCPOINTING7}) candidates per pointing of this survey with the parameters of the filtered FFT candidates using \pulsarX{}'s \psrfoldfil{}, a new software suite designed to reduce the computation costs of high candidate numbers from interferometer-based multi-beam pulsar surveys \citep{Men2023}.
\psrfoldfil{} cleans the full resolution data from RFI with \texttt{filtool} before folding and applies the \clfd{} \citep{clfd} RFI removal software on the folded data.  In this search, we downsampled the raw data into folded archives as follows: 64 time samples (bins) per pulse profile (128 for signals with $P > 100$\,ms from SMC pointings 1 and 2 second pass onwards), each sub-integration in time of the observation is 20 seconds long, and the number of frequency channels is reduced to 64.  The folds were optimised to the highest S/N pulse profile over the default \psrfoldfil{} period and period derivative range around the  \peasoup{}-detected topocentric spin frequency so that the maximum smearing is 1 pulse period over the whole observation time, and over a DM range so that the maximum dispersion delay is 1 pulse period over the whole bandwidth. Due to electricity failures during processing, 18 and 12 segments could not be folded in the second pass of pointings 3 and 4 respectively. 

Due to an oversight, the segment candidates from the first pass of pointing 1 and 2 were folded with opposite acceleration from the \peasoup{} value.  Candidates with an acceleration greater than twice the optimisation range of \psrfoldfil{} were thus not folded correctly. These data were deleted before the bug was noticed and the first pass accelerated segment candidates could not be refolded. The pointings were observed and searched anew in their second pass. The acceleration convention was fixed after the candidates of the second pass were already folded. We refolded candidates with a S/N greater than 10 from the segmented acceleration searches of these second passes.

\subsubsection{Candidate selection}
As survey sensitivities and processing capabilities advance, the number of pulsar candidates is ever increasing at all stages of a pulsar search. Thanks to the high performance software employed in this survey, we were able to retrieve and fold hundreds of thousands of candidates from each observation. However, this is too many to inspect by eye -- the so-called `candidate selection problem' \citep{Lyon2016}. Part of this issue originates from being sensitive to weaker sources of RFI. However, the main source of spurious folds are so-called `noise candidates'. These arise from the background noise levels and are simply random noise fluctuations \citep{Groth1975}. With a large number of samples (long high-resolution observations) and search trials, along with strict RFI removal, our candidates are noise-dominated.

We partially classified the folded candidates  with \pics{} \citep[a Pulsar Image-based Classification System based on Machine Learning]{PICS}. We used a conservative minimum score of 10 per cent pulsar-like for both the original and TRAPUM training sets \citep{Padmanabh2023}, as well as a minimum\footnote{Pointing 3 second pass had too many candidates to view, so they were limited to an AI score of 15 per cent and a folded S/N of 7.8.} folded S/N of 7. This returned a handful of candidates for each beam, or of the order of a thousand candidates per pointing, with few RFI candidates.

We used a Graphical User Interface candidate classifier tool, \candyjar{}\footnote{\href{https://github.com/vivekvenkris/CandyJar}{https://github.com/vivekvenkris/CandyJar} by Vivek Venkatraman Krishnan}, to classify the folded candidate plots by eye and check for harmonically related candidates or known pulsars. Two or three volunteers viewed and classified the candidates as noise, RFI, known pulsar, new pulsar, or candidate pulsar. In the entire survey, seven candidates were classified as new pulsars. They are described in \autoref{pulsar_discoveries}. The folded discovery data were refolded at multiples of the period (2, 3, 1/2 and 1/3) to check that the pipeline did not retrieve a harmonic as the best detection, but this did not happen. Usually, around 50 low S/N folded candidate plots were clasified as pulsar candidates. These signals were cross-matched between segments and passes, but none was detected and classified as pulsar candidates more than once. Furthermore, from the second pass of pointings 1 and 2, weak pulsar candidates found in segments were folded over double their detection integration time, with the default \psrfoldfil{} range search around their candidate period, DM and acceleration (the DM search was removed in later pointings). This did not strenghthen any segment candidates.
Due to this low success rate, pointings 5 and 7 were not re-observed and the time was used to observe \mbox{SMCPOINTING9} and the Globular Cluster NGC\,121 (\autoref{NGC121}). Pointings 3 and 4 were only partially re-observed due to a failed observation setup (\autoref{tab:observations}). All pulsar candidates that were not observed twice are noted for potential re-observation.

\subsubsection{Multi-beam localisation method}

\label{localisation}
When a pulsar was discovered in a coherent beam, the beams in neighbouring positions were folded with the discovery parameters to look for more detections. As seen in the previous section, only the strongest detection would have been folded by the TRAPUM pipeline, as the duplicates are clustered spatially by the multi-beam filter. We thus obtained the pulsar discovery's S/N in each detected beam with \psrfoldfil{}. These detection S/Ns, beam positions and point spread functions (generated by \mosaic{}) were input into the  \seeKAT{} software \citep{SeeKAT}. \seeKAT{} can then constrain the localisation of the pulsar to better than the detected beams' combined area (each beam having an area of about 1 square arcminute at 50 per cent sensitivity). 

However, while beam positions are tiled by the FBFUSE beamformer to attain a chosen fractional sensitivity overlap when reaching half of the observation's integration time\footnote{This was in place from pointing 3. For both passes of pointings 1 and 2, the beams were immediately tiled with the requested overlap.}, the beams have a different overlap at the start and end of the observation as the observed source moves through the sky. To account for this in \seeKAT{}, we split in time the strong detections of pulsars and their S/Ns into four segments with \psrchive{} \citep{psrchive_psrfits} and generated a PSF for each segment. We could then generate a localisation for each segment and combine them with \seeKAT{}. Weaker observations could not be split into segments, as the decrease in S/N would result in unreliable detections. In these cases, we assumed that using a PSF simulated for the middle of the observation time is sufficient. We also generally removed weak segment and/or beam detections that would not strengthen the localisation.  Usually, discovery multi-beam localisations resulted in position errors of a few square arcseconds.

\subsubsection{Single pulse search}
We also performed a single pulse search on every beam\footnote{Due to network losses, two beams could not be searched for transients  in pointing 7.} from pointing 3 onwards, aiming to find giant pulses from e.g. young pulsars that are too faint to be detected in our survey, or FRBs. As described in \autoref{introduction}, two LMC pulsars are known to emit giant radio pulses, but no single pulse emission has been detected from radio pulsars in the SMC. With the two magnetars located in the Magellanic Clouds, it is worth searching for transient radio emission like that of the Milky Way FRB-emitting magnetar SGR\,1935$+$2154 \citep{Bochenek2020b}. Moreover, dwarf galaxies are known to host FRBs (see e.g. the repeating FRB\,121102, \citealt{Tendulkar2017}), even if it is unlikely that one will be active presently. 

We used the software \transientX{} \citep{men2024} to perform the searches on the raw data downsampled to a sampling time of 306\,$\upmu$s, RFI-cleaned by a \texttt{rfifind} or standard channel mask and \pulsarX{}'s \texttt{filtool}. We used a range of widths from 306\,$\upmu$s to 100\,ms over DMs 0 to 5000\dmunits{}  (using DM trial step sizes from \presto{}'s \texttt{DDplan.py} script) and a minimum single pulse S/N of 8. Depending on the RFI during the observation, between tens of candidates to several tens of thousands per observation could be returned.   Inspection of  the candidates indicated they were consistent with RFI or noise, except for two single pulses from the known SMC pulsar \fortyfive{} with a S/N of about 8. The flux density of these pulses is 17 times stronger than the average single pulse flux over the 2h \mbox{SMCPOINTING3} observation.  No single pulses were detected from any of our discoveries.

As well as this offline processing using \transientX{}, a commensal search for transients was carried out on the incoherent beam and coherent beams using the MeerTRAP backend TUSE (Transient User Supplied Equipment). This means the raw data of both passes of Pointings 1 and 2  was searched by MeerTRAP, despite being deleted before \transientX{} was run. The capabilities and pipelines for MeerTRAP are described in \cite{Rajwade2022}. All identified candidates were consistent with RFI or noise or were from the known pulsar.

\subsubsection{Future reprocessing}
Once the processing of each pointing was finished,  a downsampled version was stored for future reprocessing. The data were first RFI cleaned by \iqrm{}. A typical pointing was then reduced by a factor of four in frequency (512 channels each de-dispersed at 115\,\dmunits{}), and by a factor of eight in time (1.225\,ms samples). This reduces the storage space occupied by a typical pointing from 74 terabytes to 2.3 terabytes. We plan to search these data with a Fast Folding Algorithm (using e.g.  \citealt{riptide})  in the future, with the possibility of extending the search to higher DMs. Due to electricity failures during processing, 72 and 48 beams could not be kept in reduced resolution for the second pass of pointings 3 and 4 respectively. The raw data of all pointings were deleted immediately after processing as they were too large to store. All secondary processing data products were retained for possible future inspection.

\subsection{Sensitivity}
\label{sensitivity}
In this section, we calculate radio flux density sensitivity limits  using the radiometer equation (\autoref{eq:radiometer}) applied to pulsar observations \citep[p. 265]{handbook}:

\begin{equation}
S_{\text{min}}  = \frac{S/N_{\text{min}}  \times  ( T_{\text{sys}}  + T_{\text{sky}})  \times  \beta }{\epsilon \times G \times  \sqrt{n_{\text{pol}}  \times  t_{\text{obs}}  \times  \Delta\nu }}  \times  \sqrt{\frac{\delta}{1 - \delta}}
\label{eq:radiometer}
\end{equation} 

In \autoref{tab:sensitivity_parameters}, we list the parameters used in the sensitivity calculations for this TRAPUM survey. Additionally, the MeerKAT L-band specifications are given in \autoref{tab:MeerKAT_parameters}. We use a FFT efficiency factor $\epsilon = 0.7$ to perform a spectral to folded S/N conversion \citep{riptide}. Thus, a minimum S/N cut of 9 in the FFT corresponds to a folded S/N of approximately 13.

\begin{table}
\centering
\caption{This survey's sensitivity calculations parameters. The sky temperature in the direction of the SMC was retrieved from \protect\cite{skytemperature} using \pygsm{} \protect\citep{pygsm}. The given value of $\delta$ is the median intrinsic duty cycle from the ATNF pulsar catalogue, for radio-emitting pulsars with a measured FHWM,  excluding Globular Cluster pulsars, Millisecond Pulsars, Rapidly Rotating Radio Transients, magnetars and binary pulsars.}
\label{tab:sensitivity_parameters}
\begin{tabular}{ll}
\hline
\textbf{Parameter}                            & \textbf{Value} \\ \hline
Sky temperature $T_{\text{sky}}$ (K)                     & 7.1   \\ 
Number of frequency channels         & 2048  \\ 
Sampling time ($\upmu$s)               & 153   \\ 
Assumed pulsar duty cycle $\delta$ (per cent) & 2.5     \\ 
Minimum spectral S/N     $S/N_{\text{min}}$       & 9     \\ 
Digitisation correction factor $\beta$ (8-bit)      & 1.0     \\ \hline
\end{tabular}
\end{table}

\begin{figure}
\centering
\includegraphics[width=\columnwidth]{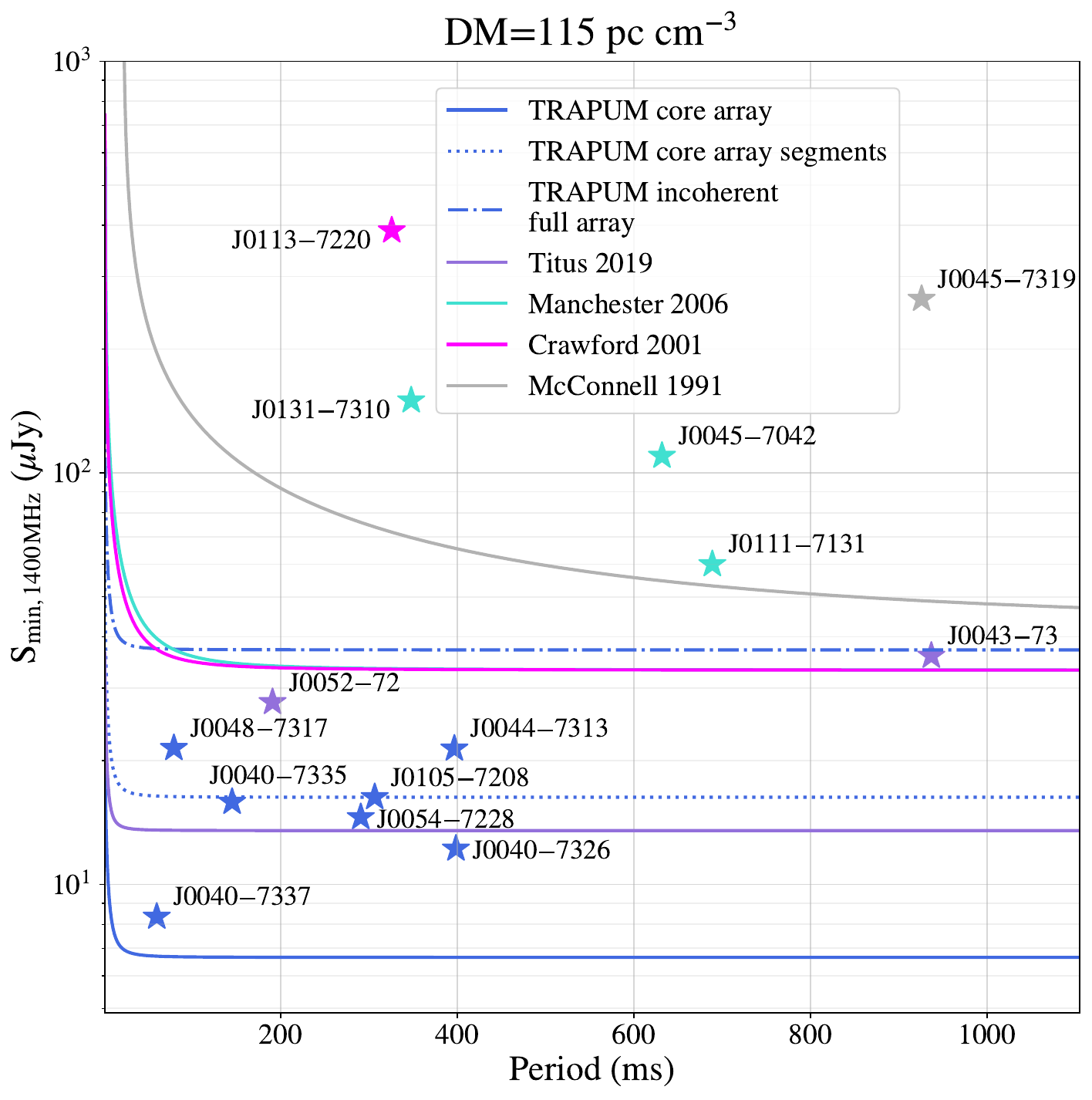}
\caption{The minimum flux density sensitivity curves and the pulsar discovery flux densities (modified by beam losses) for the pulsar surveys of the SMC. All 14 known SMC radio pulsars are plotted. The calculations used to produce this figure and interpretations are detailed in \autoref{sensitivity}.} 
\label{fig:survey-sensitivities}
\end{figure}

The best flux density sensitivity limit of this survey's full integration periodicity searches  is $S_{\text{1284\,MHz,full}} =$ 7.6\,$\upmu$Jy. We assume an integration time t$_{\text{obs}}$ of 7200\,s with the core array (44 antennas), a pulsar with a rotation period of 100\,ms (so that pulse broadening effects are negligible) placed at the centre of the pointing (incoherent beam location), and at the centre of a coherent beam (where it has maximum sensitivity).  Assuming a power law radio spectral index of $-$1.60$\pm$0.54  \citep{Jankowski2018}, our radio flux density sensitivity upper limit rescaled to 1400\,MHz is $S_{\text{1400\,MHz,full}} =$ 6.6$\pm$0.3\,$\upmu$Jy. 
This translates to a radio pseudo-luminosity upper limit of $L_{\text{pseudo,1400\,MHz}}= S_{\text{1400\,MHz}} \times D^{2} = $ 24.1$\pm$1.0\,mJy\,kpc$^{2}$, assuming an approximate distance to the SMC of $D=$ 60\,kpc \citep{Karachentsev2004}.
As seen later in \autoref{fig:survey-sensitivities}, this survey is thus twice as sensitive as the latest SMC pulsar survey by \cite{Titus2019}, conducted with Murriyang.
For the acceleration search, the segmenting of our observations reduces our sensitivity by a factor of $\sqrt{6}$, yielding  $S_{\text{1284\,MHz,segments}} =$ 18.7\,$\upmu$Jy and $S_{\text{1400\,MHz,segments}} =$ 16.3$\pm$0.8\,$\upmu$Jy with the same assumptions as before.  
Further from the pointing boresight, the sensitivity of coherent beams is reduced by the response of the incoherent beam. This is taken into account for the upper limits on the pulsed flux density of various targets presented later in \autoref{upperlimits}. Note that the incoherent beam constitutes a parallel survey with a large area and reduced gain. The full integration incoherent observations have a  flux density sensitivity limit of $S_{\text{1284\,MHz,incoherent}} =$ 42.7\,$\upmu$Jy and $S_{\text{1400\,MHz,incoherent}} =$ 37.2$\pm$1.8\,$\upmu$Jy with the same assumptions as before,  which is a similar sensitivity to the \cite{Manchester2006} and \cite{Crawford2001} surveys.

\begin{table*}
\caption{The parameters of Murriyang SMC surveys used in this work's flux density  sensitivity upper limit calculations. The system temperature in \protect\cite{McCulloch1983} is given with the sky temperature in the direction of the LMC added, that we removed using a \pygsm{} value. The spectral S/N cut in \protect\cite{Titus2019} is 4, but we assumed a minimum folded S/N value of 7 to be able to distinguish a signal by eye.} 
\begin{tabular}{ccccc}
\hline
                                             & \cite{Titus2019} & \cite{Manchester2006} & \cite{Crawford2001} & \cite{McConnell1991} \\ \hline
\textbf{Minimum S/N }    $S/N_{\text{min}}$ & 7 (folded, assumed)         & 7 (assumed spectral)              & 7 (spectral)                     & 10  (folded, assumed)                    \\ 
\textbf{Centre frequency} $\nu$ (MHz)                         & 1382                & 1374                     & 1374                   & 610                     \\ 
\textbf{Bandwidth} (MHz)      $\Delta\nu$                   & 400                 & 288                      & 288                    & 120                     \\ 
\textbf{Number of frequency channels}                           & 1024                & 96                       & 96                     & 24                      \\ 
\textbf{System temperature} $T_{\text{sys}}$ (K)                & 23  \citep{Keith2010}                & 21           \citep{Manchester2001}            & 21      \citep{Manchester2001}              & 44            \citep{McCulloch1983}           \\ 
\textbf{Sampling time} ($\upmu$s)                               & 64                  & 1000                     & 250                    & 5000                    \\ 
\textbf{Digitisation correction factor} $\beta$                       & 1.06                & 1.25                     & 1.25                   & 1.25                    \\ 
\textbf{Integration time }                                      & 15000               & 8400                     & 8400                   & 5000                    \\ 
\textbf{Sky temperature} $T_{\text{sky}}$ (K)                   & 6.6                 & 6.6                      & 6.6                    & 19.3                    \\ \hline
\end{tabular}
\label{tab:other_surveys_parameters}
\end{table*}

In \autoref{fig:survey-sensitivities}, we show the flux density sensitivity of all the SMC pulsar surveys. The parameters for these surveys, including their S/N cut, are taken from their respective papers, which we list in \autoref{tab:other_surveys_parameters}. We assume the gain of Murriyang to be 0.735 K\,Jy$^{-1}$ in all cases, taking the boresight value \citep{Manchester2001}. Two polarisations are recorded for all the Murriyang SMC surveys.  The digitisation correction factors are taken from \cite{digitisationfactors}: 1.0 for 8-bit digitisation, 1.25 for 1-bit, and 1.06 for 2-bit.  We derived the sky temperature in the direction of the SMC at each survey's central frequency using the \cite{skytemperature} model using \pygsm{} \citep{pygsm}. We rescale all survey flux density sensitivities to 1400\,MHz assuming a $-$1.60 pulsar spectral index \citep{Jankowski2018}. Note that the \cite{McConnell1991} survey had a central frequency outside of L-band, at 610\,MHz. This means that the spectral index conversion is speculative as there is a range of pulsar spectral  indices that may vary the resulting flux density upper limit. The other surveys compared are all conducted at L-band, close to a central frequency of 1400\,MHz.
We use the median DM of the SMC, 115\dmunits{}. We assume an intrinsic pulsar duty cycle of 2.5 per cent. We take into account pulse broadening due to sampling time, DM smearing \citep[p. 109]{handbook} in a frequency channel (2048 channels for this survey), and due to DM scattering -- i.e. multipath propagation through the interstellar medium \citep[using a formula from][]{Bhat2004}. Due to these effects, at shorter periods, the duty cycle $\delta$ departs from the assumed value. Temperature contributions from the ground and the atmosphere, scintillation effects, effective bandwidth (due to RFI masking and band characteristics), de-dispersion step size, and harmonic summing contributions  are not taken into account, as they are not all provided for previous surveys.

Similarly, the limiting fluence for this survey's single pulse search down to S/N$=$8 is $S_{\text{pulse,1284\,MHz}} =$ 80.4\,mJy\,ms, for a 1\,ms pulse width (and a 1\,ms integration time). We use the same assumptions as before, for a pulse originating from the centre of the pointing and at the centre of a coherent beam observed with the core array (44 antennas).

We also show in \autoref{fig:survey-sensitivities} the discovery flux densities of the SMC pulsars, scaled to 1400\,MHz assuming a $-$1.60 pulsar spectral index. This is calculated using \autoref{eq:radiometer} as before, with the values in tables \ref{tab:MeerKAT_parameters}, \ref{tab:sensitivity_parameters} and \ref{tab:other_surveys_parameters}. We use the folded S/N (thus not needing the spectral to folded S/N conversion factor $\epsilon$) and duty cycle (pulse width divided by period) of the discovery observation (see \autoref{pulsar_discoveries} and \autoref{tab:discovery_parameters}). The MeerKAT core gain is adjusted depending on the number of antennas used in the discovery observation. The observation length is approximated to 7200\,s despite the variation between pointings. Observed pulsar flux densities can be highly variable, therefore the values plotted are only valid for the discovery observation. Furthermore, the pulsar can be discovered away from the beam centre, which results in a decreased gain. To reflect discovery values, this is not taken into account in this plot. This results in an artificially decreased pulsar flux density.  Finally, the S/N can be greatly impacted by the folded data resolution (as seen later in sections \ref{PWN0048-7317} and \ref{PWN0040-7337}). We do not correct for this to reflect the search process. For the \cite{McConnell1991,Crawford2001} and \cite{Manchester2006} surveys, the pulse width and S/N in the discovery observations are not provided, thus we use directly their flux density value as an approximation. The values these authors used in \autoref{eq:radiometer} to calculate the pulsar flux densities are unknown and different from \autoref{tab:other_surveys_parameters}, and thus are not directly comparable with the sensitivity curves.
Most SMC pulsar discoveries were bright enough to be discovered in the previous survey, but were not due to survey coverage, beam losses and flux density variability.  We discuss the detectability of our discoveries in the \cite{Titus2019} survey in \autoref{pulsar_discoveries}, and the detectability of their discoveries in this survey in \autoref{titus_localisations}.

 \section{Globular Cluster NGC 121}
 \label{NGC121}
Millisecond pulsars are the fastest spinning and oldest pulsars known. They have been `recycled', i.e. spun up by accretion from a binary companion. They have science applications such as gravitational wave detection and tests of relativistic gravity \citep[see][]{Lorimer2008}. Globular Clusters are known to harbour exotic compact objects and binaries, thanks to their old age and  the high density of high-mass stars in their core that favours interactions. About 60 per cent of millisecond pulsars (here conventionally assumed to be pulsars with a period of under 10\,ms) and over 40 per cent of binary pulsars are located in Globular Clusters of the Milky Way \citep{ATNF}. As there are no known extragalactic millisecond pulsars (MSPs), the detection of the first rotationally stable extragalactic MSP could improve Pulsar Timing Arrays  by greatly increasing their longest baseline distance \citep{Lee2008}. The first extragalactic pulsar in a double neutron star would inform the extragalactic neutron star merger rate.

NGC\,121 is the only Globular Cluster of the SMC, and its oldest stellar cluster, with an age of 11 billion years. Situated 65\,kpc away on the outskirts of the dwarf galaxy, it is a mass-segregated, but not core-collapsed cluster. Its core, half-light and tidal radii are 4.2, 8.5 and 51.9\,pc respectively \citep[][and references therein]{glatt2008,Glatt2009}. The expected Milky Way Dispersion Measure contribution in its direction is as expected for the SMC (30-40\dmunits{}) according to both the \cite{YMW2016} and \cite{NE2001} electron density models. Since it is situated away from the main body of the dwarf galaxy, we can expect that the SMC DM contribution will be low. We conclude it is an ideal target to search for extragalatic MSPs or relativistic binaries.

We conducted two observations of NGC\,121 with the same backends introduced in \autoref{observations_introduction}. However, the observation parameters, shown in \autoref{tab:observations_NGC121}, are different to those of the rest of the survey (described in  \autoref{observations}). Thanks to the small angular size of the cluster, we were able to use the full array of MeerKAT, despite its much reduced beam size. We used 63 coherent beams to cover an area that extended beyond the half-light radius of the cluster in the first pass observation, and seven to cover its core in the second pass. Thanks to the low number of beams recorded, we were able to double the frequency and time resolution to 4096 channels and a sampling time of 76\,$\upmu$s. This also reduces the pulse smearing in a frequency channel due to dispersion, and the sampling time pulse smearing, that strongly affect millisecond pulsars due to their extremely short pulse length.

\begin{table*}

	\caption{The NGC\,121 observations' parameters. The pointing position is given in Right Ascension and Declination. The coherent beam ellipse semi-axes are given in arcseconds in the Right Ascension and Declination directions, with the ellipse position angle.}
	\label{tab:observations_NGC121}

\resizebox{\textwidth}{!}{
\

\begin{tabular}{ccccccc}
\hline
\begin{tabular}[c]{@{}c@{}}\textbf{Pointing name}\\\textbf{Date}\end{tabular}                                                & \textbf{Observation length (s)} & \begin{tabular}[c]{@{}c@{}}\textbf{Pointing position}\\\textbf{(J2000)}\end{tabular}                                                   & \begin{tabular}[c]{@{}c@{}}\textbf{Number of antennas}\\(coherent, incoherent)\end{tabular} & \textbf{Number of coherent beams} & \begin{tabular}[c]{@{}c@{}}\textbf{Coherent beam size}\\(50 per cent sensitivity)\end{tabular}                                                                                           \\ \hline
\begin{tabular}[c]{@{}c@{}}NGC\,121\\First pass\\19 Mar 2023\end{tabular}    & 7188                     & \begin{tabular}[c]{@{}c@{}}00\h26\m49\fs00\\ -71\textdegree{}32\arcmin10\farcs0\end{tabular} & \begin{tabular}[c]{@{}c@{}}60, 62\end{tabular}     & 63                               &                                                                        27\farcs3 8\farcs0 $-$16\fdg1                                                                                   \\[0.6cm] 
\begin{tabular}[c]{@{}c@{}}NGC\,121\\Second pass\\31 Aug 2023\end{tabular}                & 7193                     & \begin{tabular}[c]{@{}c@{}}00\h26\m49\fs00\\ -71\textdegree{}32\arcmin10\farcs0\end{tabular} & \begin{tabular}[c]{@{}c@{}}64, 64\end{tabular}     & 7                               &                                                                        19\farcs5 6\farcs7 $-$1\fdg6                                                                                        \\ \hline
\end{tabular}
}
\end{table*}
    
The central beam of the first pass observation (see \autoref{tab:observations_NGC121}) was RFI-cleaned with \texttt{filtool} then processed with \pulsarminer{}\footnote{\href{https://github.com/alex88ridolfi/PULSAR_MINER}{https://github.com/alex88ridolfi/PULSAR\_MINER} by Alessandro Ridolfi, a wrapper of \presto{} 2.1  and its GPU version: \href{https://github.com/jintaoluo/presto_on_gpu}{https://github.com/jintaoluo/presto\_on\_gpu}} v1.1.5, on the four manual processing nodes of the APSUSE cluster (\autoref{observations_introduction}). We followed the same \pulsarminer{} processing steps as detailed in \cite{Ridolfi2021}. We produced just over 6000 de-dispersed timeseries with  \texttt{prepsubband}  (using 64 subbands) from 30 to 335\dmunits{}. We chose this maximum DM as the dispersion smearing within a frequency channel becomes larger than twice our sampling time at higher DMs. We ran \texttt{accelsearch} with $z_{\text{max}} = 0$ and 200 to target isolated and binary pulsars, on segments of 15, 30, 60 minutes and the full observation, from periods spanning 0.91\,ms to 10\,s. This results in a maximum acceleration searched of 1.2 or 74\accelunits{} for a 1\,ms pulsar and 116 or 7.4$\times10^{4}$\accelunits{} for a 100\,ms pulsar, in the full observation and 15-minute segments respectively. Candidates were harmonically summed up to the eighth harmonic. A list of common local RFI fluctuation frequencies was removed from the candidates. After sifting, candidates with a DM above 10\dmunits{} and a minimum Fourier significance of 6\,$\upsigma$ were kept. The raw data were folded with these candidates' parameters, using 64 time subintegrations, 128 frequency channels, and 128 pulse profile bins.
All candidates (about 2000) were visually classified by three viewers  and about 40 were identified as faint pulsar candidates in segments or full observations. None of the segment candidates was confirmed by doubling their integration time in a \psrfoldfil{} fold optimised over the default range, or by appearing in several segments.
10 full observation candidates remained. To confirm or dismiss these candidates, a second pass observation of the core of the cluster was conducted (\autoref{tab:observations_NGC121}). All seven beams of the second pass were folded using \psrfoldfil{} at the parameters of the 10 candidates, and optimised over twice the default range. This did not yield any pulsations and the candidates were dismissed as noise. 

The full tiling of 63 coherent beams of the first pass observation was then searched following the steps described in \autoref{processing}. Some adjustments were made to the \peasoup{}  parameters: segments of 20 and 40 minutes, as well as the full observation, were searched with no acceleration, using the same de-dispersion plan described in the previous paragraph. 
The parameters of \psrfoldfil{} were also altered:  we folded candidates with S/Ns greater than 8 and subintegrations were reduced to 10\,s for the segment searches. 
These searches cumulatively returned around 800 candidates after multi-beam filtering and AI scoring. These were classified by two or three viewers. Around 15 low confidence segment candidates were found. Half of these were refolded with double their integration time, which did not strenghthen their S/N. The other half were refolded at the wrong parameters due to an oversight and could not be confirmed or discarded before the raw data were deleted. The reduced resolution stored data (subbanded at a DM of 100\dmunits{}) did not have a sufficiently high time resolution to detect them or was not available (due to an electricity failure, three beams of the first pass could not be stored). The second pass observation did not cover their position in the edge of the cluster and was also stored in reduced resolution. Consequently, no candidates were retained from the \peasoup{} segmented search of the 63 coherent beams of the first pass observation.
Single pulse searches were performed on data downsampled to 153\,$\upmu$s (first pass); or 306\,$\upmu$s and 2048 frequency channels (second pass, incoherent beam not searched). They returned candidates consistent with RFI or noise. 
    
\begin{figure}
\centering
\includegraphics[width=0.92\columnwidth]{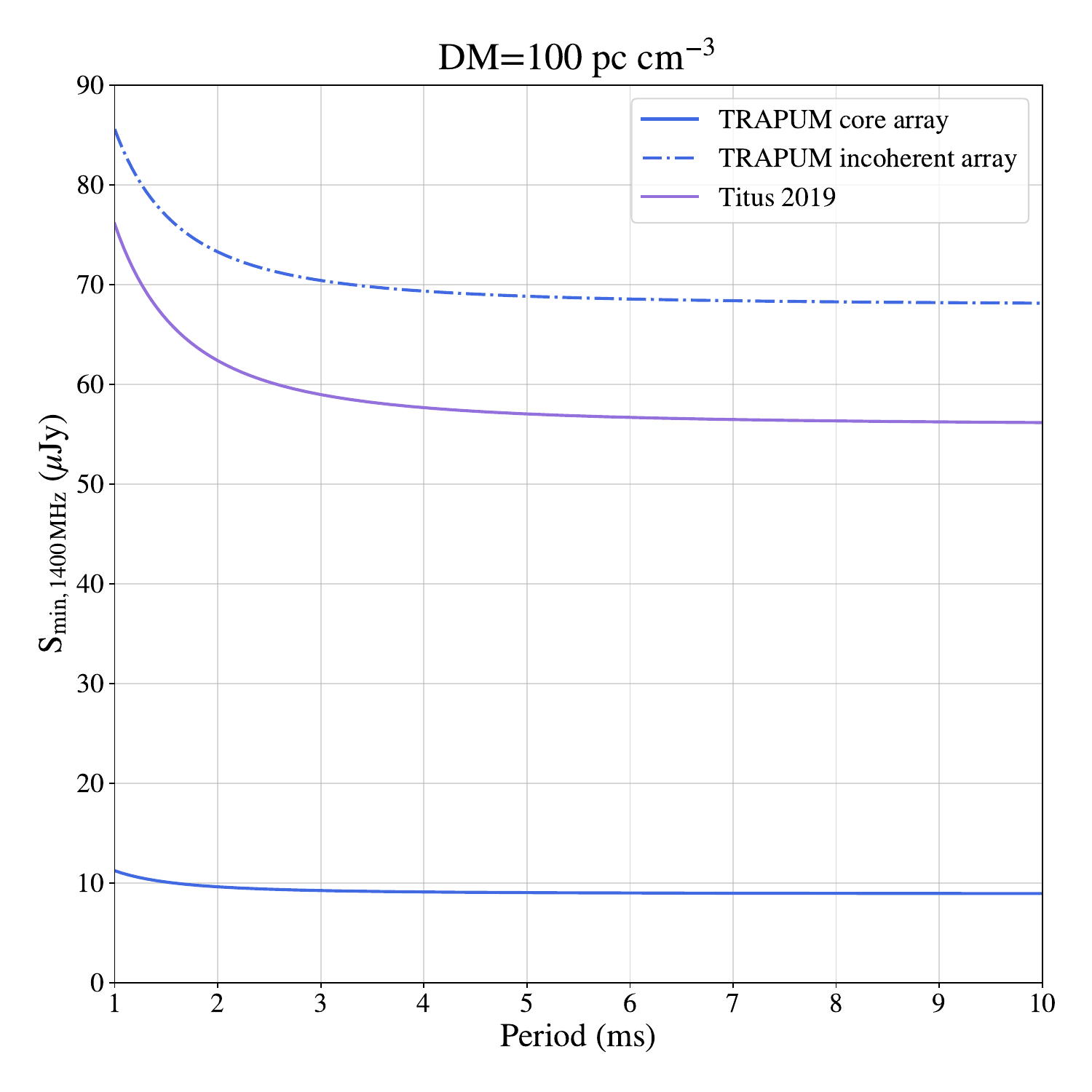}
\caption{The minimum flux density sensitivity curves for the two most recent pulsar surveys of the SMC Globular Cluster NGC\,121. The calculations used to produce this figure and interpretations are detailed in \autoref{NGC121}.} 
\label{fig:survey-sensitivities_NGC121}
\end{figure}

Using the values and assumptions detailed in \autoref{sensitivity}, we can calculate a flux density sensitivity limit with \autoref{eq:radiometer} for our \peasoup{} periodicity (no acceleration) search of the full integration, full tiling of the first pass observation of NGC\,121: $S_{\text{1284\,MHz, NGC\,121}}=10.2$\,$\upmu$Jy. We used the appropriate gain, S/N cut and sky temperature to reflect our observation. We set the DM to 100\dmunits{} and the duty cycle to 10 per cent (approximately the median duty cycle of pulsars in Globular Clusters, \citealt{ATNF}). 
Due to the very small size of the tiling (see Figure \ref{fig:NGC121}), the incoherent beam loss is negligible (less than 1 per cent) at the edge of the tiling which extends beyond the half-light radius of the cluster. This limit is valid for a pulsar with a rotation period of 50\,ms (so that pulse broadening effects are negligible).  For a pulsar with a rotation period of 1\,ms, the limit is  $S_{\text{1284\,MHz, NGC\,121, MSP}}=12.9$\,$\upmu$Jy, taking into account pulse broadening as detailed in \autoref{sensitivity}. These upper limits rescaled at 1400\,MHz are $S_{\text{1400\,MHz, NGC\,121}}=8.9\pm0.4$\,$\upmu$Jy and $S_{\text{1400\,MHz, NGC\,121, MSP}}=11.2\pm0.5$\,$\upmu$Jy. This is six times more sensitive than the latest pulsar search of this cluster, \cite{Titus2019}. 
Their beam loss for this target was 50 per cent, which is taken into account here, but this was still the most sensitive search of NGC\,121 previously.  We show the sensitivity curves for both surveys in \autoref{fig:survey-sensitivities_NGC121}.

Our flux density upper limit translates to a radio pseudo-luminosity upper limit of $L_{\text{pseudo, 1400\,MHz, MSP}}= S_{\text{1400\,MHz, NGC\,121, MSP}} \times 65$\,kpc$^{2} = $ 47$\pm$3\,mJy\,kpc$^{2}$. Only a handful of Milky Way pulsars with a period under 10\,ms have a pseudo-luminosity higher than this value \citep{ATNF}, thus potential NGC\,121 MSPs could still be too faint to be detected in this search. We reserve a more in-depth population limit of this cluster for Paper III.
The limiting fluence for the first pass single pulse search is $S_{\text{pulse,1284\,MHz}} =$ 58\,mJy\,ms, for a 1\,ms pulse with a S/N of 8 (and a 1\,ms integration time).

\begin{landscape}

\begin{figure}
\centering
\includegraphics[width=0.9\textwidth]{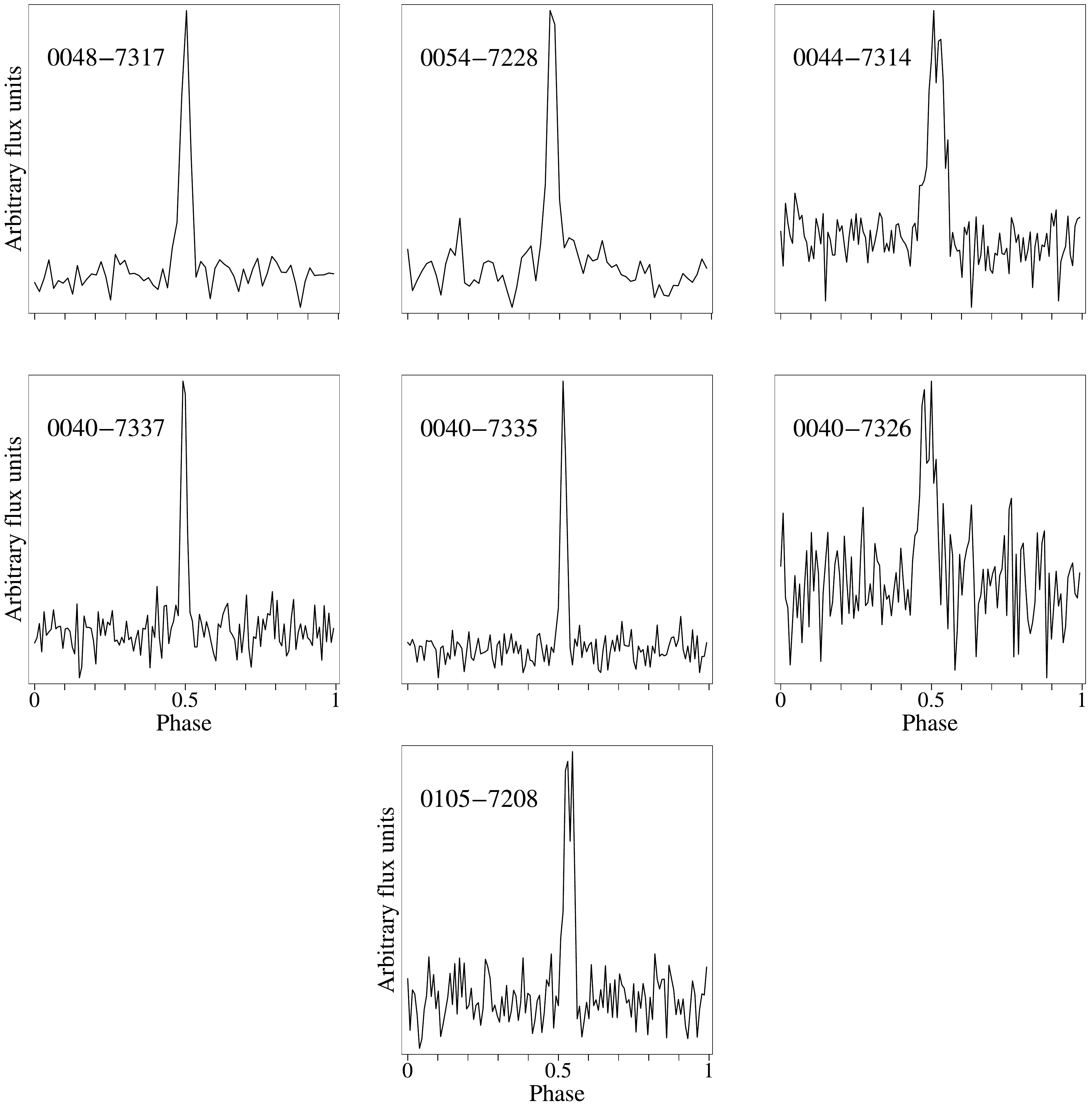}
\caption{ The discovery pulse profiles from the folded data of the \pulsarX{} pipeline candidate. The pulse smearing is due to the number of phase bins: the raw data smearing due to sampling time or dispersion within a channel is negligible compared to the  bin size. The data were folded with 64 phase bins for the first pass of pointings 1 and 2, then this increased to 128 phase bins for periods over 100\,ms. For \psrfortyfour{}  the original candidate data were corrupted by an  electricity failure, so we display the RFI-cleaned raw discovery data folded to the resolution of the pipeline.} 
\label{fig:discovery_pulse_profiles}
\end{figure}
\end{landscape}

\begin{table*}

	\caption{Archival Murriyang data from which we attempted to obtain detections of our pulsar discoveries.}
	\label{tab:archival-folds}

\resizebox{\textwidth}{!}{
\

\begin{tabular}{ccccccc}
\hline
\textbf{Pulsars}                                                       & \textbf{Survey}                                                                                        & \begin{tabular}[c]{@{}c@{}}\textbf{Project}\\\textbf{Code}\end{tabular} & \textbf{Dates}                                                                             & \begin{tabular}[c]{@{}c@{}}\textbf{Pointing}\\\textbf{Position}\end{tabular}                                                               & \begin{tabular}[c]{@{}c@{}}\textbf{Period range}\\\textbf{searched}\end{tabular} & \textbf{Notes}                                                                                                                                                      \\ \hline
\begin{tabular}[c]{@{}c@{}}\thirtyseven{}\\ \thirtyfive{}\end{tabular} & \cite{Alsaberi2019}                                                                                    & PX044                 & 14 September 2018                                                                          & \begin{tabular}[c]{@{}c@{}}00\h40\m46\fs49\\ $-$73\textdegree37\m05\farcs8\end{tabular} & 1\,ms                          &       See \autoref{PWN0040-7337}                                                                                                                                                              \\ \hline
\begin{tabular}[c]{@{}c@{}}\thirtyseven{}\\ \thirtyfive{}\end{tabular} & \cite{Titus2019}                                                                                       & P944                  & \begin{tabular}[c]{@{}c@{}}25 August 2017\\ 11 September 2017\end{tabular}                 & \begin{tabular}[c]{@{}c@{}}00\h41\m24\s\\ $-$73\textdegree39\m25\arcsec\end{tabular}     & 1\,ms                          & \begin{tabular}[c]{@{}c@{}}Viewed some of the candidates from\\the \cite{Titus2019} search.\end{tabular}                                                                                                     \\ \hline
\fortyfour{}                                                           & 70-cm pulsar survey                                                                                    & P050                  & 1991-1993                                                                                  & Multiple nearby positions                                                                & 3\,ms                          & \begin{tabular}[c]{@{}c@{}}36 observations searched\\Some compromised by RFI\end{tabular}                                                                                                                  \\ \hline
\fortyfour{}                                                           & \cite{Titus2019}                                                                                       & P944                  & \begin{tabular}[c]{@{}c@{}}25 August 2017\\ 11 September 2017\end{tabular}                 & \begin{tabular}[c]{@{}c@{}}00\h43\m26\s \\ $-$73\textdegree11\m19\arcsec\end{tabular}    & 1\,ms                          & \begin{tabular}[c]{@{}c@{}}Viewed some of the candidates from \\the \cite{Titus2019} search. \\ Over 50 per cent sensitivity los \\due to beam placement.\end{tabular} \\ \hline
\onezerofive{}                                                         & \begin{tabular}[c]{@{}c@{}}Survey subsequent to \\ \cite{Ridley2013},\\ not yet published\end{tabular} & P682                  & 25 February 2015                                                                           & \begin{tabular}[c]{@{}c@{}}01\h06\m47\fs33 \\ $-$72\textdegree07\m44\farcs5\end{tabular} & 1\,ms                          & Compromised by RFI                                                                                                                                                  \\ \hline
\onezerofive{}                                                         & \cite{Titus2019}                                                                                       & P944                  & \begin{tabular}[c]{@{}c@{}}28 August 2017\\ 12 September 2017\\ 20 March 2018\end{tabular} & \begin{tabular}[c]{@{}c@{}}01\h07\m24\s\\ $-$72\textdegree06\m42\arcsec\end{tabular}     & 1\,ms                          & \begin{tabular}[c]{@{}c@{}}Over 50 per cent sensitivity loss\\due to beam placement. \\ One pass compromised by RFI\end{tabular}                              \\ \hline
\end{tabular}
}
\end{table*}

\begin{table*}

\caption{The discovery parameters of the new pulsars of this survey in order of discovery. The errors on the \seeKAT{} localisation are given at the 2$\upsigma$ level. When an error is asymmetrical, the largest error is used, resulting in an position error region approximated to an ellipse. This approximates the localisation probability to a Gaussian distribution  \citep[see][]{SeeKAT}. For each pulsar, a preliminary estimate of the pulse width is obtained with \psrchive{}'s \texttt{pdmp} tool applied on high resolution \dspsr{} folded discovery observation data.}
\label{tab:discovery_parameters}
\resizebox{\textwidth}{!}{
\

\begin{tabular}{ccccccc}
\hline
 & \begin{tabular}[c]{@{}c@{}}\textbf{SeeKAT Right Ascension}\\\textbf{(J2000)}\end{tabular} & \begin{tabular}[c]{@{}c@{}}\textbf{SeeKAT Declination}\\\textbf{(J2000)}\end{tabular}
                  & \textbf{Period (ms)} & \textbf{Pulse width (ms)} & \begin{tabular}[c]{@{}c@{}}\textbf{Dispersion measure}\\\textbf{(\dmunits{})}\end{tabular} & \textbf{Association}                                                                                       \\ \hline
\psrfortyeight{}     & 00\h48\m56\fs2 $\pm$ 0\fs7       & $-$73\textdegree17\m46\farcs7 $\pm$ 1\farcs3 & 79.3                 & 3.3                                               & 292                                      & \begin{tabular}[c]{@{}c@{}}New PWN.\\ See \autoref{PWN0048-7317}\end{tabular}                                   \\ 
\psrfiftyfour{}      & 00\h54\m54\fs3 $\pm$ 0\fs6       & $-$72\textdegree28\m34\farcs7 $\pm$ 1\farcs3 & 290.9                & 8.0                                               & 92                                       &                                                                                                            \\ 
\psrfortyfour{}      & 00\h44\m58\fs2 $\pm$ 1\fs2        & $-$73\textdegree14\m03\farcs6 $\pm$ 1\farcs8 & 396.9                & 24.4                                              & 78                                       &                                                                                                            \\ 
\psrthirtyseven{}    & 00\h40\m48\fs6 $\pm$ 1\fs3       & $-$73\textdegree37\m07\farcs3 $\pm$ 2\farcs0 & 59.9                 & 0.8                                               & 102                                      & \begin{tabular}[c]{@{}c@{}}SNR\,DEM\,S\,5\\ and associated PWN.\\  See \autoref{PWN0040-7337}\end{tabular} \\ 
\psrthirtyfive{}     & 00\h40\m53\fs9 $\pm$ 0\fs8       & $-$73\textdegree35\m57\farcs4 $\pm$ 4\farcs5 & 145.2                & 4.0                                               & 199                                      &                                                                                                     See \autoref{PWN0040-7337}        \\ 
\psrtwentysix{}      & 00\h40\m24\fs3 $\pm$ 3\fs2       & $-$73\textdegree26\m18\farcs8 $\pm$ 8\farcs3 & 398.7                & 36.6                                              & 85                                       &                                                                                                            \\ 
\psronezerofive{}    & 01\h05\m37\fs7 $\pm$ 1\fs3       & $-$72\textdegree08\m53\farcs4 $\pm$ 2\farcs3 & 306.7                & 15.3                                              & 120                                      & \begin{tabular}[c]{@{}c@{}}See \autoref{DEMS128}\end{tabular}          \\ \hline
\end{tabular}
}

\end{table*}

\section{Results}
\label{results}
    \subsection{Pulsar discoveries}
    \label{pulsar_discoveries}
We present the discovery parameters of the seven new pulsars discovered in this survey in \autoref{tab:discovery_parameters}. The period and dispersion measure are approximate values from the \pulsarX{} discovery plots. A short description of the discovery of each pulsar is provided in the following sections, and their discovery pulse profiles are shown in \autoref{fig:discovery_pulse_profiles}. These parameters and others will be robustly estimated in Paper II of this series. All Parkes (Murriyang) Pulsar Data Archive observations\footnote{Accessed through CSIRO's Data Access Portal \href{https://data.csiro.au/domain/atnf}{https://data.csiro.au/domain/atnf}} near the position of the discoveries  were folded with \dspsr{} \citep{DSPSR} at the pulsars' period and dispersion measure. RFI was removed from the folded data with \clfd{} and optimised over a large period search range with \psrchive{}'s \texttt{pdmp} tool \citep{psrchive_psrfits}.  None of the pulsars could be detected. Notably, none of our pulsar discoveries was located within any of the survey beams of \cite{Titus2019} (considering the beam area with over 50 per cent of the full sensitivity), except for \psrthirtyseven{} which we discuss in \autoref{PWN0040-7337}. This explains why our discoveries that appear detectable in this latest Murriyang SMC survey in \autoref{fig:survey-sensitivities} were not found then. We provide the details of the folded archival observations in \autoref{tab:archival-folds}.

    \subsubsection{\psrfortyeight}
    \label{0048-7317}
\psrfortyeight{} was discovered in the full observation search of the central tiling of the first pass of \mbox{SMCPOINTING1} (see figure \ref{fig:pointing1}) with a \pulsarX{} folded S/N of 32.1. It was also detected by the search pipeline in a segment with a folded S/N of 11.2. Following the procedure described in \autoref{localisation}, the pulsar was localised using segments of six coherent beam detections. 
In the second pass of \mbox{SMCPOINTING1}, this SeeKAT localisation position was targeted with a single beam (see figure \ref{fig:pointing1-2ndpass}) and resulted in a pipeline detection in the full observation with a S/N of 55.2 and in all six segments. It was also detected by the pipeline in \mbox{SMCPOINTING9} as a high S/N harmonic detection in the full observation and in 5 segments.
We note that this pulsar has the highest Dispersion Measure known in the SMC. The DM range of the SMC previously spanned 71.3 to 205.2\dmunits{} and now extends to 292.3\dmunits{}. We discuss a new PWN association for this pulsar in \autoref{PWN0048-7317}. This pulsar has a short spin period of 79.3\,ms. No single pulses were detected from this pulsar.
    \subsubsection{\psrfiftyfour}
\psrfiftyfour{} was detected in in the full observation search of the central tiling of the first pass of \mbox{SMCPOINTING2} (see figure \ref{fig:pointing2}) with a \pulsarX{} folded S/N of 26.9. The pulsar was localised using segments of five coherent beam detections (one was removed as its S/N was too weak).
In the second pass of \mbox{SMCPOINTING2}, this position was targeted with a single beam (see figure \ref{fig:pointing2-2ndpass}) and resulted in a pipeline detection in the full observation with a S/N of 45.5 and in all six segments. It was also detected by the pipeline in the full integration of \mbox{SMCPOINTING9} with a low S/N due to being in a low sensitivity area of the incoherent beam.
    \subsubsection{\psrfortyfour}
\psrfortyfour{} was detected in the full observation search of the second pass of \mbox{SMCPOINTING1}, in a coherent beam tiling covering the position error of known pulsar \psrfortythree{} (see figure \ref{fig:pointing1-2ndpass}) with a \pulsarX{} folded S/N of 23.5. The pulsar was localised using full integrations of three coherent beams. In the first pass of \mbox{SMCPOINTING3}, this position was targeted with a single beam (see figure \ref{fig:pointing3}) but did not result in any significant detections, due to a combination of flux variability and being further away from the pointing centre (i.e. the gain at the pulsar location was reduced). Similarly, no significant detection was made in \mbox{SMCPOINTING9}. The pulsar was later confirmed in follow-up observations.

    \subsubsection{\psrthirtyseven}
    \label{0040-7337}
\psrthirtyseven{}  was detected in the full observation search of the first pass of \mbox{SMCPOINTING3}, in a beam targeted at the PWN of DEM\,S5 (see figure \ref{fig:pointing3}) with a \pulsarX{} folded S/N of 22.8.   We discuss the new PWN association for this pulsar in \autoref{PWN0040-7337}. The pulsar was localised using full integrations of four coherent beam detections (the additional beams came from the untargeted tiling the targeted beam was overlapping with) and was confirmed in follow-up observations. We show the output \seeKAT{} plot as described in \cite{SeeKAT} in \autoref{fig:0040-7337_seeKAT}. This pulsar is now the fastest spinning known radio pulsar in the SMC with a spin period of 59.9\,ms. No single pulses were detected from this pulsar.
    \subsubsection{\psrthirtyfive}
    \label{0040-7335}
\psrthirtyfive{} was detected in the full observation search of the first pass of \mbox{SMCPOINTING3} with a \pulsarX{} folded S/N of 29.3. It was detected in a beam neighbouring \psrthirtyseven{} as seen in \ref{fig:pointing3}. Owing to their very different DMs and different (not harmonically related) periods, we do not believe these two pulsars are associated, as detailed in \autoref{PWN0040-7337}. The pulsar was localised using full integrations of five coherent beams and was confirmed in follow-up observations. 
    \subsubsection{\psrtwentysix}
\psrtwentysix{}  was detected in the full observation search of the first pass of \mbox{SMCPOINTING3} (in the central tiling, see figure \ref{fig:pointing3}) with a \pulsarX{} folded S/N of 11.9, making it the weakest pipeline detection. The pulsar was localised using the full integrations of three coherent beams and was confirmed in follow-up observations.
    \subsubsection{\psronezerofive}
\psronezerofive{}  was detected in the full observation search of the first pass of \mbox{SMCPOINTING5} (in the central tiling, see figure \ref{fig:pointing5}) with a \pulsarX{} folded S/N of 22.0. The pulsar was localised using the full integrations of three coherent beams and was confirmed in follow-up observations.  Its vicinity to a SNR is discussed in \autoref{DEMS128}. 

\begin{figure}
\centering
\includegraphics[width=\columnwidth]{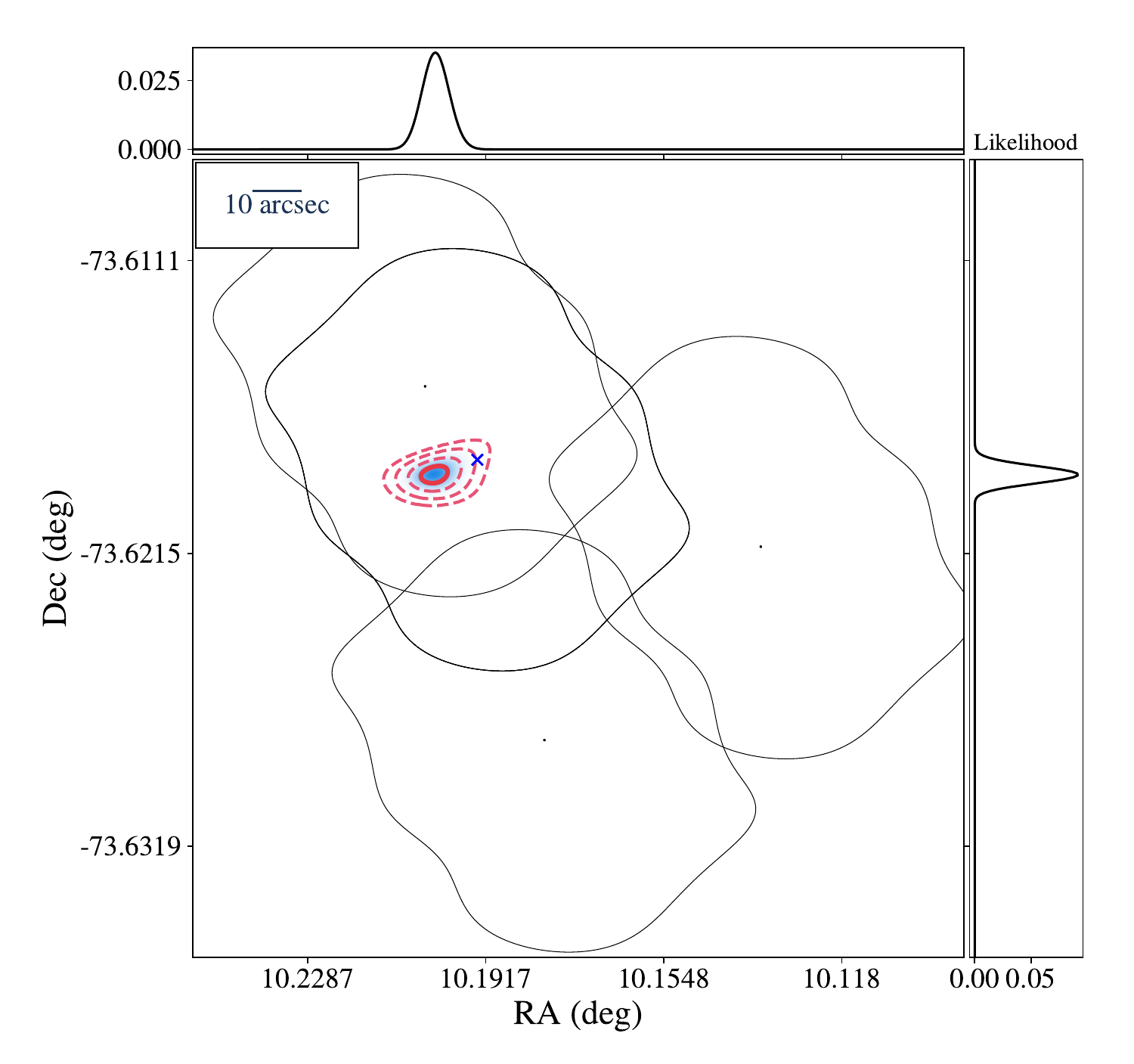}
\caption{The \seeKAT{} localisation output plot of \psrthirtyseven{}. The localisation's  1 to 4$\upsigma$ error regions are depicted in red. The coherent beams in which  \psrthirtyseven{} was detected are shown (50 per cent sensitivity contour).  The position of the radio point source resolved by \protect\cite{Alsaberi2019} in a 5500\,MHz high-resolution ATCA image (see \autoref{PWN0040-7337}) is shown as a dark blue cross.} 
\label{fig:0040-7337_seeKAT}
\end{figure}

        \subsection{Associations}
        \label{Associations}
The latest radio continuum images of the SMC \citep{Joseph2019,Cotton2024} were checked for extended emission at the locations of our new pulsars. This revealed the associations detailed in the next sections. 
            \subsubsection{A new Pulsar Wind Nebula associated with \psrfortyeight{}}
            \label{PWN0048-7317}

\begin{figure}
\centering
\includegraphics[width=\columnwidth]{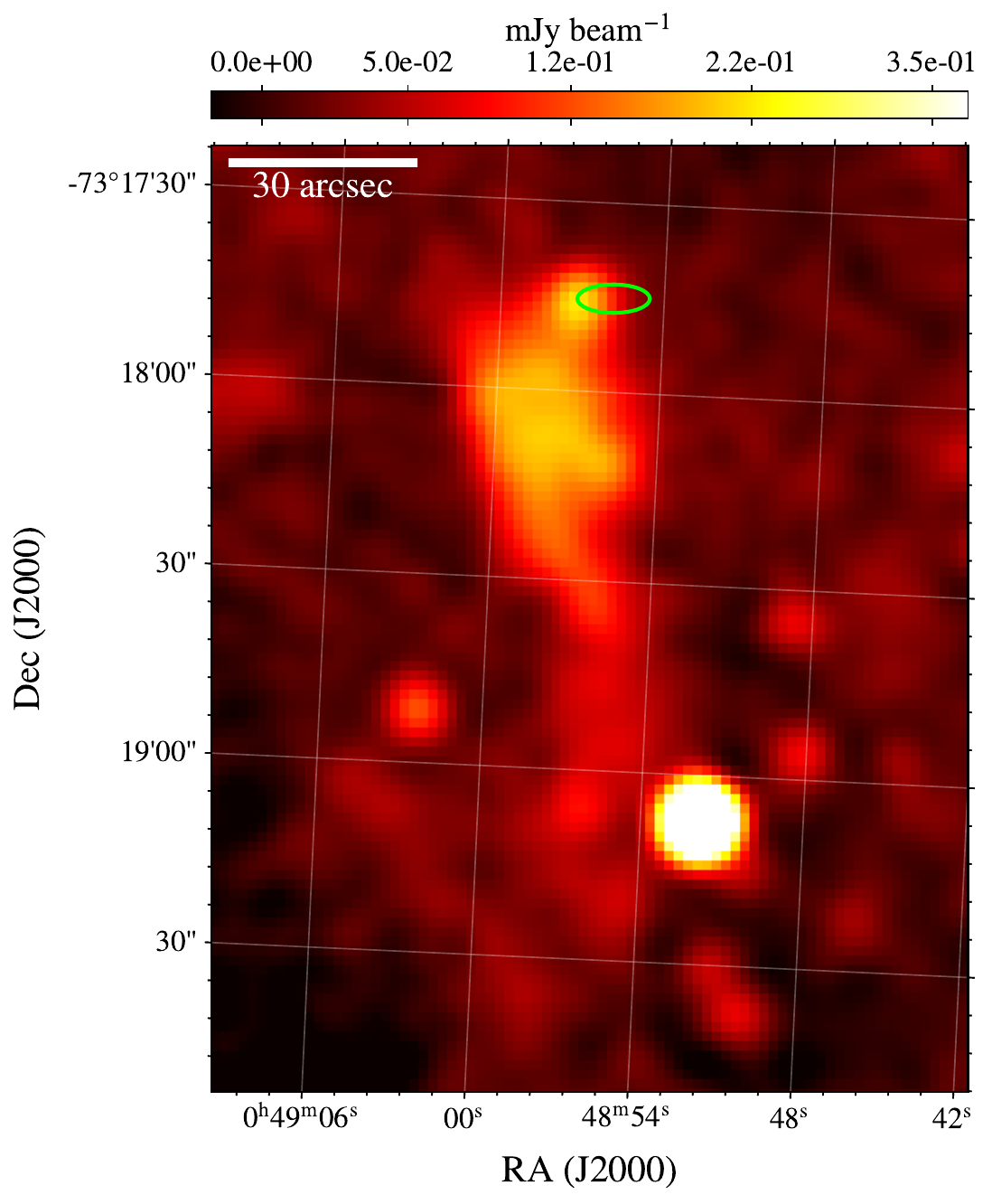}
\caption{The new Pulsar Wind Nebula associated with \psrfortyeight{} from the \protect\cite{Cotton2024} MeerKAT image of the SMC and the \seeKAT{} localisation region with its 3$\upsigma$ error approximated to a green ellipse. This localisation is consistent with the brightest emission in the head of the nebula at the top of the figure. The southern tail of the nebula overlaps with the brightest source in this figure, a steep spectrum background source.}
\label{fig:0048_new_PWN}
\end{figure}

In \cite{Cotton2024}, a new extended source was identified in the SMC. We localised  (see \autoref{0048-7317}) \psrfortyeight{} to the northern tip of the feature\footnote{A preliminary pulsar timing position was used for \psrfortyeight{} in \cite{Cotton2024}. This position is consistent with the 3$\upsigma$ error region of the \seeKAT{} localisation used in this work (\autoref{tab:discovery_parameters}). The timing position stated in \cite{Cotton2024} is centred closer to the shock front, a few arcseconds away from the brightest emission in the nebula head.}, as seen in \autoref{fig:0048_new_PWN}. We note that the beam this pulsar was discovered in was part of an untargeted tiling and so overlays this new PWN by chance. \cite{Cotton2024} suggest the pulsar could be situated at the front of a bow-shock PWN with a tail extending south of the pulsar away from the direction of motion. This local environment could be contributing to this pulsar's DM, now the largest in the SMC \citep{Straal2020}.

We can estimate the age of \psrfortyeight{} using the periods from the detections in the two passes of \mbox{SMCPOINTING1} and assuming that the difference due to positional uncertainties is small, we get a preliminary period derivative of about $3\times10^{-14}$\,s\,s$^{-1}$.
Using the standard relation we get a preliminary characteristic age of about 38\,ky. This would make \psrfortyeight{} the first known young (<100\,kyr) radio pulsar in the SMC and there are only two young radio pulsars in the LMC  (see \autoref{introduction}). We reserve a detailed analysis of this pulsar's age and energetics to Paper II.

Due to the suggested young age of the pulsar and its localisation, we classify this nebula as a PWN. This is the third such object in the SMC after IKT\,16 \citep{Maitra2015,Carli2022}  and DEM\,S5 \citep[][discussed in the next section]{Alsaberi2019}. 
With \psrthirtyseven{} described in the next section, these discoveries are the first radio pulsar-PWN systems in the SMC and only one other such system was previously known outside our galaxy: the `Crab pulsar twin' \crabtwin{} (see \autoref{introduction}).

\cite{Cotton2024} measure the size of the nebula to be about 100\,arcseconds or 29\,parsec at a distance of 60\,kpc. They also state that there is no visible SNR on their radio continuum image near the pulsar. Thus, the PWN and pulsar system has no known parent SNR, like several others, including the `Potoroo' PWN \citep{lazarevic2023}. The nearest known SNRs are IKT\,4 and IKT\,5. However, to travel from their centres to its current position in 38\,ky, the pulsar would have needed to have a projected velocity of about 1500\,km\,s$^{-1}$, which is at the limit of the known distribution of pulsar velocities \citep{Lyne1994,verbunt2017}. The tail of emission extending to the south likely marks the direction the pulsar is moving in. It points to neither of the nearby remnants. Morever, IKT\,4 is a Core-Collapse supernova, IKT\,5 is of possible Type Ia supernova origin, but both are estimated to be younger than the pulsar \citep{Maggi2019,Leahy2022}. This further favours the suggestion that the parent SNR is not visible, but a better estimate of this pulsar's age will inform this further. This is the only known extragalactic PWN with no SNR association (see \autoref{introduction}).

The SMCPOINTING1 second pass detection of \psrfortyeight{} was obtained from a coherent beam targeted at our \seeKAT{} localisation from the first pass (see \autoref{0048-7317}). We created high resolution folded data  with \dspsr{} and optimised the folding parameters with  \psrchive{}'s \texttt{pdmp} tool to obtain a S/N of 69.1 and a pulse width of 2.9\,ms. This second pass S/N is  different to the \pulsarX{} S/N in \autoref{0048-7317} due to the higher resolution data improving the signal. Similarly, the pulse width is slightly different to the discovery value in \autoref{tab:discovery_parameters}. We can estimate a preliminary flux density for \psrfortyeight{} of 56\,$\upmu$Jy at 1284\,MHz, noting however that it can vary significantly between observations\footnote{For example, this value is different to the discovery value in \autoref{fig:survey-sensitivities} due to a combination of frequency conversion, beam losses, folded data resolution, and intrinsic variability.}. We used  \autoref{eq:radiometer} and corrected the MeerKAT gain for the incoherent beam sensitivity loss (2 per cent) due to the pulsar being slightly away from the centre of the beam (see figure \ref{fig:pointing1-2ndpass}). No FFT efficiency factor $\epsilon$ is needed as we do not use a spectral S/N value.

 \cite{Cotton2024} state that no emission from the nebula could be found in images of the region in any other wavelength, suggesting the nebula is non-thermal. No high energy gamma-ray source was detected in a recent survey of the SMC by HESS \citep{Haupt2020}. A nearby X-ray source, 4XMM\,J004855.2$-$731753, is identified in the \textit{XMM-Newton} DR13 catalogue \citep{Webb2020}, but it has a low significance of about 4\,$\upsigma$. \psrfortyeight{} is located off-axis of all nearby \textit{XMM} observations: stronger emission may be revealed with targeted observations.

            \subsubsection{DEM\,S5 PWN association for \psrthirtyseven{}, and nearby pulsar \thirtyfive{}}
            \label{PWN0040-7337}
\begin{figure*}
\centering
\includegraphics[width=\textwidth]{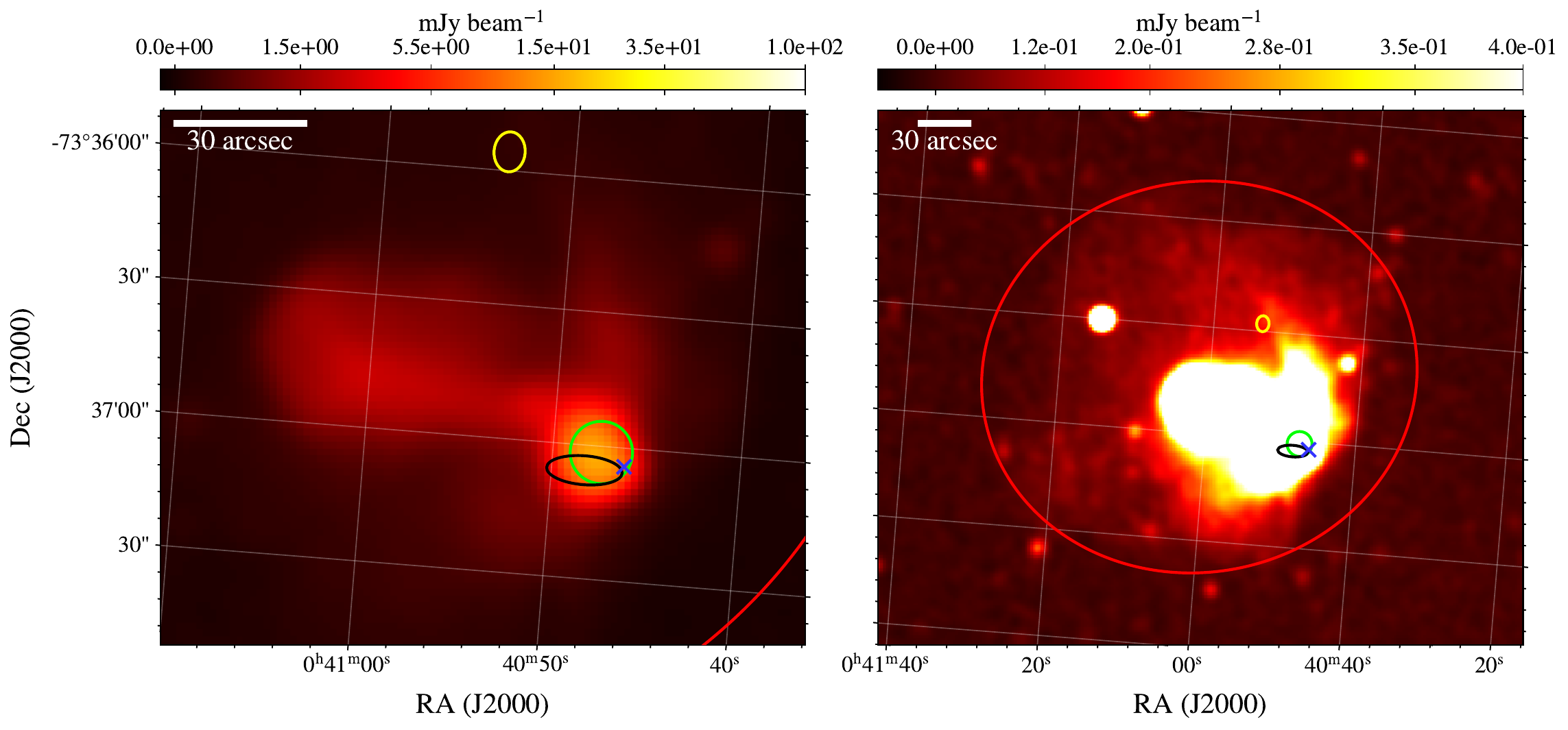}
\caption{We show the \protect\cite{Cotton2024} MeerKAT radio continuum image of the SMC around the SNR DEM\,S5. On the left, we use a color scale that shows the bow-shock PWN with a shock front, and a pulsar displacement tail as discovered in \protect\cite{Alsaberi2019}. The soft X-ray extent of the PWN from \protect\cite{Alsaberi2019} is shown in green. On the right, we use a color scale that reveals the fainter emission from the SNR, with its extent as stated in \protect\cite{Maggi2019} shown as a large red ellipse. In \mbox{SMCPOINTING3}, a coherent beam was centred on the position of the radio point source resolved by \protect\cite{Alsaberi2019} in a 5500\,MHz high-resolution ATCA image (see sections \ref{0040-7337} and \ref{PWN0040-7337}). This position is shown  as a blue cross -- the ATCA beam size is too small to display here. Our \psrthirtyseven{} \seeKAT{} discovery localisation 3$\upsigma$ error contour (approximated as a symmetrical ellipse) is shown in black. Our background pulsar discovery \psrthirtyfive{} is shown in yellow (\seeKAT{} discovery localisation 2$\upsigma$ error contour, approximated as a symmetrical ellipse).
} 
\label{fig:PWN0040-7337}
\end{figure*}

\cite{Alsaberi2019} discovered a Pulsar Wind Nebula in the SNR DEM\,S5 \citep{Haberl2000,Filipovic2008} using radio and \textit{XMM-Newton} X-ray images. In both the X-ray and radio data, they were able to resolve the PWN's soft thermal emission from the SNR's. Within the 14\,arcsecond soft X-ray PWN, they detected a confused, unresolved  hard X-ray point source; coincident with a steep spectrum, non-thermal, resolved radio point source: a putative pulsar's magnetospheric emission.  Indeed, steep spectrum radio emission from pulsars' magnetospheres is powered by their rapid rotation, which can also power hard X-ray emission when the pulsar is young and most rapidly rotating, with a characteristic hard power-law spectrum \citep[and references therein]{Becker1997}. They also resolved a 10\,pc radio bow shock nebula and tail due to the putative pulsar's fast displacement through the SNR medium from its centre, and estimated the age of the PWN-SNR system to be between 10 and 28\,kyr. The latest radio continuum image \citep{Cotton2024}  of this PWN-SNR system is shown in  \autoref{fig:PWN0040-7337}. \cite{Alsaberi2019} searched for the pulsar using Murriyang observations but could not find any pulsations. No gamma-ray source was detected in a recent survey of the SMC by HESS \citep{Haupt2020}. 

In the first pass of \mbox{SMCPOINTING3}, as detailed in \autoref{0040-7337}, we discovered \psrthirtyseven{} in a beam targeted at the position of the steep spectrum radio point source resolved by \cite{Alsaberi2019} in a 5500\,MHz ATCA image: \mbox{RA(J2000) $=$ 00\h40\m46\fs49} \mbox{Dec (J2000) $= -$73\textdegree{}37\arcmin05\farcs8}. This targeted beam was overlapping with an untargeted tiling which allowed detections in multiple coherent beams and a \seeKAT{} localisation. The localisation is shown in figures \ref{fig:PWN0040-7337} and \ref{fig:0040-7337_seeKAT}, and overlaps the position of the ATCA radio point source within 3\,$\upsigma$ (with the position error region approximated as Gaussian, see \citealt{SeeKAT}).
As the second pass of \mbox{SMCPOINTING3} failed (see \autoref{fig:pointing3-2ndpass}), this survey only has one observation of the new \psrthirtyseven{} and we cannot estimate its period derivative and age from survey observations. However, the period of the pulsar is indicative of a young pulsar, with a characteristic age of order 100\,ky or less \citep{ATNF}. We therefore believe this pulsar is associated with DEM\,S5 and powers its PWN. We reserve a detailed analysis of this pulsar's age and energetics that could strengthen this association to Paper II.

This is the first radio pulsar-SNR system in the SMC and only one other such system was previously known outside our galaxy (the `Crab pulsar twin' \crabtwin{}). There are only two known young (<100\,kyr) radio pulsars  outside our galaxy, in the LMC  (see \autoref{introduction}). If the age of this pulsar was to match that of the PWN, this would be, with \psrfortyeight{}, the first two young radio pulsars in the SMC.  
        
We created high resolution folded data of the discovery observation  with \dspsr{} and optimised the folding parameters with  \psrchive{}'s \texttt{pdmp} tool to a S/N\footnote{This discovery S/N is higher than the \pulsarX{} S/N given in \autoref{0040-7337} due to the higher resolution data improving the signal. Also, this value is different to the discovery value in \autoref{fig:survey-sensitivities} from the same observation due to a combination of frequency conversion, folded data resolution, approximating the observation time, and not taking into account incoherent beam loss.} of 37.7.  We can estimate a preliminary flux density for \psrthirtyseven{} of 16\,$\upmu$Jy at 1284\,MHz, noting however that it can vary significantly between similar observations. Using \autoref{eq:radiometer}, we corrected the MeerKAT gain for the incoherent beam sensitivity loss (2 per cent) due to the pulsar being slightly away from the centre of the beam (see figure \ref{fig:pointing3}). No FFT efficiency factor $\epsilon$ is needed as we do not use a spectral S/N value. This is much lower than the steep spectrum radio point source flux\footnote{220$\pm$7\,$\upmu$Jy at 9000\,MHz, 420$\pm$3\,$\upmu$Jy at 5500\,MHz, and a possible 3$\pm$1\,mJy measurement at 2100\,MHz.} in \cite{Alsaberi2019}, therefore it is possible that foreground emission, likely from the shock between the pulsar wind and the surrounding medium, contributed to the point source flux they measured.

We downloaded the targeted Murriyang observations conducted by \cite{Alsaberi2019} from the Parkes Pulsar Data Archive\footnote{CSIRO's Data Access Portal \href{https://data.csiro.au/domain/atnf}{https://data.csiro.au/domain/atnf}}. The bandwidth (256\,MHz), length (40 minutes) and gain were reduced compared to our discovery observation. We folded the data with high resolution using our discovery period and dispersion measure with \dspsr{}. We then optimised the S/N of the folded profile over a large period search range with \psrchive{}'s \texttt{pdmp}, but the pulsar could not be detected. As seen in \autoref{fig:survey-sensitivities}, \psrthirtyseven{} could also be below the detection threshold of the latest SMC survey, \cite{Titus2019}. We followed the same procedure as above on these observations, and the pulsar could not be detected. We note also that it was placed off-centre of the Murriyang beam, resulting in further gain loss.
         
The localisation of our new pulsar \thirtyfive{} (see \autoref{tab:discovery_parameters} and \autoref{0040-7335}) is coincident with the SNR DEM\,S5, just north of the PWN of \psrthirtyseven{} as seen in \autoref{fig:PWN0040-7337}. However, the DM of \psrthirtyfive{} is nearly double that of \psrthirtyseven{}, so we believe it is a chance alignment and not an association, with \psrthirtyfive{} in the background of the SNR. We checked the images in \cite{Alsaberi2019} for multiwavelength emission from this pulsar within the DEM S\,5 emission but none could be found.

        \subsubsection{A pulsar in DEM\,S128 but not associated?}
        \label{DEMS128}
We have noted, using  recent SMC radio continuum images \citep{Joseph2019, Cotton2024}, that our pulsar discovery \onezerofive{} is situated just at the edge of the radio extent of the supernova remnant DEM\,S128 \citep{filipovic1998a}. It is not known whether this remnant originated from a Type Ia or core-collapse supernova \citep{Roper2015} but evidence leans towards the former \citep{Maggi2019,Leahy2022}. The age of the SNR is 12\,ky according to \cite{Leahy2022}. To travel from the centre of the SNR to its position in that time, \psronezerofive{} would have needed to have a projected velocity of over 2000\,km\,s$^{-1}$, which is well beyond the known distribution of supernova birth kicks \citep{Lyne1994,verbunt2017}. Therefore, we suggest that the pulsar is not associated with the SNR. There is a central radio and X-ray source in the SNR (XMMU\,J010532.0$-$721017) that is tentatively classified as a background Active Galactic Nucleus but not excluded as a PWN \citep{Sturm2013}. There are no known X-ray sources at the pulsar position. The determination of the age of this pulsar in Paper II will inform further this potential association. 

    \subsection{Upper limits}
    \label{upperlimits}
No other pulsars were found in SNRs in this survey. In \autoref{tab:sensitivity_targets}, we show the full observation flux density upper limits on radio pulsations for all the SNRs observed and also for the magnetar \smcmagnetar{}. We use the same method as in \autoref{sensitivity}, taking into account the integration time and number of antennas of each observation (see \autoref{tab:observations}
) as well as the decrease in sensitivity away from boresight. We note that these values are valid at the centre of each coherent beam. The coherent beams were tiled to overlap at a decreased sensitivity level, usually 75 per cent.  Our upper limit on the rotation-powered X-ray pulsar \fiftyeight{} in a PWN of the supernova remnant SNR\,IKT\,16 \citep{Maitra2021} is detailed in \cite{Carli2022}.

\begin{table*}

	\caption{Flux density upper limits for various targets observed in this survey. The targets' Right Ascension and Declination coordinates are given. The most sensitive observation is used. The parameters of the observations are given in \protect\autoref{tab:observations}. The method used to calculate the limits is detailed in \protect\autoref{sensitivity}. The gain is decreased away from the centre of the pointing by the incoherent beam loss, using a beam model from \protect\cite{Asad2021}. Reference 1 is \protect\cite{Lamb2002}, 2 is \protect\cite{Maggi2019}, 3 is \protect\cite{Titus2019} and their proposal, and 4 is \protect\cite{Cotton2024}. }
	\label{tab:sensitivity_targets}

\resizebox{\textwidth}{!}{
\

\begin{tabular}{cccccc}
\hline
\textbf{\textbf{Target}}                                              & \textbf{\textbf{Best observation}}                                 & \textbf{\begin{tabular}[c]{@{}c@{}}\textbf{Target position}\\\textbf{(J2000)}\end{tabular}}    & \textbf{\textbf{Incoherent beam loss}} & \textbf{\begin{tabular}[c]{@{}c@{}}\textbf{Flux density  upper}\\\textbf{limit at 1400\,MHz ($\upmu$Jy)}\end{tabular}} & \textbf{\textbf{References and notes}}                                                                                                                        \\ \hline
Magnetar \smcmagnetar   & SMCPOINTING5                                                       & \begin{tabular}[c]{@{}c@{}}01\h00\m43\fs03\\ $-$72\textdegree{}11\arcmin33\farcs6\end{tabular} & 0.81                                   & 9.5                                                                                                        & 1                                                                                                                                                             \\[0.3cm]
SNR\,J0049$-$7314                                                     & \begin{tabular}[c]{@{}c@{}}SMCPOINTING1\\First pass\end{tabular}   & \begin{tabular}[c]{@{}c@{}}00\h49\m07\fs7\\ $-$73\textdegree14\arcmin45\arcsec\end{tabular}    & 0.99                                   & 7.9                                                                                                        & 2                                                                                                                                                             \\[0.3cm]
SNR\,J0048$-$7319                                                     & \begin{tabular}[c]{@{}c@{}}SMCPOINTING1\\First pass\end{tabular}   & \begin{tabular}[c]{@{}c@{}}00\h48\m19\fs6\\ $-$73\textdegree19\arcmin40\arcsec\end{tabular}    & 0.96                                   & 8.1                                                                                                        & 2                                                                                                                                                             \\[0.3cm]
SNR\,J0051$-$7321                                                     & \begin{tabular}[c]{@{}c@{}}SMCPOINTING1\\ First pass\end{tabular}  & \begin{tabular}[c]{@{}c@{}}00\h51\m06\fs7\\ $-$73\textdegree21\arcmin26\arcsec\end{tabular}    & 0.91                                   & 8.6                                                                                                        & 2                                                                                                                                                             \\[0.3cm]
SNR\,J0047$-$7309                                                     & \begin{tabular}[c]{@{}c@{}}SMCPOINTING1\\ First pass\end{tabular}  & \begin{tabular}[c]{@{}c@{}}00\h47\m36\fs5\\ $-$73\textdegree09\arcmin20\arcsec\end{tabular}    & 0.97                                   & 8.0                                                                                                        & 2                                                                                                                                                             \\[0.3cm]
SNR\,J0047$-$7308                                                     & \begin{tabular}[c]{@{}c@{}}SMCPOINTING1\\ First pass\end{tabular}  & \begin{tabular}[c]{@{}c@{}}00\h47\m16\fs6\\ $-$73\textdegree08\arcmin36\arcsec\end{tabular}    & 0.96                                   & 8.1                                                                                                        & 2                                                                                                                                                             \\[0.3cm]
SNR\,J0046$-$7308                                                     & \begin{tabular}[c]{@{}c@{}}SMCPOINTING1\\ First pass\end{tabular}  & \begin{tabular}[c]{@{}c@{}}00\h46\m40\fs6\\ $-$73\textdegree08\arcmin15\arcsec\end{tabular}    & 0.92                                   & 8.5                                                                                                        & 2                                                                                                                                                             \\[0.3cm]
Candidate SNR\,0049$-$7306  & \begin{tabular}[c]{@{}c@{}}SMCPOINTING1\\ First pass\end{tabular}  & \begin{tabular}[c]{@{}c@{}}00\h49\m46\s\\ $-$73\textdegree06\arcmin17\arcsec\end{tabular}      & 0.97                                   & 8.0                                                                                                        & 3                                                                                                                                                             \\[0.3cm]
SNR\,J0052$-$7236                                                     & \begin{tabular}[c]{@{}c@{}}SMCPOINTING2\\ First pass\end{tabular}  & \begin{tabular}[c]{@{}c@{}}00\h52\m59\fs9\\ $-$72\textdegree36\arcmin47\arcsec\end{tabular}    & 0.87                                   & 8.8                                                                                                        & 2                                                                                                                                                             \\[0.3cm]
SNR\,J0056$-$7209                                                     & \begin{tabular}[c]{@{}c@{}}SMCPOINTING2\\ First pass\end{tabular}  & \begin{tabular}[c]{@{}c@{}}00\h56\m28\fs1\\ $-$72\textdegree09\arcmin42\farcs2\end{tabular}    & 0.83                                   & 9.2                                                                                                        & 2                                                                                                                                                             \\[0.3cm]
SNR\,J0057$-$7211                                                     & \begin{tabular}[c]{@{}c@{}}SMCPOINTING2\\ First pass\end{tabular}  & \begin{tabular}[c]{@{}c@{}}00\h57\m49\fs9\\ $-$72\textdegree11\arcmin47\farcs1\end{tabular}    & 0.81                                   & 9.4                                                                                                        & 2                                                                                                                                                             \\[0.3cm]
Candidate SNR\,J0056$-$7219 & \begin{tabular}[c]{@{}c@{}}SMCPOINTING2\\ First pass\end{tabular}  & \begin{tabular}[c]{@{}c@{}}00\h56\m25\s\\ $-$72\textdegree19\arcmin05\arcsec\end{tabular}      & 0.95                                   & 8.0                                                                                                        & 3                                                                                                                                                             \\[0.3cm]
SNR\,J0100$-$7133                                                     & \begin{tabular}[c]{@{}c@{}}SMCPOINTING4\\ First pass\end{tabular}  & \begin{tabular}[c]{@{}c@{}}01\h00\m23\fs9\\ $-$71\textdegree33\arcmin41\arcsec\end{tabular}    & 0.72                                   & 10.6                                                                                                       & 2                                                                                                                                                             \\[0.3cm]
SNR\,J0059$-$7210                                                     & \begin{tabular}[c]{@{}c@{}}SMCPOINTING4\\ First pass\end{tabular}  & \begin{tabular}[c]{@{}c@{}}00\h59\m27\fs7\\ $-$72\textdegree10\arcmin10\arcsec\end{tabular}    & 0.79                                   & 9.7                                                                                                        & 2                                                                                                                                                             \\[0.3cm]
Candidate SNR\,J0053$-$7148 & \begin{tabular}[c]{@{}c@{}}SMCPOINTING4\\ First pass\end{tabular}  & \begin{tabular}[c]{@{}c@{}}00\h53\m07\fs8\\ $-$71\textdegree48\arcmin4\farcs1\end{tabular}     & 0.79                                   & 9.7                                                                                                        & 3                                                                                                                                                             \\[0.3cm]
SNR\,J0105$-$7223                                                     & SMCPOINTING5                                                       & \begin{tabular}[c]{@{}c@{}}01\h05\m04\fs2\\ $-$72\textdegree23\arcmin10\arcsec\end{tabular}    & 0.96                                   & 8.0                                                                                                        & 2                                                                                                                                                             \\[0.3cm]
SNR\,J0105$-$7210                                                     & SMCPOINTING5                                                       & \begin{tabular}[c]{@{}c@{}}01\h05\m30\fs5\\ $-$72\textdegree10\arcmin40\arcsec\end{tabular}    & 0.98                                   & 7.9                                                                                                        & 2                                                                                                                                                             \\[0.3cm]
SNR\,J0106$-$7205                                                     & SMCPOINTING5                                                       & \begin{tabular}[c]{@{}c@{}}01\h06\m17\fs5\\ $-$72\textdegree05\arcmin34\arcsec\end{tabular}    & 0.92                                   & 8.4                                                                                                        & 2                                                                                                                                                             \\[0.3cm]
SNR\,J0103$-$7209                                                     & SMCPOINTING5                                                       & \begin{tabular}[c]{@{}c@{}}01\h03\m17\fs0\\ $-$72\textdegree09\arcmin42\arcsec\end{tabular}    & 0.96                                   & 8.0                                                                                                        & 2                                                                                                                                                             \\[0.3cm]
SNR\,J0103$-$7201                                                     & SMCPOINTING5                                                       & \begin{tabular}[c]{@{}c@{}}01\h03\m36\fs5\\ $-$72\textdegree01\arcmin35 \arcsec\end{tabular}   & 0.89                                   & 8.7                                                                                                        & 2                                                                                                                                                             \\[0.3cm]
SNR\,J0104$-$7201                                                     & SMCPOINTING5                                                       & \begin{tabular}[c]{@{}c@{}}01\h04\m01\fs2\\ $-$72\textdegree01\arcmin52\arcsec\end{tabular}    & 0.9                                    & 8.6                                                                                                        & 2                                                                                                                                                             \\[0.3cm]
SNR\,J0103$-$7247                                                     & \begin{tabular}[c]{@{}c@{}}SMCPOINTING6\\ Second pass\end{tabular} & \begin{tabular}[c]{@{}c@{}}01\h03\m29\fs1\\ $-$72\textdegree47\arcmin33\arcsec\end{tabular}    & 0.81                                   & 9.4                                                                                                        & 2                                                                                                                                                             \\[0.3cm]
Candidate SNR\,J0109$-$7318 & \begin{tabular}[c]{@{}c@{}}SMCPOINTING6\\ Second pass\end{tabular} & \begin{tabular}[c]{@{}c@{}}01\h09\m43\fs6\\ $-$73\textdegree18\arcmin46\arcsec\end{tabular}    & 0.69                                   & 11.0                                                                                                       & 2, confirmed by 4                                                                                                                                             \\[0.3cm]
Candidate SNR\,J0106$-$7242 & \begin{tabular}[c]{@{}c@{}}SMCPOINTING6\\ Second pass\end{tabular} & \begin{tabular}[c]{@{}c@{}}01\h06\m32\fs1\\ $-$72\textdegree42\arcmin17\arcsec\end{tabular}    & 0.87                                   & 8.8                                                                                                        & 2, confirmed by 4                                                                                                                                             \\[0.3cm]
Candidate SNR\,J0112$-$7326 & SMCPOINTING7                                                       & \begin{tabular}[c]{@{}c@{}}01\h12\m37\s\\ $-$73\textdegree26\arcmin05\arcsec\end{tabular}      & 0.83                                   & 9.2                                                                                                        & 3                                                                                                                                                             \\[0.3cm]
SNR\,J0127$-$7333                                                     & \begin{tabular}[c]{@{}c@{}}SMCPOINTING8\\ Second pass\end{tabular} & \begin{tabular}[c]{@{}c@{}}01\h27\m44\fs1\\ $-$73\textdegree33\arcmin01\arcsec\end{tabular}    & 0.94                                   & 8.1                                                                                                        & 2                                                                                                                                                             \\[0.3cm]
New SNR\,J0049$-$7322       & \begin{tabular}[c]{@{}c@{}}SMCPOINTING1\\ Second pass\end{tabular} & \begin{tabular}[c]{@{}c@{}}00\h49\m57\fs1\\ $-$73\textdegree22\arcmin23\farcs7\end{tabular}    & 0.94                                   & 9.1                                                                                                        & 4, partial coverage (see \autoref{fig:pointing1-2ndpass})                                                                                                     \\[0.3cm]
New SNR\,J0050$-$7238       & SMCPOINTING9                                                       & \begin{tabular}[c]{@{}c@{}}00\h50\m35\fs3\\ $-$72\textdegree38\arcmin38\farcs3\end{tabular}    & 0.89                                   & 8.6                                                                                                        & 4, partial coverage (see \autoref{fig:pointing9})                                                                                                             \\[0.3cm]
New SNR\,J0112$-$7303      & \begin{tabular}[c]{@{}c@{}}SMCPOINTING6\\ Second pass\end{tabular} & \begin{tabular}[c]{@{}c@{}}01\h12\m18\fs8\\ $-$73\textdegree04\arcmin23\farcs8\end{tabular}    & 0.72                                   & 10.6                                                                                                       & \begin{tabular}[c]{@{}c@{}}4, we only observed a steep  \\ spectrum source inside  the SNR\\identified in \cite{Joseph2019}: \\ EMU-ESP-SMC-5407\end{tabular} \\ \hline
\end{tabular}

}
\end{table*}

    \subsection{Localisation of two Murriyang discoveries}
    \label{titus_localisations}
The two pulsars discovered in \cite{Titus2019} were not localised: their position was not constrained within the 14\,arcminute diameter Murriyang beam in which they were detected. In the second pass of Pointings 1 (\psrfortythree{}, figure \ref{fig:pointing1-2ndpass}) and 2 (\psrfiftytwo{}, figure \ref{fig:pointing2-2ndpass}), we tiled these error regions with about 170 MeerKAT coherent beams.  \psrfiftytwo{}  was detected by the full integration search pipeline with a \pulsarX{} folded S/N of 15.5. Using three full integration detections in neighbouring beams, we localised the pulsar with \seeKAT{}  to \mbox{RA(J2000) $=$ 00\h51\m36\fs5 $\pm$ 1\fs5} \mbox{Dec(J2000) $= -$72\textdegree04\m26\farcs1}  $\pm$3\farcs5 (2$\upsigma$ error).  \psrfortythree{} was not detected by the pipeline. We used \citealt{Titus2019}'s discovery period and DM to fold every beam in the tiling, optimised over the default \psrfoldfil{} range. This yielded a single detection in an edge beam of the tiling with a \pulsarX{} folded S/N of 9.5 (which is below our search pipeline limit of folded S/N 13, see \autoref{sensitivity}). A second observation was necessary to obtain simultaneous detections in several coherent beams. In the first pass of \mbox{SMCPOINTING3}, we observed a tiling of 45 coherent beams overlapping at the 70 per cent sensitivity level centred on the position of the \mbox{SMCPOINTING1} detected beam. The pipeline again did not yield any detections. We used our \mbox{SMCPOINTING1} detection period and DM to fold every beam in the tiling, optimised over the default \psrfoldfil{} range. This recovered four coherent beam detections, the strongest of which had a \pulsarX{} folded S/N of 12.9. Using full integrations, we localised the pulsar with \seeKAT{} to \mbox{RA(J2000) $=$ 00\h43\m17\fs6 $\pm$ 3\fs9} \mbox{Dec(J2000) $= -$73\textdegree19\m47\farcs5 $\pm$ 15\farcs0}, just outside the Murriyang survey beam area with over 50 per cent sensitivity. We note that \psrfortythree{}'s flux density has  decreased by about a factor of 10 since its discovery in \cite{Titus2019}. Indeed, \psrfortythree{} could not be detected by our pipeline despite its discovery flux (shown in \autoref{fig:survey-sensitivities}) being several times brighter than our sensitivity limit. Furthermore, its intrinsic flux upon discovery was at least twice as strong as reported in \cite{Titus2019} as we have localised it to outside the Murriyang beam 50 per cent sensitivity contour. These pulsars are now named PSR\,J0043$-$7319 and J0051$-$7204.

\enlargethispage{\baselineskip} 
\section{Summary}
\label{summary}
This paper described a radio-domain search for accelerated pulsars and transients in the Small Magellanic Cloud, a low-metallicity galaxy with recent star formation history expected to host a population of young compact objects. 
The survey, conducted at L-band with the core array of MeerKAT in 2-h integrations, took advantage of TRAPUM's beamforming capabilities to simultaneously cover the Supernova Remnants of the \cite{Maggi2019} census, as well as point sources and untargeted areas.

About 800 coherent beams were recorded for each observation. Each observation's large dataset of high-resolution time and frequency domain records was processed locally on the TRAPUM computing cluster and yielded considerable numbers of candidates. This number was reduced using new software adapted to the sizeable outputs of interferometric surveys.
The survey is generally about twice as sensitive as the latest SMC pulsar search, conducted at Murriyang. Individual upper limits on pulsed emission from Supernova Remnants are provided where no discovery was made, reaching down to 8\,$\upmu$Jy at 1400\,MHz.

We reported the discovery of seven new SMC pulsars, doubling this galaxy's known radio pulsar population and increasing the total extragalactic population by nearly a quarter. 
Localisations were computed upon discovery using multi-beam techniques  to a position error of a few arcseconds. We surveyed the position error region of two previous Murriyang discoveries to localise them in the same manner.
Further, we conducted an expansive search for accelerated millisecond pulsars in the SMC Globular Cluster NGC\,121 using the full array of MeerKAT. This improved by a factor of six the previous upper limit on pulsed radio emission from this cluster, and started to probe into the high end of known millisecond pulsar luminosities.
    
Finally, our discoveries revealed the first radio pulsar-PWN systems in the SMC, with only one such system previously known outside our galaxy (the Crab pulsar twin in the Large Magellanic Cloud, PSR\,J0540$-$6919). In particular, we presented a new young pulsar with a 79\,ms period, \psrfortyeight{}, powering a Pulsar Wind Nebula recently discovered in a MeerKAT radio continuum image. We also associated the 59\,ms pulsar discovery \psrthirtyseven{}, now the fastest spinning pulsar in the SMC, to the bow-shock PWN of Supernova Remnant DEM\,S5.

The timing solutions and peculiarities of our discoveries will be reported in Paper II of this series. The discoveries will enable an analysis of  the impact of the low-metallicity, recent star formation environment  of the SMC on its neutron star population and comparisons with predictions. We will present this population study in Paper III of this series.

\section*{Acknowledgements}

The MeerKAT telescope is operated by the South African Radio Astronomy Observatory (SARAO), which is a facility of the National Research Foundation, an agency of the Department of Science and Innovation. SARAO acknowledges the ongoing advice and calibration of GPS systems by the National Metrology Institute of South Africa (NMISA) and the time space reference systems department of the Paris Observatory.

TRAPUM observations used the FBFUSE and APSUSE computing clusters for data acquisition, storage and analysis. These clusters were funded and installed by the Max-Planck-Institut für Radioastronomie and the Max-PlanckGesellschaft.

The Parkes `Murriyang' radio telescope is part of the Australia Telescope National Facility which is funded by the Australian Government for operation as a National Facility managed by CSIRO (the Commonwealth Scientific and Industrial Research Organisation). We acknowledge the Wiradjuri people as the Traditional Owners of the Observatory site.

EC acknowledges funding from the United Kingdom's Research and Innovation Science and Technology Facilities Council (STFC) Doctoral Training Partnership, project reference 2487536. For the purpose of open access, the author has applied a Creative Commons Attribution (CC BY) licence to any Author Accepted Manuscript version arising. 

EDB, MK, PVP, WB, VVK and AR acknowledge continuing support from the Max Planck society. 

RPB acknowledges support from the ERC under the European Union's Horizon 2020 research and innovation programme (grant agreement No. 715051; Spiders).

MB and AP acknowledge funding from the INAF Large Grant 2022 `GCjewels' (P.I. Andrea Possenti) approved by the Presidential Decree 30/2022.

AR is supported by the Italian National Institute for Astrophysics (INAF) through an `IAF - Astrophysics Fellowship in Italy' fellowship (Codice Unico di Progetto: C59J21034720001; Project `MINERS').

Some of the plots in this publication were made with APLpy, an open-source astronomy plotting package for Python \citep{Robitaille2012}. We thank Laura Driessen for her help with this package.

This research has made use of the SIMBAD database, operated at CDS, Strasbourg, France \citep{SIMBAD}, NASA’s \href{https://ui.adsabs.harvard.edu/}{Astrophysics Data System} Bibliographic Services, the \href{https://www.herta-experiment.org/frbstats/catalogue}{FRBSTATS} catalogue, and the ATNF pulsar catalogue version 1.69.

EC thanks Avishek Basu, Laila Vleeschower Calas and James Turner for their help with pulsar software.

\section*{Data Availability}
The data underlying this article will be shared upon reasonable request to the TRAPUM collaboration.

\bibliographystyle{mnras}
\bibliography{main}

\appendix
\section{Steep spectrum sources observed}

\begin{landscape}

\begin{table}

\caption{The steep spectrum continuum point sources targeted with a single beam in this survey, from the \protect\cite{Joseph2019} ASKAP EMU SMC image. We selected all radio continuum point sources in their catalogue with three or more flux density measurements that had a spectral index lower than $-1$ and observed those that fell in our pointings from \mbox{SMCPOINTING3} onwards. More steep spectrum sources were observed, with beams overlapping their position while targeted at different sources (see Appendix \ref{beam_maps}).} 
\label{tab:steep_spectrum_sources}

\resizebox{1.3\textwidth}{!}{
\

\begin{tabular}{ccccccc}
\hline
\begin{tabular}[c]{@{}c@{}}\textbf{SMCPOINTING3}\\First pass only\end{tabular} & \begin{tabular}[c]{@{}c@{}}\textbf{SMCPOINTING4}\\First pass only\end{tabular} & \begin{tabular}[c]{@{}c@{}}\textbf{SMCPOINTING5}\end{tabular} & \begin{tabular}[c]{@{}c@{}}\textbf{SMCPOINTING6}\\Two passes\end{tabular} & \begin{tabular}[c]{@{}c@{}}\textbf{SMCPOINTING7}\end{tabular} & \begin{tabular}[c]{@{}c@{}}\textbf{SMCPOINTING8}\\Two passes\end{tabular} & \begin{tabular}[c]{@{}c@{}}\textbf{SMCPOINTING9}\end{tabular} \\ \hline
EMU-ESP-SMC-2115                                                               & EMU-ESP-SMC-4199                                                               & EMU-ESP-SMC-4972                                              & EMU-ESP-SMC-5171                                                          & EMU-ESP-SMC-5818                                              & EMU-ESP-SMC-6724                                                          & EMU-ESP-SMC-3650                                              \\ 
EMU-ESP-SMC-2087                                                               & EMU-ESP-SMC-3872                                                               & EMU-ESP-SMC-5026                                              & EMU-ESP-SMC-5283                                                          & EMU-ESP-SMC-6104                                              & EMU-ESP-SMC-6919                                                          & EMU-ESP-SMC-3591                                              \\ 
EMU-ESP-SMC-2277                                                               & EMU-ESP-SMC-4115                                                               & EMU-ESP-SMC-5110                                              & EMU-ESP-SMC-5345                                                          & EMU-ESP-SMC-6088                                              & EMU-ESP-SMC-6832                                                          & EMU-ESP-SMC-3370                                              \\ 
EMU-ESP-SMC-1936                                                               & EMU-ESP-SMC-3939                                                               & EMU-ESP-SMC-4335                                              & EMU-ESP-SMC-4855                                                          & EMU-ESP-SMC-6340                                              & EMU-ESP-SMC-6682                                                          & EMU-ESP-SMC-3500                                              \\ 
EMU-ESP-SMC-2310                                                               & EMU-ESP-SMC-3668                                                               &                                                               & EMU-ESP-SMC-5407                                                          & EMU-ESP-SMC-6286                                              & EMU-ESP-SMC-6639                                                          & EMU-ESP-SMC-2992                                              \\ 
EMU-ESP-SMC-1868                                                               & EMU-ESP-SMC-3836                                                               &                                                               & EMU-ESP-SMC-4677                                                          & EMU-ESP-SMC-6491                                              & EMU-ESP-SMC-6631                                                          & EMU-ESP-SMC-2643                                              \\ 
EMU-ESP-SMC-2177                                                               & EMU-ESP-SMC-3484                                                               &                                                               & EMU-ESP-SMC-4560                                                          & EMU-ESP-SMC-6412                                              & EMU-ESP-SMC-6719                                                          & EMU-ESP-SMC-3746                                              \\ 
                                                                               & EMU-ESP-SMC-3203                                                               &                                                               & EMU-ESP-SMC-4419                                                          & EMU-ESP-SMC-6362                                              & EMU-ESP-SMC-7146                                                          & EMU-ESP-SMC-2602                                              \\ 
                                                                               & EMU-ESP-SMC-3558                                                               &                                                               & EMU-ESP-SMC-4444                                                          & EMU-ESP-SMC-6338                                              & EMU-ESP-SMC-7133                                                          & EMU-ESP-SMC-3375                                              \\ 
                                                                               &                                                                                &                                                               & EMU-ESP-SMC-5238                                                          &                                                               & EMU-ESP-SMC-7212                                                          & EMU-ESP-SMC-3232                                              \\ 
                                                                               &                                                                                &                                                               &                                                                           &                                                               &                                                                           & EMU-ESP-SMC-2575                                              \\ 
                                                                               &                                                                                &                                                               &                                                                           &                                                               &                                                                           & EMU-ESP-SMC-2552                                              \\ 
                                                                               &                                                                                &                                                               &                                                                           &                                                               &                                                                           & EMU-ESP-SMC-2981                                              \\ \hline
\end{tabular}
}
\end{table}
\end{landscape}

\section{Beam layout maps}
\label{beam_maps}
In the following figures \ref{fig:pointing1} to \ref{fig:NGC121_2ndpass}, we show the layout of the MeerKAT coherent beams for each observation, aimed at targets detailed in \autoref{observations}. We will show the full survey coverage of the SMC in Paper III. The regions are overlaid on the \cite{Joseph2019} ASKAP EMU radio continuum image of the SMC shown in greyscale.  The beam positions were retrieved from the FBFUSE record of the observation. A high resolution coherent beam PSF was simulated with the latest version of \mosaic{} at the centre coordinates of the pointing, at the central time of the observation, with the antennas used during the observation and at the central frequency of L-band. An ellipse was then fitted by \mosaic{} to the PSF fractional sensitivity level chosen for the tiling overlap. This is the beam size displayed on the figures.

The beam placement and packing is not optimal for all observations. This can be due to several reasons. First, for both passes of pointings 1 and 2, the beams were immediately tiled with the requested overlap. The tilings' overlap departs from the requested value as the observed sources move through the sky during the observation. From pointing 3 onwards, the beams were tiled so that they would reach the desired overlap in the middle of the observation, giving minimal deviation from this optimal tiling. Secondly, the beam shape is simulated with \mosaic{} at very low resolution during the observation setup by FBFUSE, which results in an approximation of the true beam shape. Finally, for both passes of pointings 1 and 2, \mosaic{} used a Gaussian fitting technique which was later replaced by a more performant contour ellipse fitting.

\clearpage

\begin{landscape}

\begin{figure}

\begin{subfigure}{.67\textwidth}
    \includegraphics[width=\linewidth]{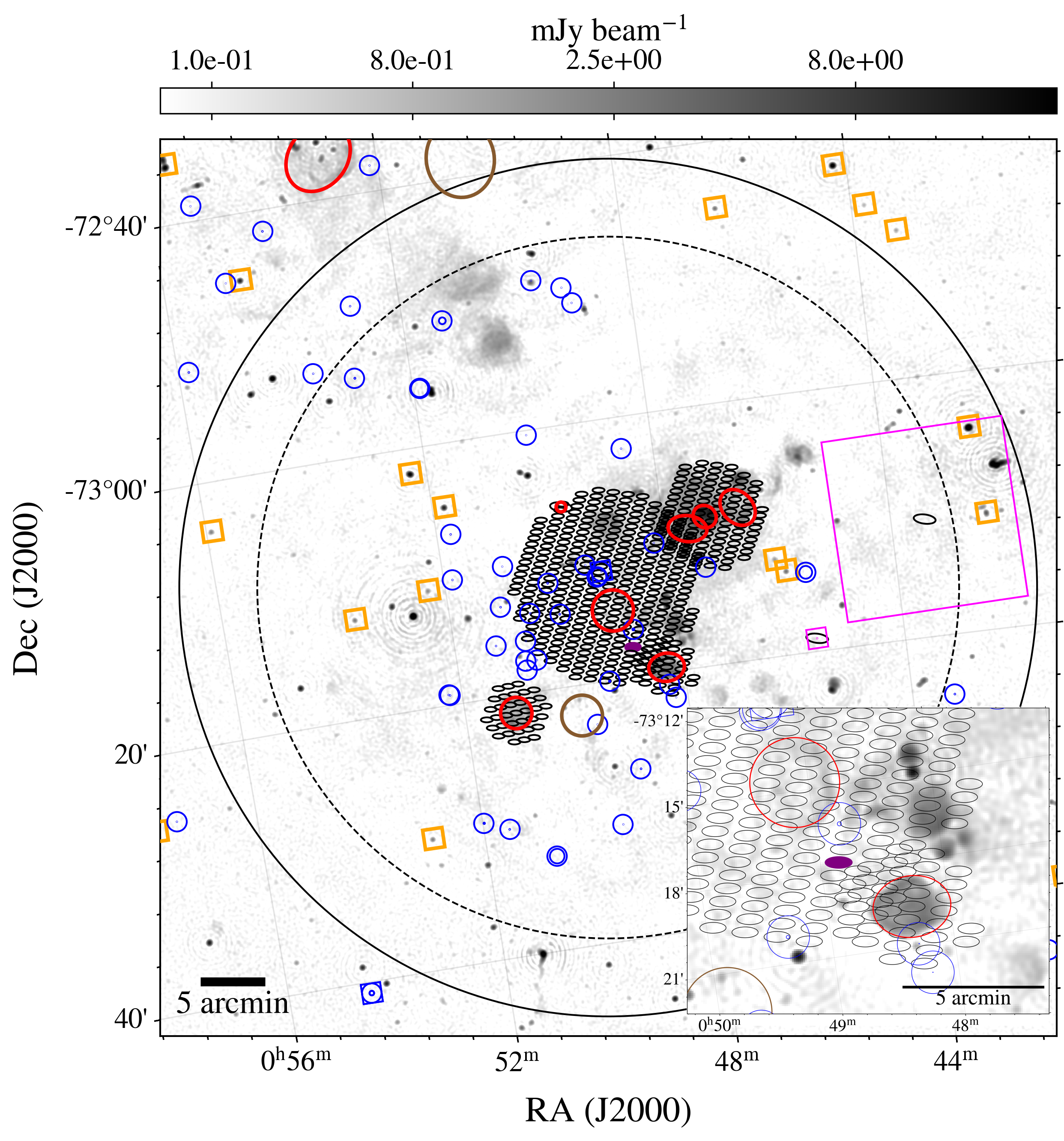}
    \caption{The first pass of \mbox{SMCPOINTING1}.}
    \label{fig:pointing1}
\end{subfigure}%
\begin{subfigure}{.67\textwidth}
    \includegraphics[width=\linewidth]{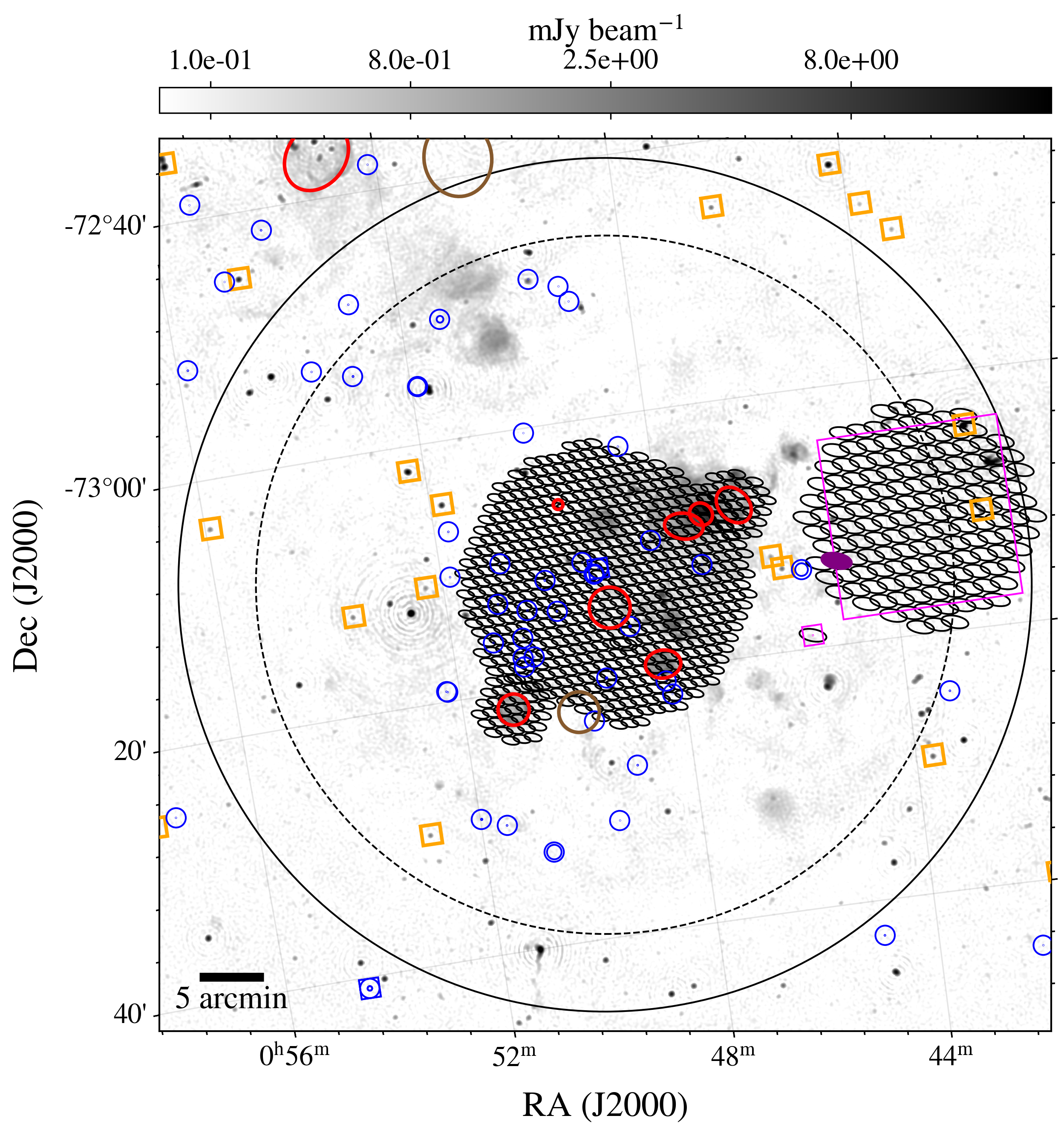}
    \caption{The second pass of \mbox{SMCPOINTING1}.}
    \label{fig:pointing1-2ndpass}
\end{subfigure}

\label{fig:pointing1-both-passes}

\caption{(a): The MeerKAT core array coherent beams observed in the first pass of  \mbox{SMCPOINTING1} are shown as small black ellipses. The large black circles show the extent of the full array MeerKAT incoherent beam with a FWHM of 1.1\,degrees (solid line) at the centre frequency of L-band  and 0.9\,degrees (dashed line) at the  highest frequency of the band \protect\citep{Asad2021}. Regions are overlaid on the \protect\cite{Joseph2019} ASKAP EMU radio continuum image of the SMC at 1320\,MHz. Supernova remnants (bona fide or candidate, see \autoref{tab:sensitivity_targets}) are displayed as red ellipses. Five supernova remnants are observed with a 75 per cent coherent beam overlap requested and a slightly lower overlap obtained (see Appendix \ref{beam_maps}) and one with a single beam displayed at 50 per cent contour. Two SNRs from \protect\cite{Cotton2024} are depicted as brown ellipses. Most of these new SNRs discovered with MeerKAT were not observed as they were not known at the time of observation. The remaining untargeted beams were tiled at boresight where sensitivity is highest, and overlapped with one SNR. One pulsar was discovered in these beams: \psrfortyeight{} (see \autoref{pulsar_discoveries} and \autoref{PWN0048-7317}). The strongest S/N discovery beam is highlighted in purple. Single beams (shown at 50 per cent sensitivity contour) were placed on the known pulsar \fortyfive{} (small pink box with position error box inside) and the published position of \psrfortythree{}. However, the error on the position of the latter (large pink box covering the Murriyang beam area) was not taken into account, and the pulsar was not detected in this pass of \mbox{SMCPOINTING1}. Steep spectrum point sources from \protect\cite{Joseph2019} (orange boxes) were not yet targeted and none was covered by our coherent beam tilings. HXMBs from the \protect\cite{Haberl2016} census are displayed as dark blue circles with their position error box inside. Those with a position error larger than 100\arcsec are omitted. HMXBs with a period of under 10\,s (dark blue boxes, see \autoref{tab:fast_HMXBs}) were not yet targeted but one was observed in the central tiling: SXP\,9.13. In the top left corner, one SNR (red ellipse) observed in \mbox{SMCPOINTING2} is visible. This figure was generated with the \href{http://aplpy.github.io/}{\aplpy{}} Python package.
\\
(b): The second pass of \mbox{SMCPOINTING1}. The regions displayed are as detailed in \autoref{fig:pointing1}. Due to the larger beam sizes, a candidate SNR (red circle) and part of a SNR newly discovered in \protect\cite{Cotton2024} (brown circle) were observed in the central tiling. The position error area of known pulsar \fortythree{} was tiled with beams overlapping at 50 per cent sensitivity (covering two untargeted steep spectrum point sources). This resulted in the detection and localisation of the latter (see \autoref{titus_localisations}) as well as the discovery of \psrfortyfour{} (purple beam, see \autoref{pulsar_discoveries}). A single beam, pictured at 50 per cent sensitivity, was placed on the \seeKAT{} localisation position of \psrfortyeight{}. } 
\end{figure}

\end{landscape}

\begin{landscape}

\begin{figure}

\begin{subfigure}{.68\textwidth}
    \includegraphics[width=\linewidth]{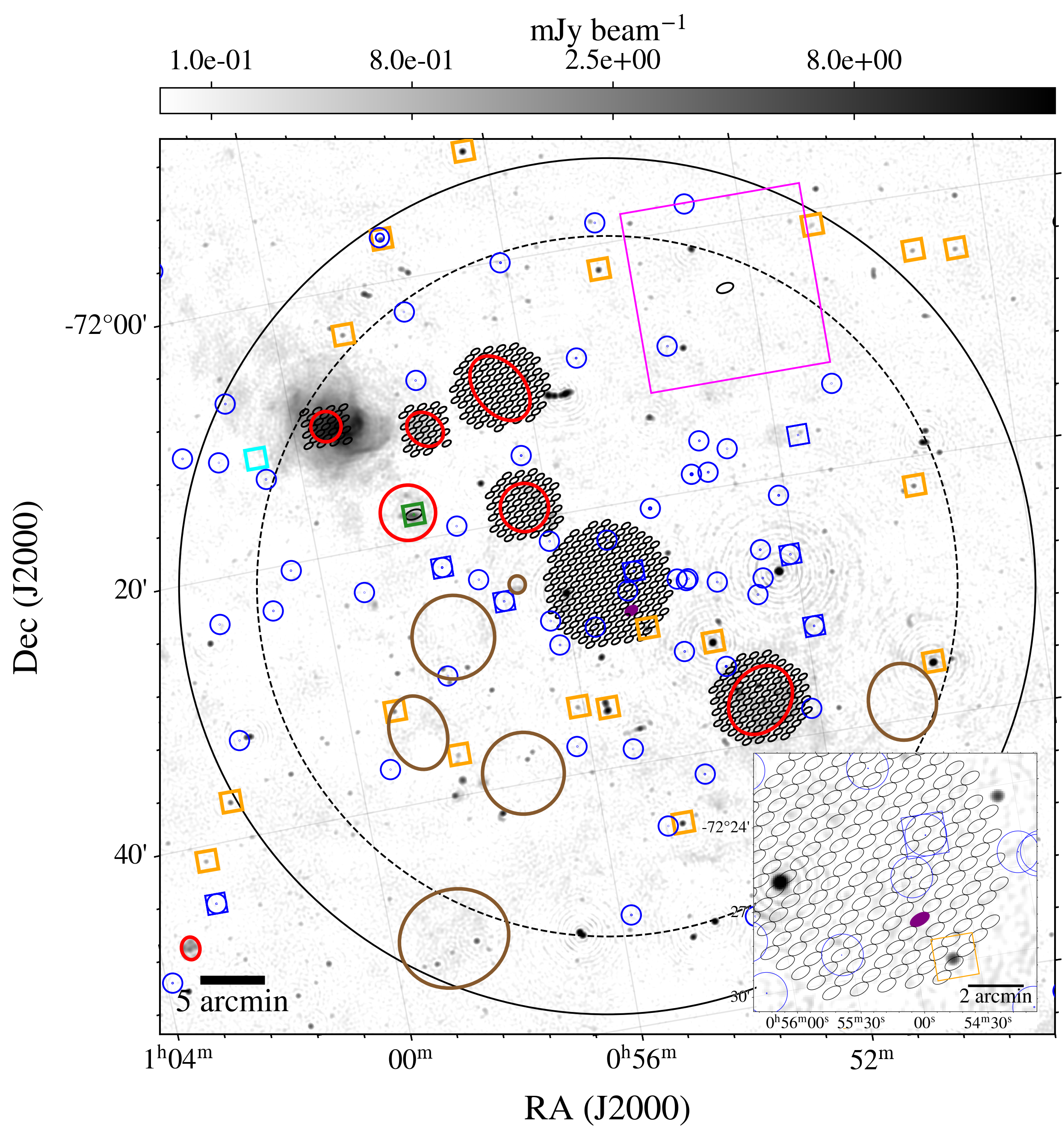}
    \caption{The first pass of \mbox{SMCPOINTING2}.}
    \label{fig:pointing2}
\end{subfigure}%
\begin{subfigure}{.68\textwidth}
    \includegraphics[width=\linewidth]{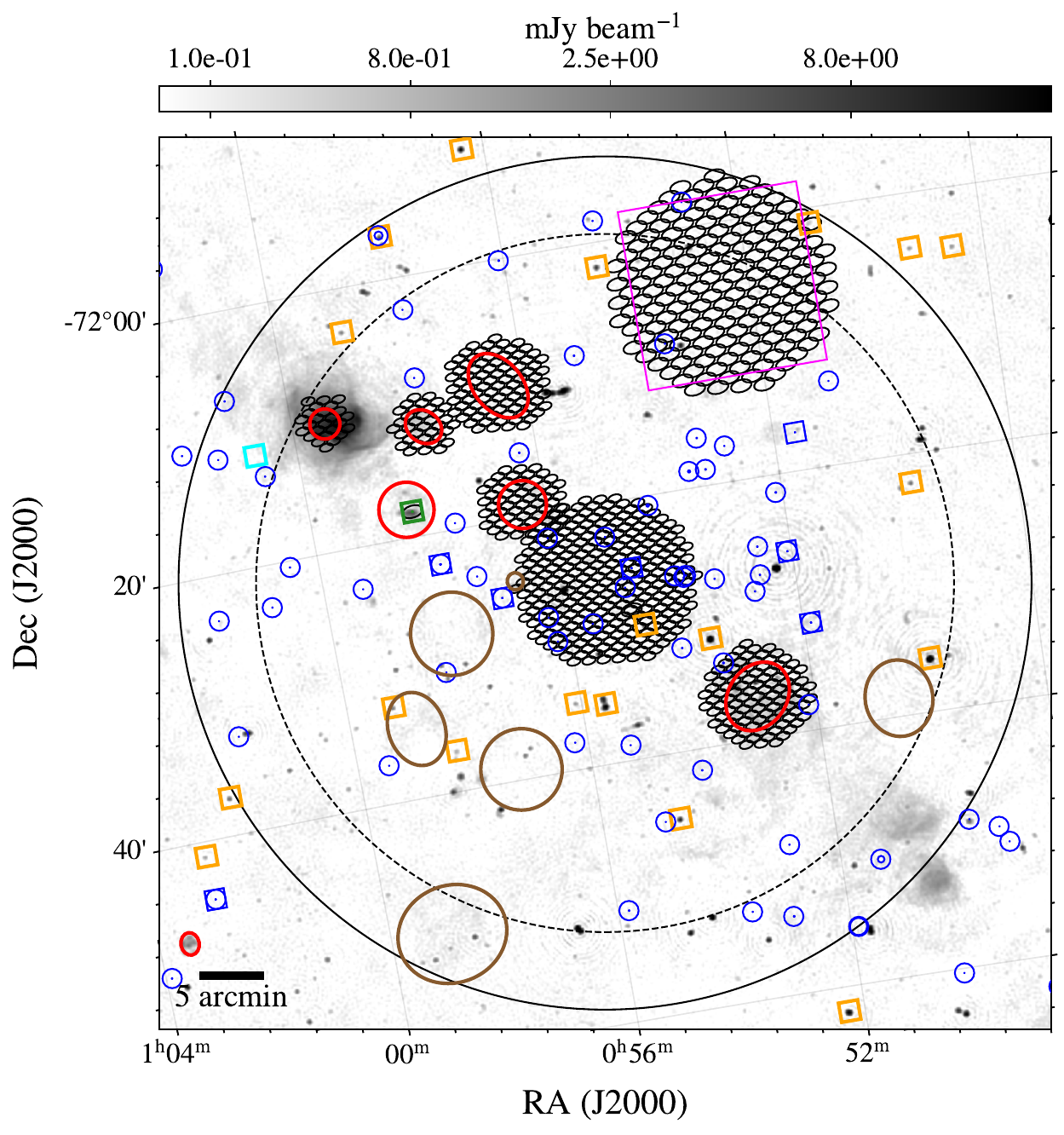}
    \caption{The second pass of \mbox{SMCPOINTING2}.}
    \label{fig:pointing2-2ndpass}
\end{subfigure}

\label{fig:pointing2-both-passes}

\caption{(a): The first pass of \mbox{SMCPOINTING2}. The regions displayed are as detailed in \autoref{fig:pointing1}. Beam tilings overlapping at 75 per cent sensitivity are positioned on five SNRs (red ellipses). The remaining beams were tiled with the same overlap at boresight, which resulted in the discovery of \psrfiftyfour{} (purple coherent beam, see \autoref{pulsar_discoveries}). One SNR observed in \mbox{SMCPOINTING6} is visible in the bottom left corner as a small red ellipse. One steep spectrum point source and one HMXB with period under 10\,s (SXP\,6.88) were observed without being targeted (see \autoref{tab:fast_HMXBs}). The two SMC X-ray pulsars that are not accretion-powered are visible on this plot: the magnetar \smcmagnetar{} (cyan box, observed in later pointings) and \psrfiftyeight{} \protect\citep{Maitra2021} in PWN IKT\,16 (green box with PWN size as a green ellipse inside, \protect\citealt{Maitra2015}). The latter was targeted with a single beam, for which the 50 per cent sensitivity contour is shown  \protect\citep[see][]{Carli2022}. Similarly to the first pass of \mbox{SMCPOINTING1}, a beam was placed on the published position of \psrfiftytwo{} but its position error due to the Murriyang beam size (large pink box) was not taken into account, resulting in its non-detection in this pass of the pointing. 
\\
(b): The second pass of \mbox{SMCPOINTING2}. The regions displayed are as detailed in \autoref{fig:pointing2}. Due to the larger beam sizes,  part of one SNR from \protect\cite{Cotton2024} (brown circle) was observed in the central tiling. The position error area of known pulsar \fiftytwo{} was tiled with beams overlapping at 50 per cent sensitivity (covering an untargeted steep spectrum point source). This resulted in the detection and localisation of this pulsar (see \autoref{titus_localisations}). A single beam, pictured at 50 per cent sensitivity, was placed on the \seeKAT{} localisation position of \psrfiftyfour{}. } 
\end{figure}

\end{landscape}

\begin{landscape}

\begin{figure}

\begin{subfigure}{.68\textwidth}
    \includegraphics[width=\linewidth]{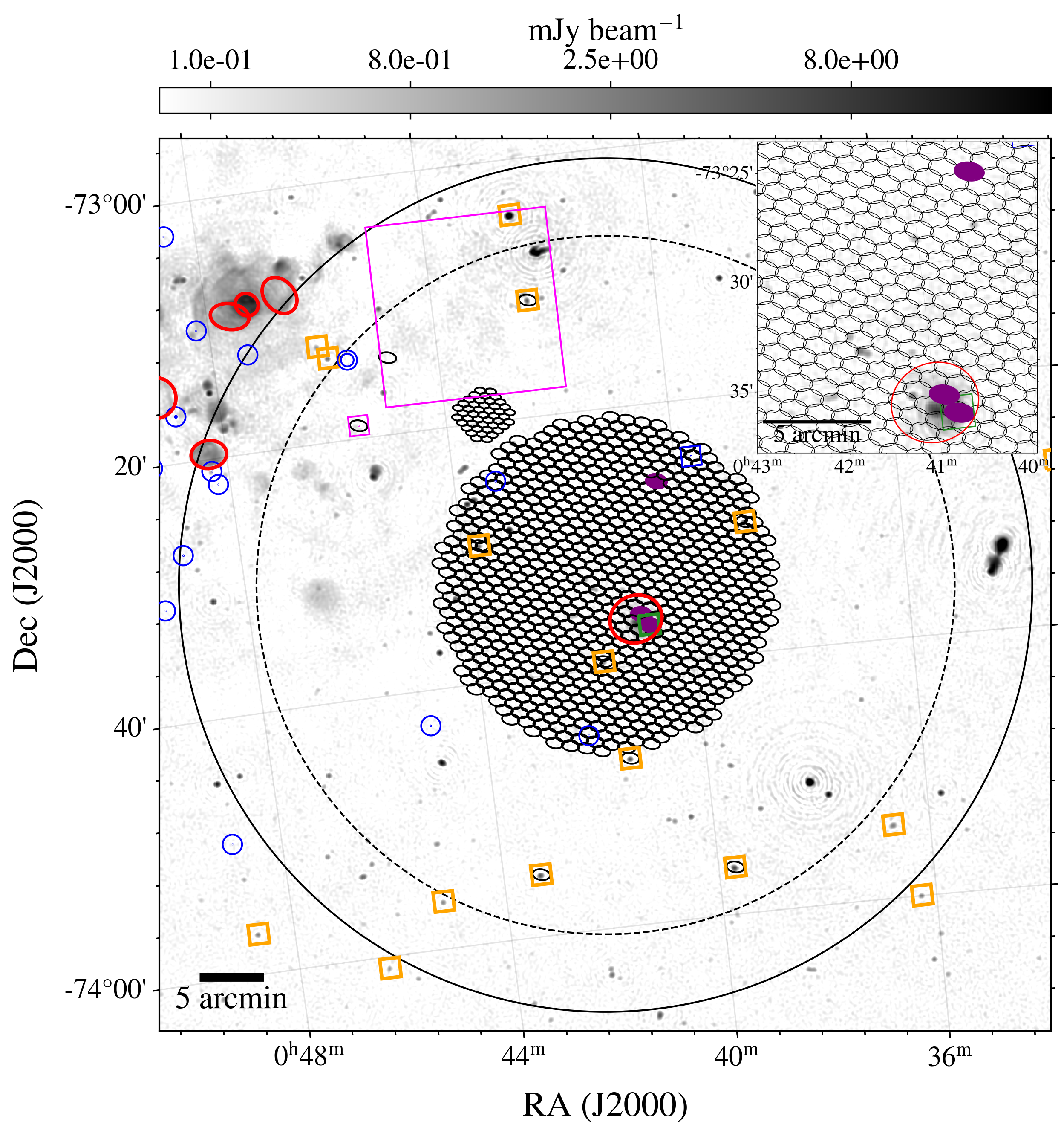}
    \caption{The first pass of \mbox{SMCPOINTING3}.}
    \label{fig:pointing3}
\end{subfigure}%
\begin{subfigure}{.68\textwidth}
    \includegraphics[width=\linewidth]{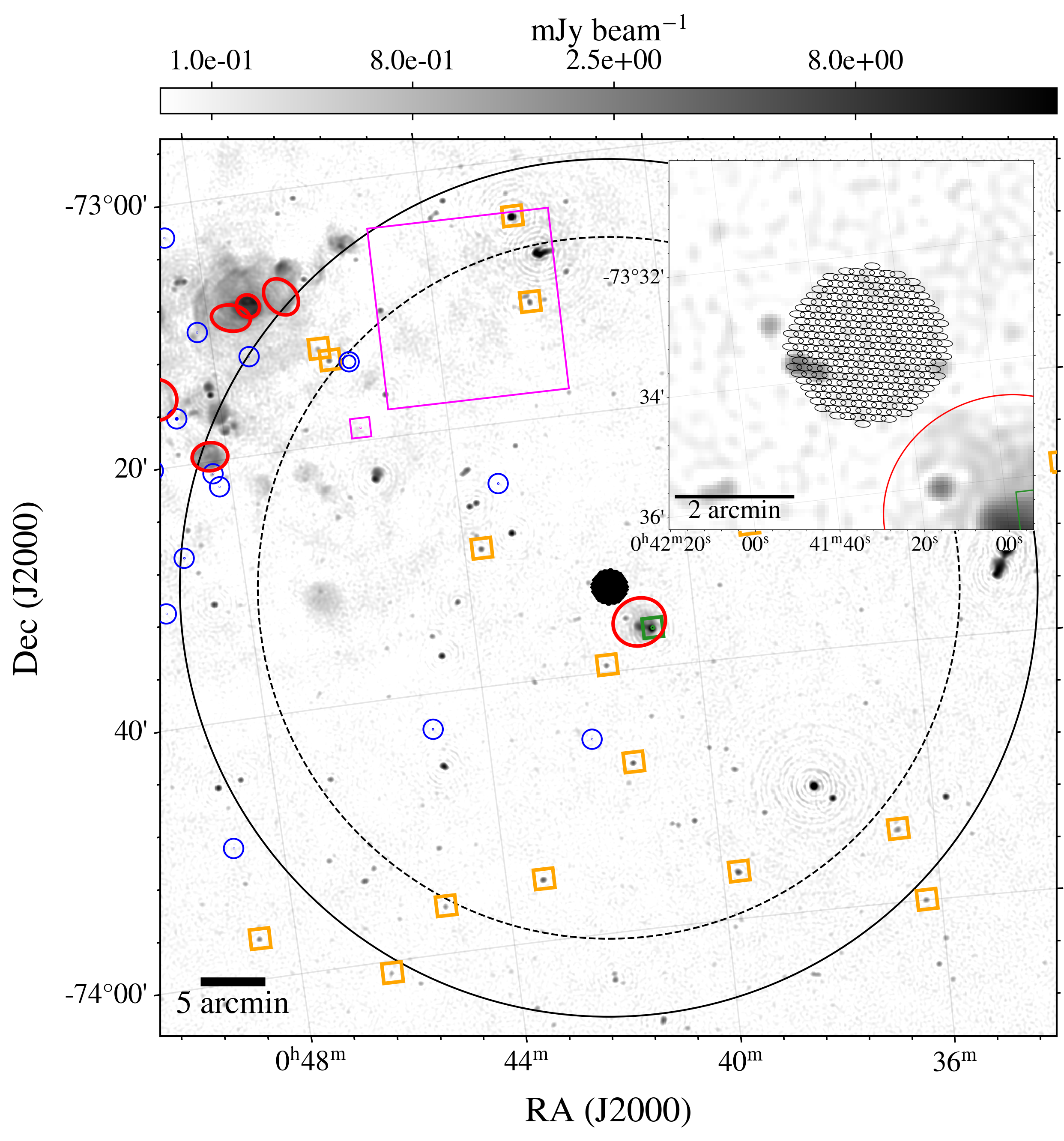}
    \caption{The second pass of \mbox{SMCPOINTING3}.}
    \label{fig:pointing3-2ndpass}
\end{subfigure}

\label{fig:pointing3-both-passes}

\caption{(a): The first pass of \mbox{SMCPOINTING3}. The regions displayed are as detailed in \autoref{fig:pointing1}. The central tiling covers the SNR DEM\,S5 (central red ellipse). A single beam was nevertheless placed on its associated PWN \protect\citep[green box,][]{Alsaberi2019}. The PWN, \psrfortyfive{} (small pink box), \psrfortyfour{} (discovered in the second pass of \mbox{SMCPOINTING1}, see \autoref{fig:pointing1-2ndpass}) and seven steep spectrum point sources (orange boxes, see \autoref{tab:steep_spectrum_sources}) were targeted with single beams shown with their 50 per cent sensitivity contour, even those that fell in the central tiling for this pointing. Three pulsars were discovered in this pointing (purple beams): \psrtwentysix{}, \psrthirtyseven{} (with the single beam on the PWN the strongest detection), and nearby pulsar \thirtyfive{} in the main tiling. A smaller coherent beam tiling overlapping at 70 per cent sensitivity was placed around the detected position of \psrfortythree{} (from the second pass of \mbox{SMCPOINTING2}, see \autoref{fig:pointing2-2ndpass}) to localise it further than its large Murriyang position error (large pink box, see \autoref{titus_localisations}). SXP\,7.59 \protect\citep{Hong2017} was observed without being targeted. SNRs observed in both passes of \mbox{SMCPOINTING1} are seen in the top left corner. 
\\
(b): The second pass of \mbox{SMCPOINTING3}. The regions displayed are as detailed in \autoref{fig:pointing3}. The setup of this observation failed, and used the full array instead of the core, resulting in a single small coherent beam tiling overlapping at 80 per cent. } 
\end{figure}

\end{landscape}

\begin{landscape}

\begin{figure}

\begin{subfigure}{.68\textwidth}
    \includegraphics[width=\linewidth]{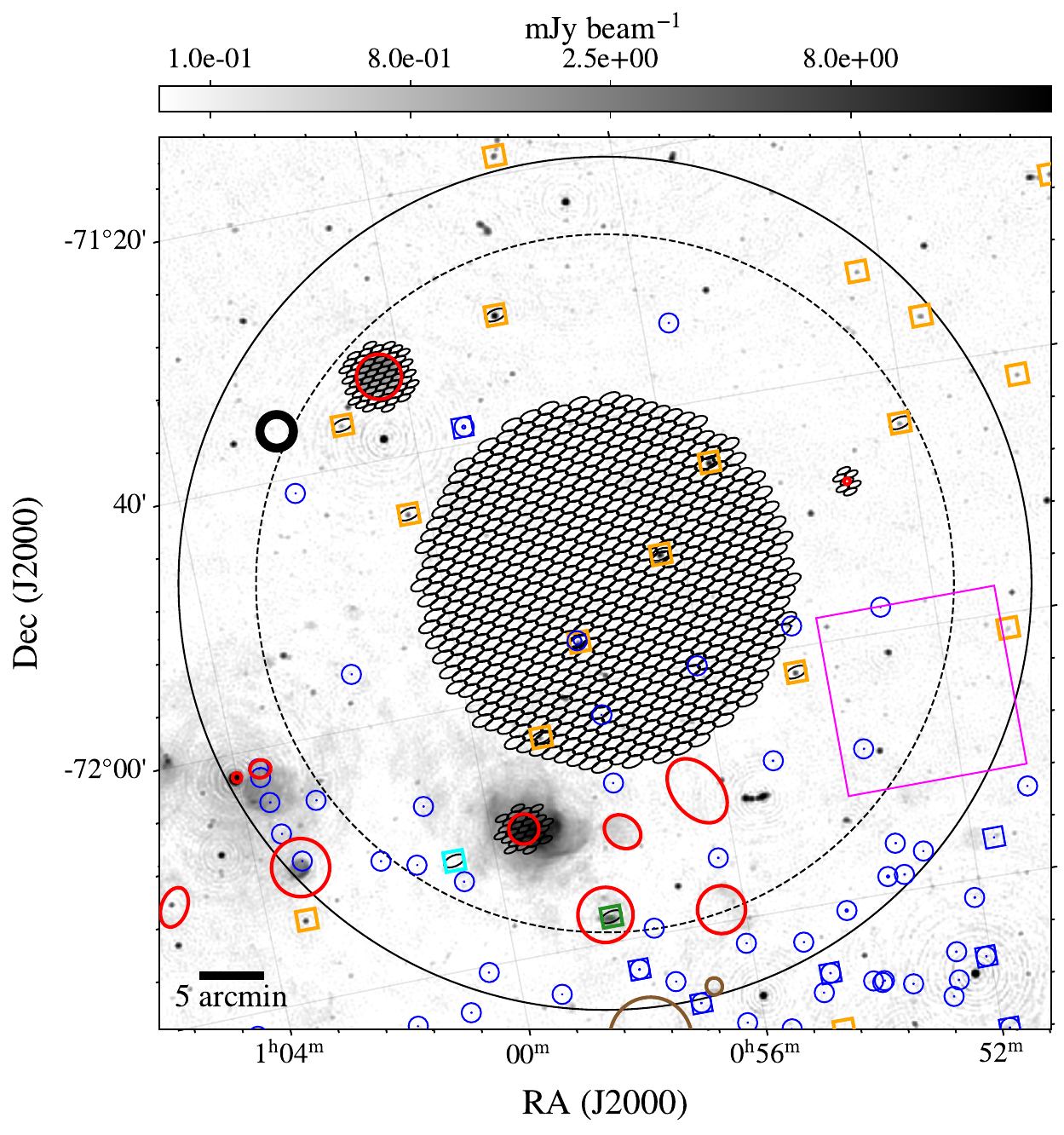}
    \caption{The first pass of \mbox{SMCPOINTING4}.}
    \label{fig:pointing4}
\end{subfigure}%
\begin{subfigure}{.68\textwidth}
    \includegraphics[width=\linewidth]{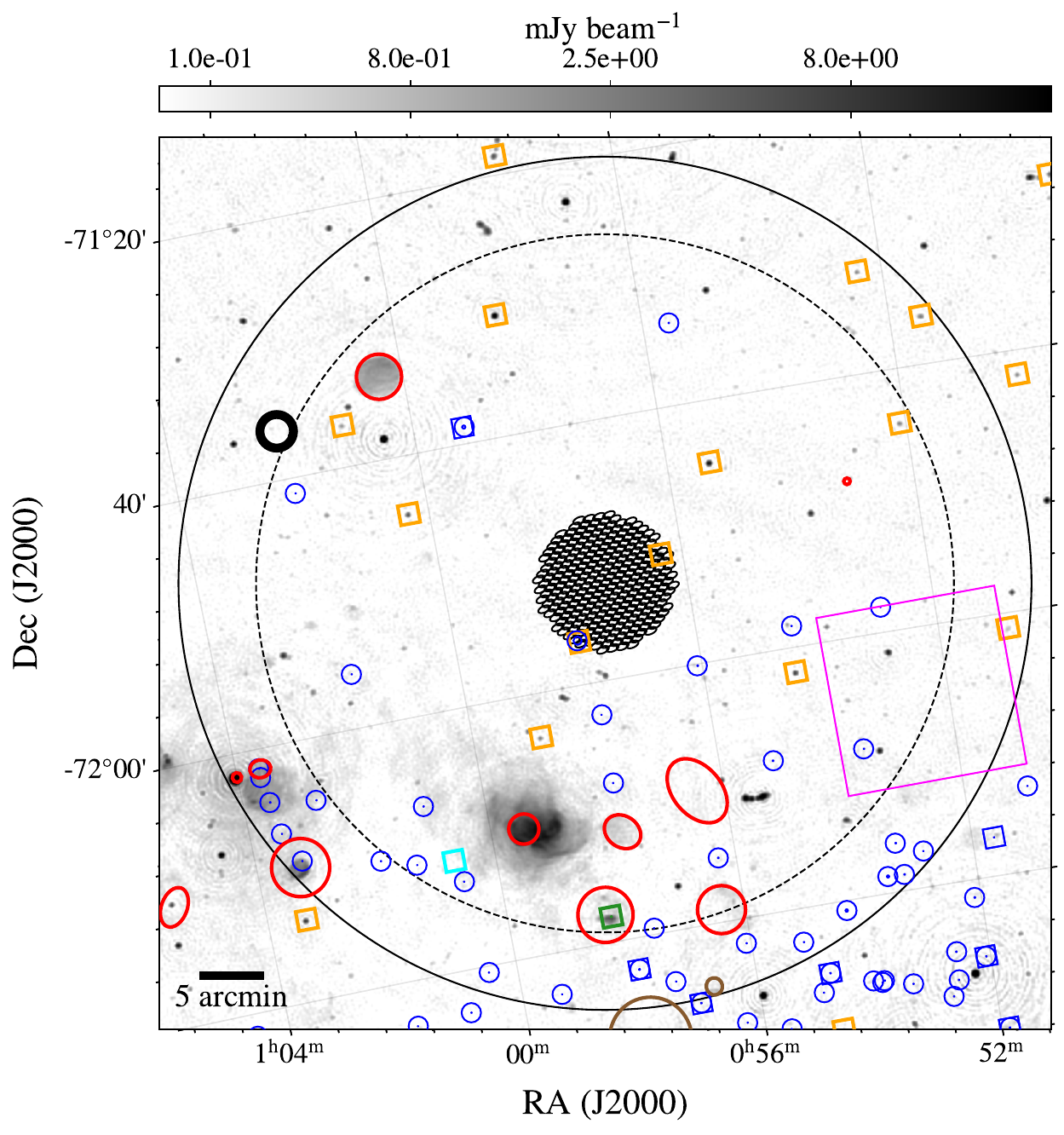}
    \caption{The second pass of \mbox{SMCPOINTING4}.}
    \label{fig:pointing4-2ndpass}
\end{subfigure}

\label{fig:pointing4-both-passes}

\caption{(a): The first pass of \mbox{SMCPOINTING4}. The regions displayed are as detailed in \autoref{fig:pointing1} and \autoref{fig:pointing2}. Three SNRs (red) are tiled with beams overlapping at 70 per cent sensitivity. The PWN within IKT\,16 (green box with PWN size inside), the magnetar \smcmagnetar{} (cyan box), and nine steep spectrum point sources (orange boxes, see \autoref{tab:steep_spectrum_sources}) were targeted with single beams shown at their 50 per cent sensitivity contour, even those that fell in the central tiling for this pointing. NGC\,361 (thick black circle) was omitted due to being in a lower sensitivity area of the incoherent beam. At the bottom centre and right of the figure, SNRs and the error box of \psrfiftytwo{}, both observed in \mbox{SMCPOINTING2} are shown. At the bottom left, SNRs observed in \mbox{SMCPOINTING5} (see \autoref{fig:pointing5}) appear. HMXBs with a period of under 10\,s (dark blue boxes) were not yet targeted and none was observed in this pointing. 
\\
(b): The second pass of \mbox{SMCPOINTING4}. The regions displayed are as detailed in \autoref{fig:pointing3}. The setup of this observation failed, resulting in a small core array coherent beam tiling overlapping at 80 per cent. } 
\end{figure}

\end{landscape}

\begin{landscape}

\begin{figure}

\begin{subfigure}{.68\textwidth}
    \includegraphics[width=\linewidth]{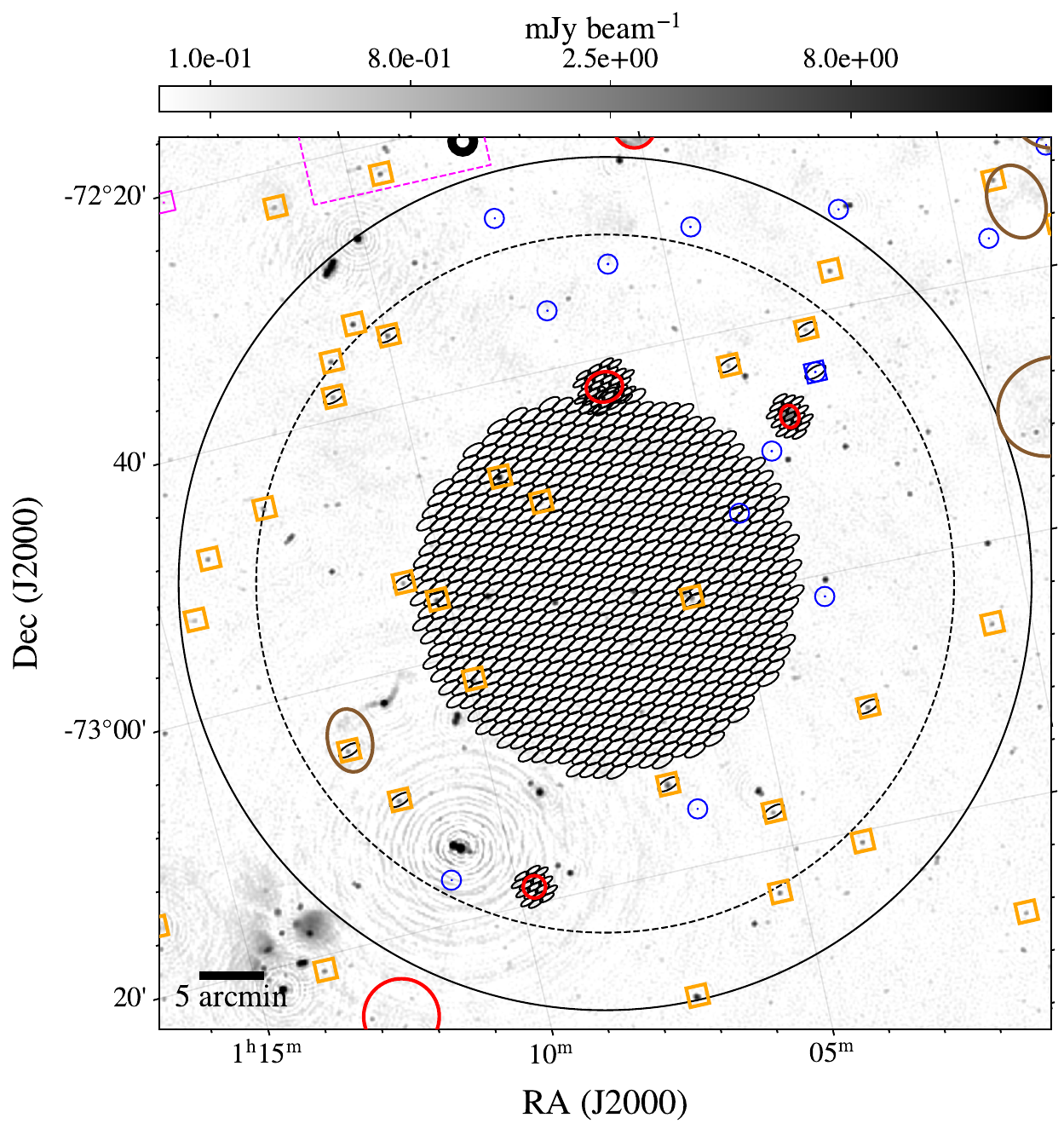}
    \caption{The first pass of \mbox{SMCPOINTING6}.}
    \label{fig:pointing6}
\end{subfigure}%
\begin{subfigure}{.68\textwidth}
    \includegraphics[width=\linewidth]{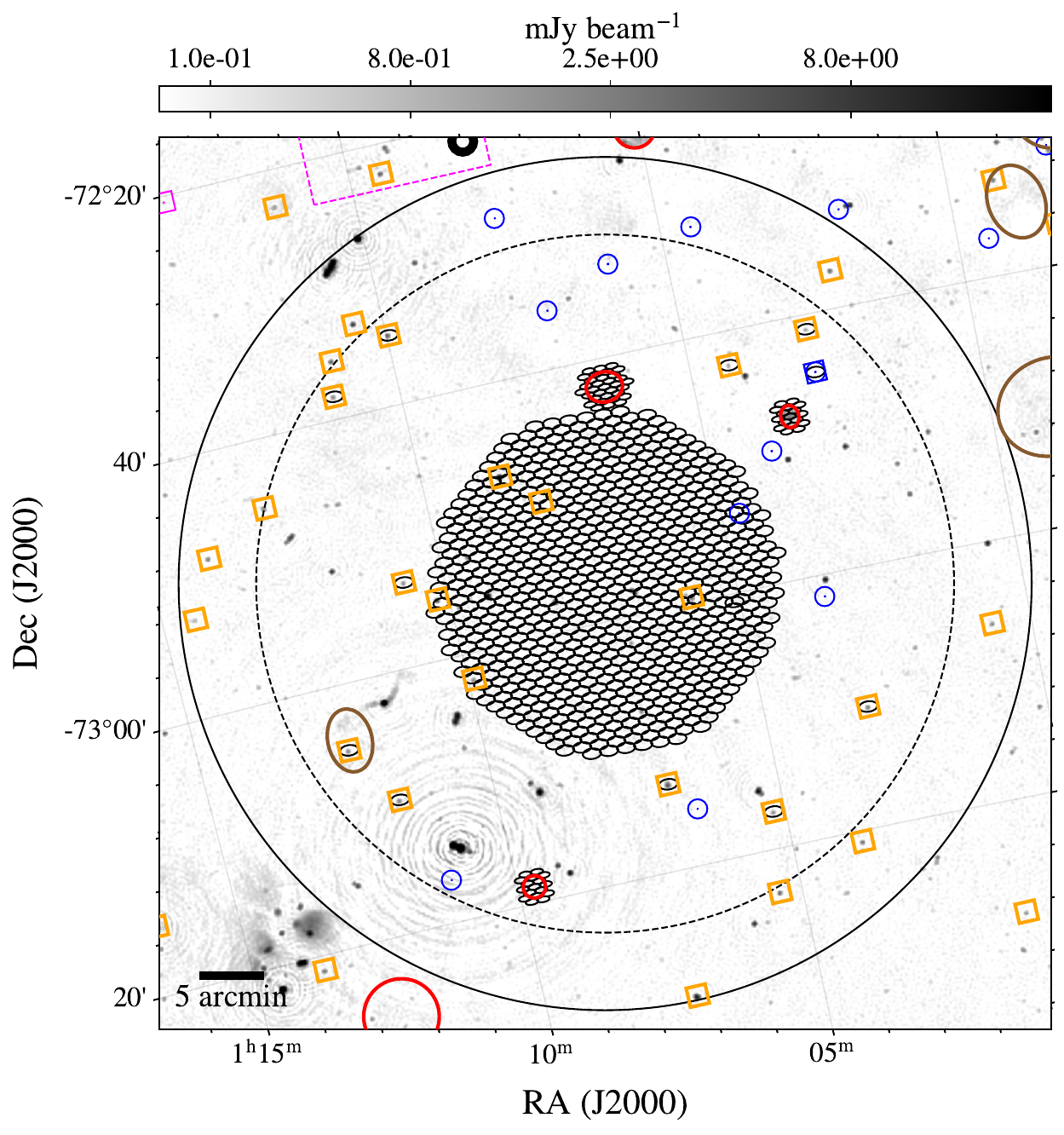}
    \caption{The second pass of \mbox{SMCPOINTING6}.}
    \label{fig:pointing6-2ndpass}
\end{subfigure}

\label{fig:pointing6-both-passes}

\caption{(a): The first pass of \mbox{SMCPOINTING6}. The regions displayed are as detailed in \autoref{fig:pointing1} and \autoref{fig:pointing2}. Three SNRs (red) are targeted with coherent beam tilings overlapping at 70 per cent sensitivity. One HMXB with a period of under 10\,s (dark blue box)  and 10 steep spectrum point sources (orange boxes) are targeted with single beams that are outside tilings (50 per cent sensitivity contours displayed). The central coherent beam tiling, overlapping at 50 per cent sensitivity, covers a further five untargeted steep spectrum point sources. It is interesting to note that steep spectrum point source EMU-ESP-SMC-5407 is observed inside SNR J0112$-$7304, newly discovered by \protect\citealt{Cotton2024} (see \autoref{tab:sensitivity_targets}). Other new SNRs from \protect\cite{Cotton2024} are visible in the top right corner but were not observed in this survey due to not being known at the time. A candidate SNR from \protect\cite{Titus2019} observed in \mbox{SMCPOINTING7} is visible at the bottom of the figure. In the top left of the figure, a small pink box shows the known pulsar J0113$-$7220, with its position error inside.  Other targets observed in SMPOINTING5 (see \autoref{fig:pointing5}) are visible in the top middle of the figure. 
\\
(b): The second pass of \mbox{SMCPOINTING6}. The regions displayed are as detailed in \autoref{fig:pointing6}. } 
\end{figure}

\end{landscape}

\begin{landscape}

\begin{figure}

\begin{subfigure}{.68\textwidth}
    \includegraphics[width=\linewidth]{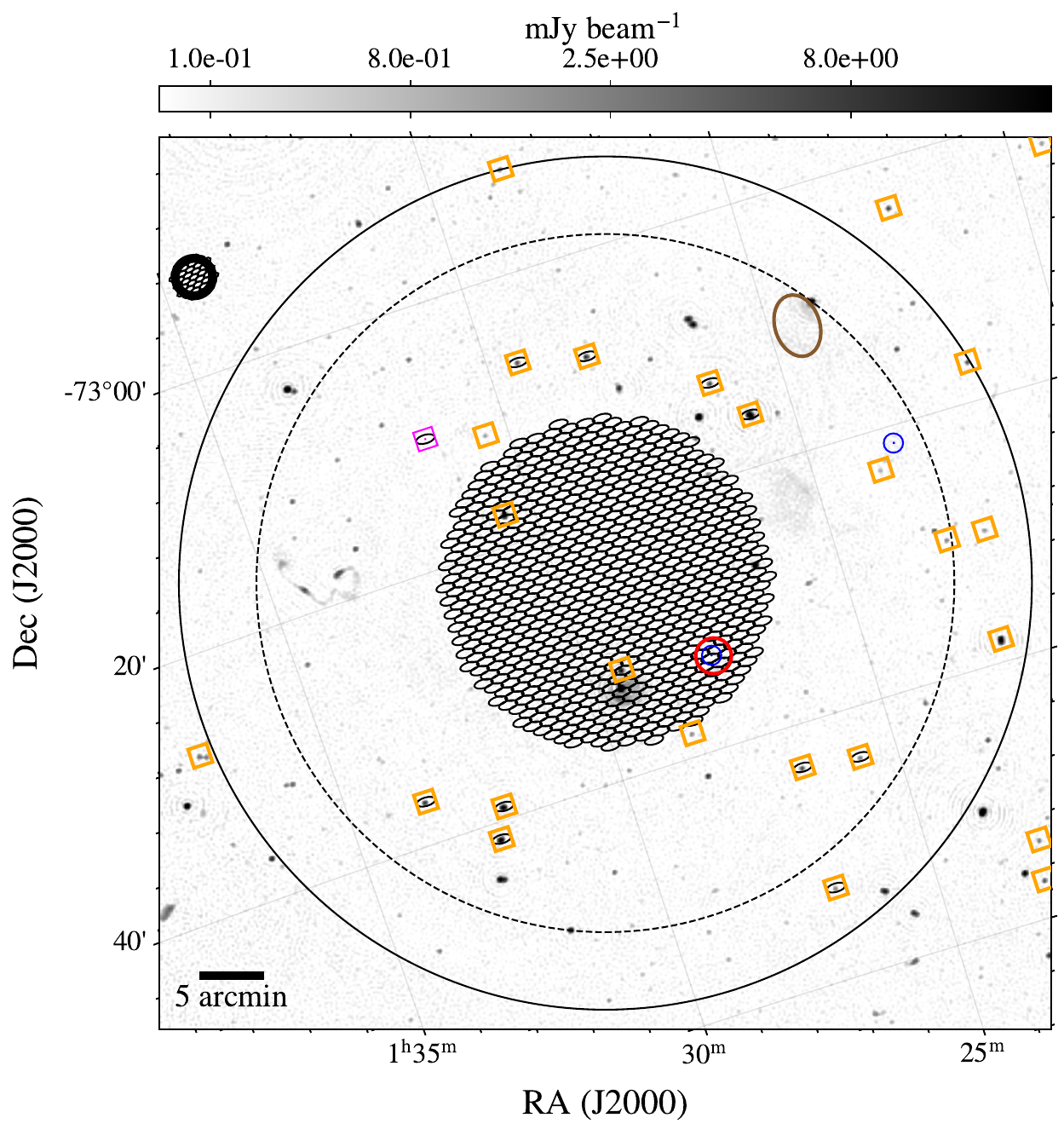}
    \caption{The first pass of \mbox{SMCPOINTING8}.}
    \label{fig:pointing8}
\end{subfigure}%
\begin{subfigure}{.68\textwidth}
    \includegraphics[width=\linewidth]{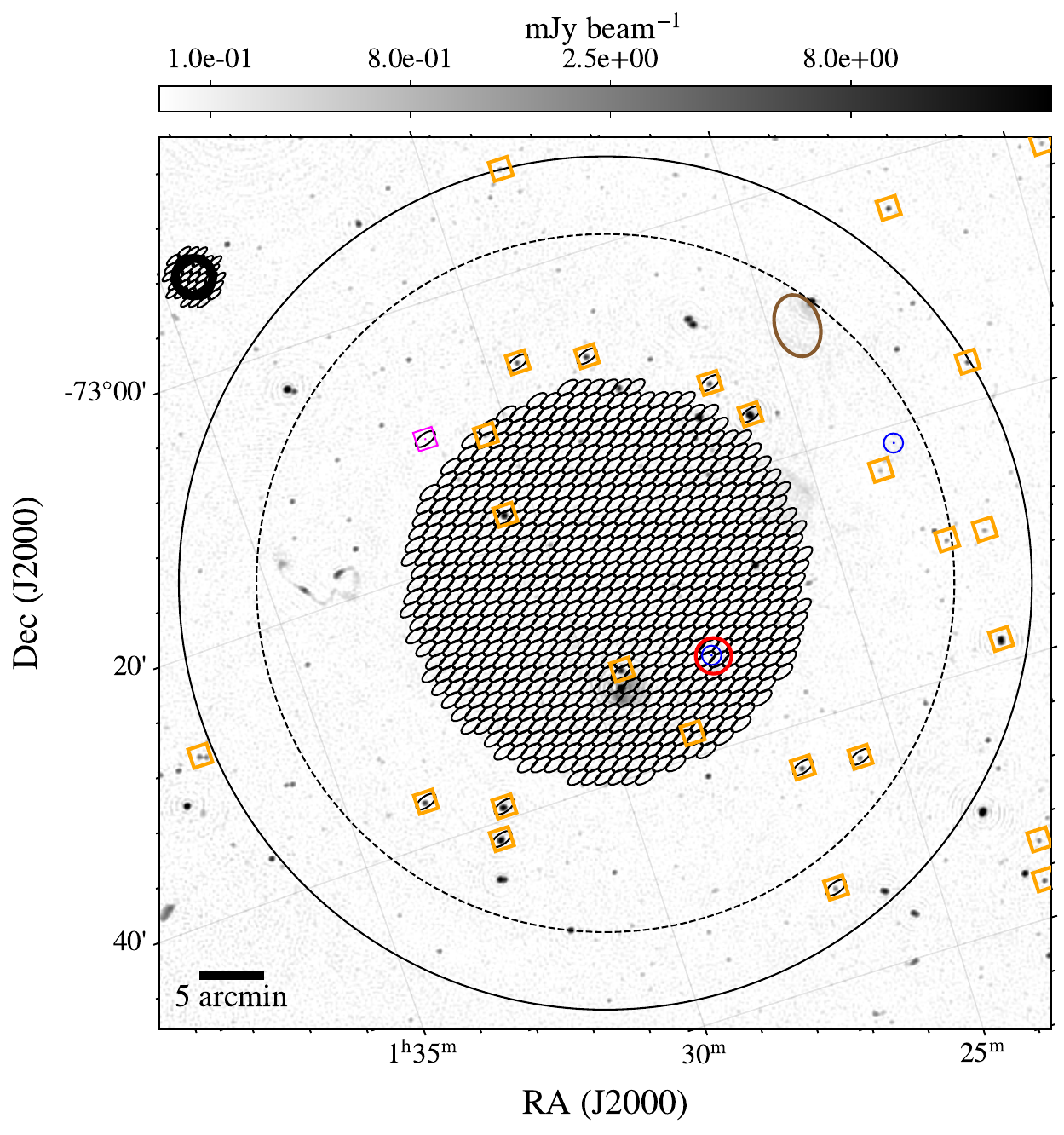}
    \caption{The second pass of \mbox{SMCPOINTING8}.}
    \label{fig:pointing8-2ndpass}
\end{subfigure}

\label{fig:pointing8-both-passes}

\caption{(a): The first pass of \mbox{SMCPOINTING8}. The regions displayed are as detailed in \autoref{fig:pointing1}. A SNR (red) is observed in the central tiling overlapping at 50 per cent sensitivity. Ten steep spectrum point sources (orange) outside this tiling were targeted with single beams (50 per cent sensitivity contour displayed). Two were covered by the central tiling. PSR\,J0131$-$7310 (pink box with position error inside) was observed with a single coherent beam (50 per cent sensitivity contour shown). The cluster Lindsay\,110 (thick black circle) is tiled with 70 per cent sensitivity overlap. However, sensitivity is limited in this area, as it is beyond the 50 per cent sensitivity contour of the incoherent beam at the central L-band frequency (large thin black circle).  A new \protect\cite{Cotton2024} SNR is displayed as a brown ellipse but was not known at the time.  
\\
(b): The second pass of \mbox{SMCPOINTING8}. The regions displayed are as detailed in \autoref{fig:pointing8} with two additional steep spectrum point sources observed in the central beam tiling due to the larger beam size. } 
\end{figure}

\end{landscape}

\begin{landscape}

\begin{figure}

\begin{subfigure}{.68\textwidth}
    \includegraphics[width=\linewidth]{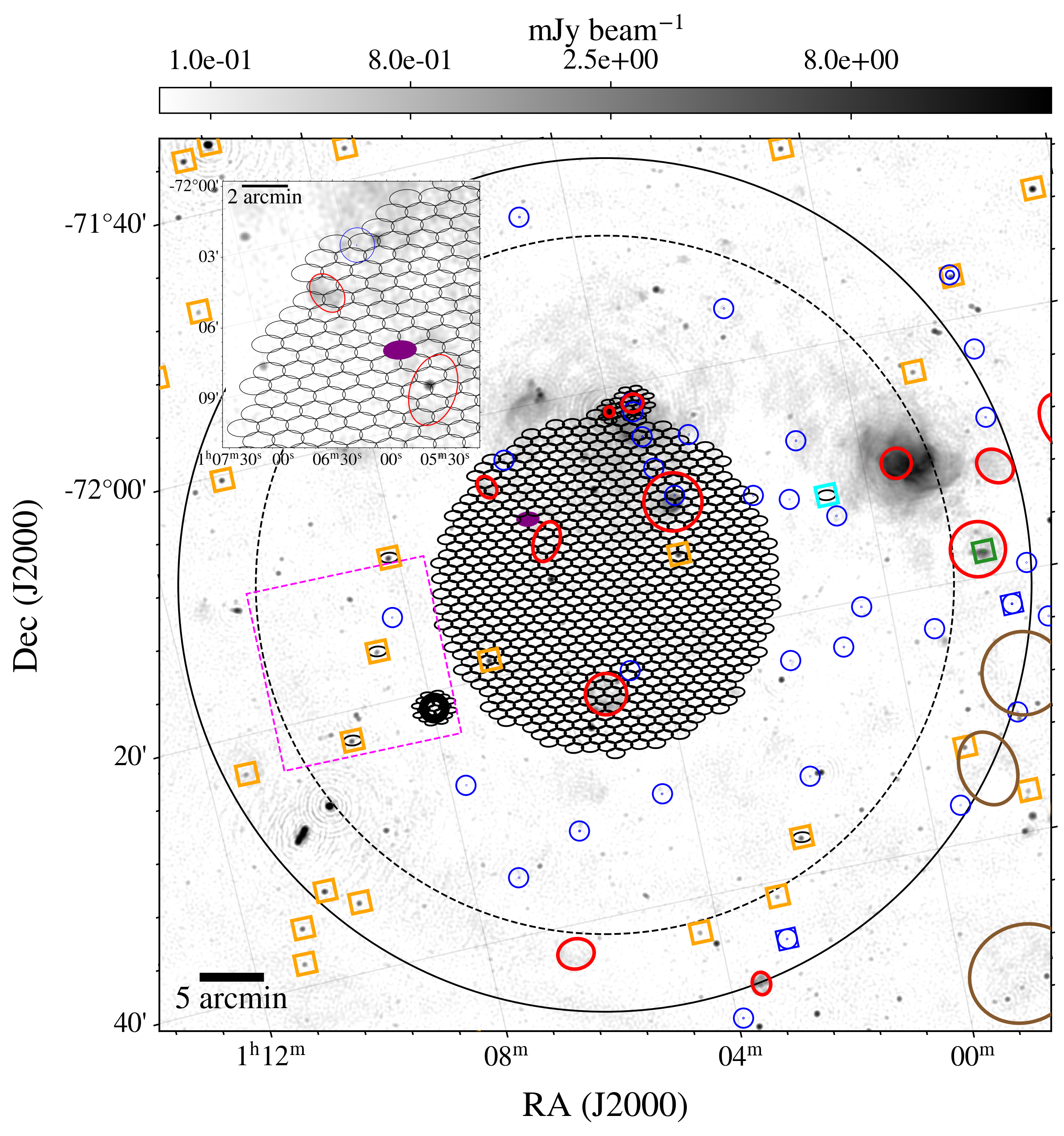}
    \caption{The single observation of \mbox{SMCPOINTING5}.}
    \label{fig:pointing5}
\end{subfigure}%
\begin{subfigure}{.68\textwidth}
    \includegraphics[width=\linewidth]{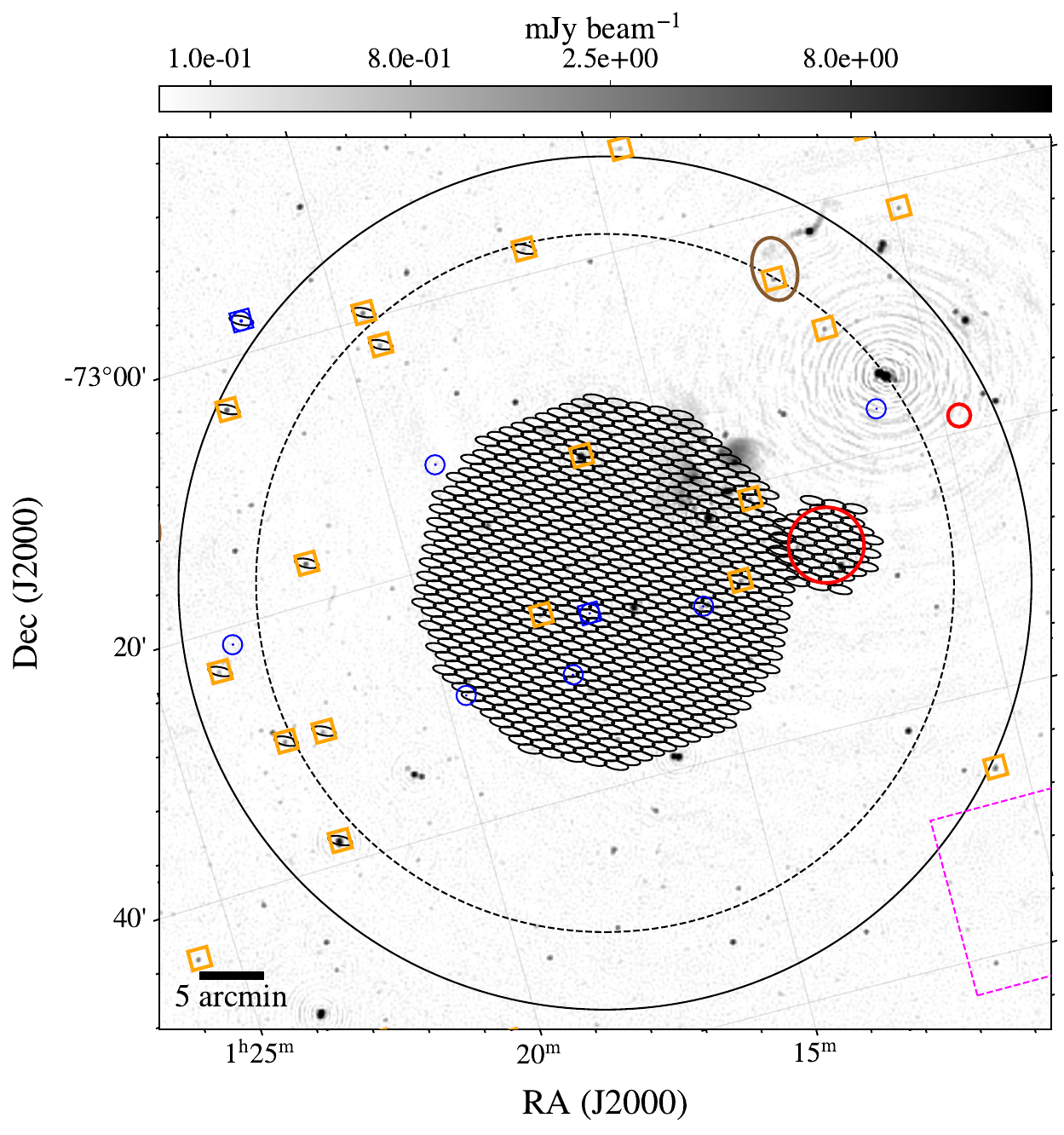}
    \caption{The single observation of \mbox{SMCPOINTING7}.}
    \label{fig:pointing7}
\end{subfigure}

\label{fig:pointing5-and-7}

\caption{(a): The single observation of \mbox{SMCPOINTING5}. The regions displayed are as detailed in \autoref{fig:pointing1} and \autoref{fig:pointing2}. Two SNRs (red) were observed with tilings of coherent beams overlapping at the 70 per cent sensitivity level, while four were observed with the central tiling overlapping at 50 per cent sensitivity. \psronezerofive{} was discovered with the strongest S/N in the coherent beam highlighted in purple (see \autoref{pulsar_discoveries}). The SMC magnetar \smcmagnetar{} and four steep spectrum point sources (orange) were targeted with a single beam (50 per cent sensitivity contour shown). No fast-spinning HMXBs (blue boxes) were thought close enough to boresight to target. Two steep spectrum point sources were observed in the main tiling without being targeted. The large position error area of a pulsar candidate from \protect\cite{Titus2019} (pink dashed box) was not entirely observed, instead focusing on the cluster NGC\,416 (thick black circle) with a tiling of coherent beams overlapping at the 70 per cent sensitivity level. Two SNRs (red) observed in \mbox{SMCPOINTING6} are visible in the bottom centre of the figure, while the right side of the figure contains several targets from \mbox{SMCPOINTING2}. 
\\
(b): The single observation of \mbox{SMCPOINTING7}. The regions displayed are as detailed in \autoref{fig:pointing1} and \autoref{fig:pointing2}. All beams are tiled at 50 per cent sensitivity overlap, covering a central zone, a candidate SNR (red), four steep spectrum point sources (orange), and one fast-spinning HMXB, SXP\,0.72 also known as SMC X-1 (blue boxes). In the top right corner, targets observed in \mbox{SMCPOINTING6} are visible. Single beams are placed on one fast-spinning HMXB and nine steep spectrum point sources. } 
\end{figure}

\end{landscape}

\begin{landscape}

\begin{figure}
\centering

    \includegraphics[width=.8\textwidth]{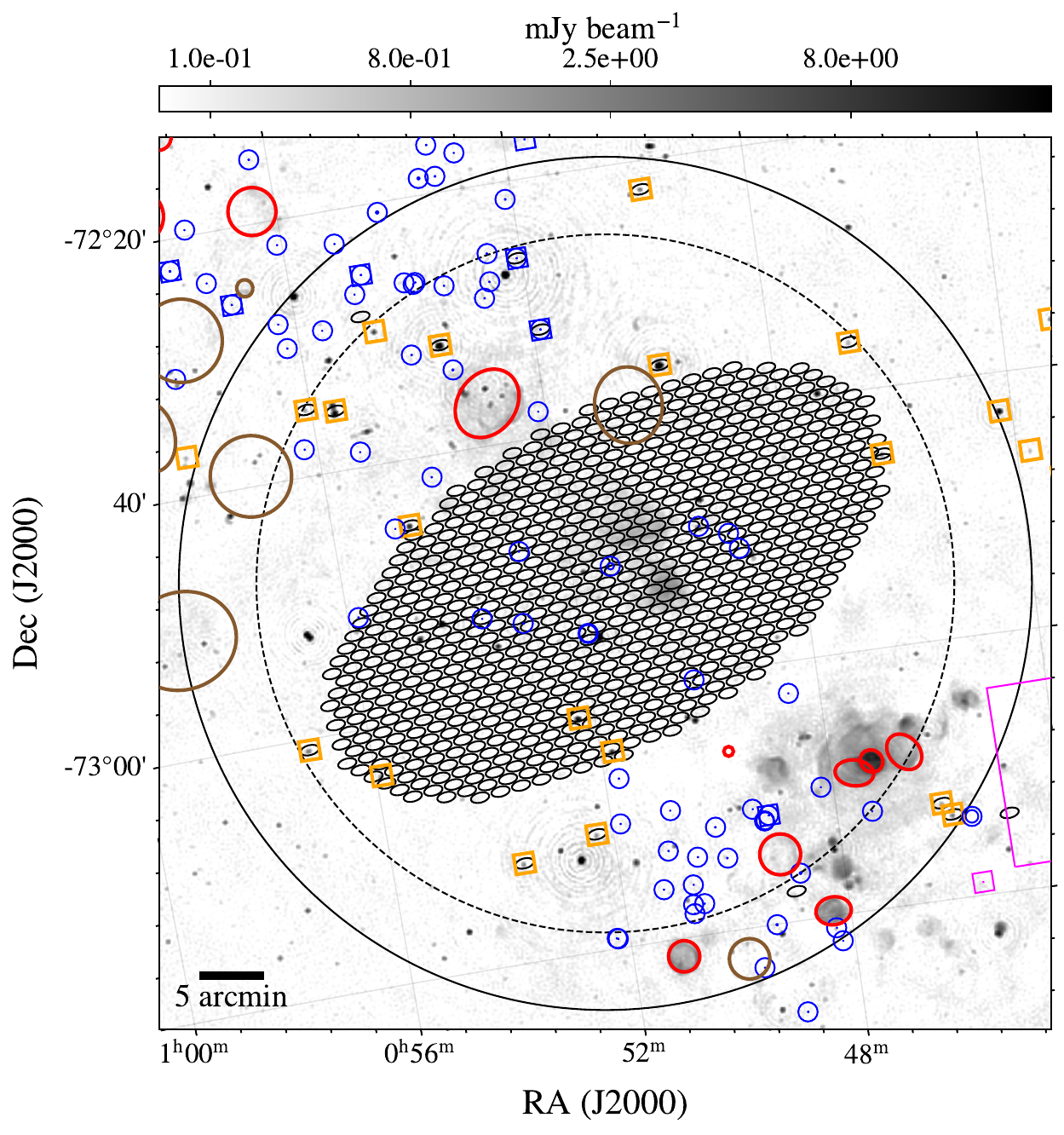}

\caption{The extra pointing \mbox{SMCPOINTING9} consists of an ellipse tiling of coherent beams which 50 per cent sensitivity contour is shown, covering three steep spectrum point sources and part of a SNR newly discovered in \protect\cite{Cotton2024} (brown ellipse). The regions displayed are as detailed in \autoref{fig:pointing1}. Single beams were placed on 13 steep spectrum point sources (orange) and two fast-spinning HMXBs (dark blue boxes), as well as \psrfortyfour{}, \psrfortyeight{} and \psrfiftyfour{} discovered earlier in the survey. The SNR above the tiling (red) was observed in \mbox{SMCPOINTING2} while the sources in the bottom right corner were observed in \mbox{SMCPOINTING1}.} 
    \label{fig:pointing9}

\end{figure}

\end{landscape}

\begin{landscape}

\begin{figure}

\begin{subfigure}{.68\textwidth}
    \includegraphics[width=\linewidth]{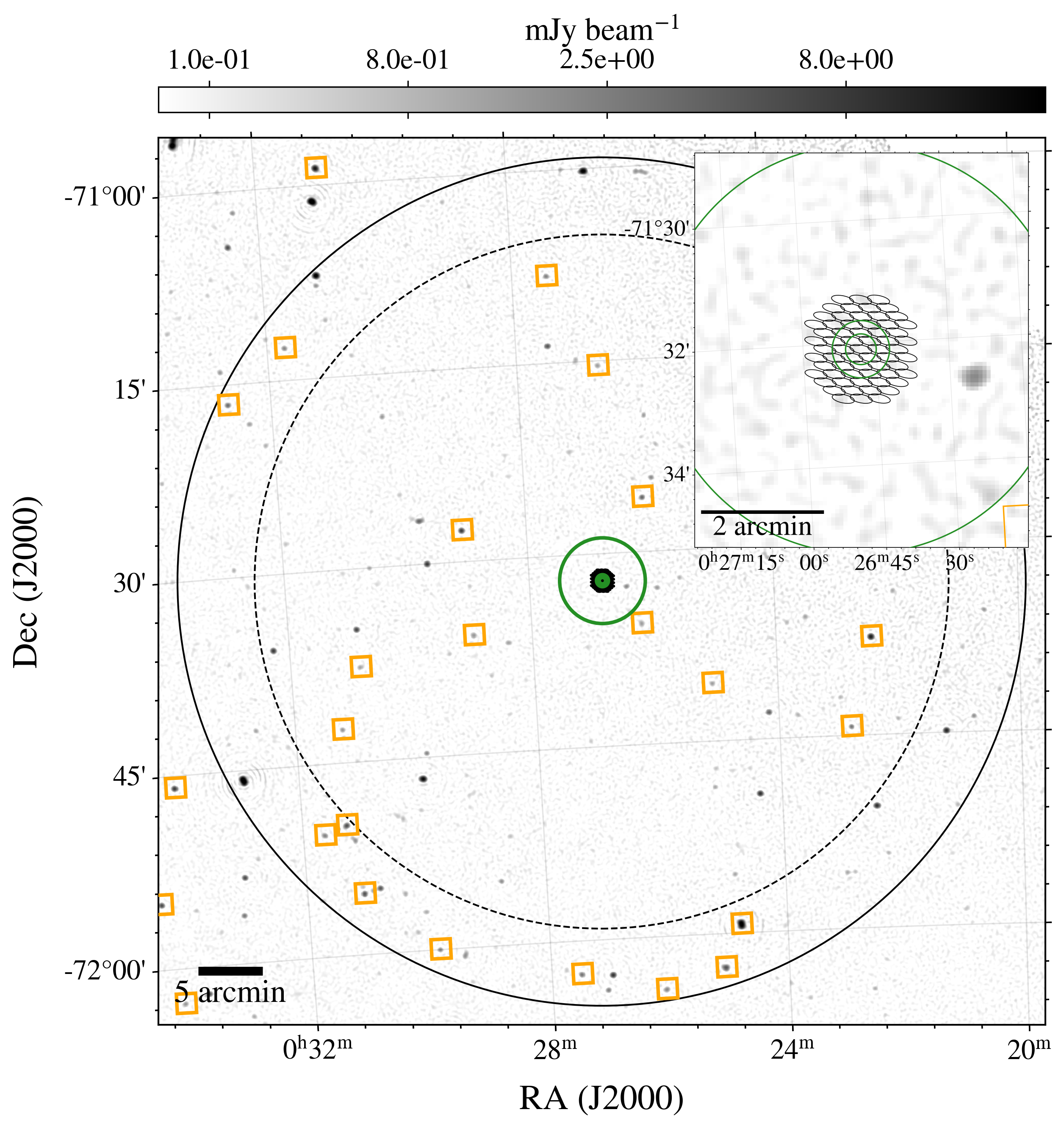}
    \caption{The first pass of NGC\,121.}
    \label{fig:NGC121}
\end{subfigure}%
\begin{subfigure}{.68\textwidth}
    \includegraphics[width=\linewidth]{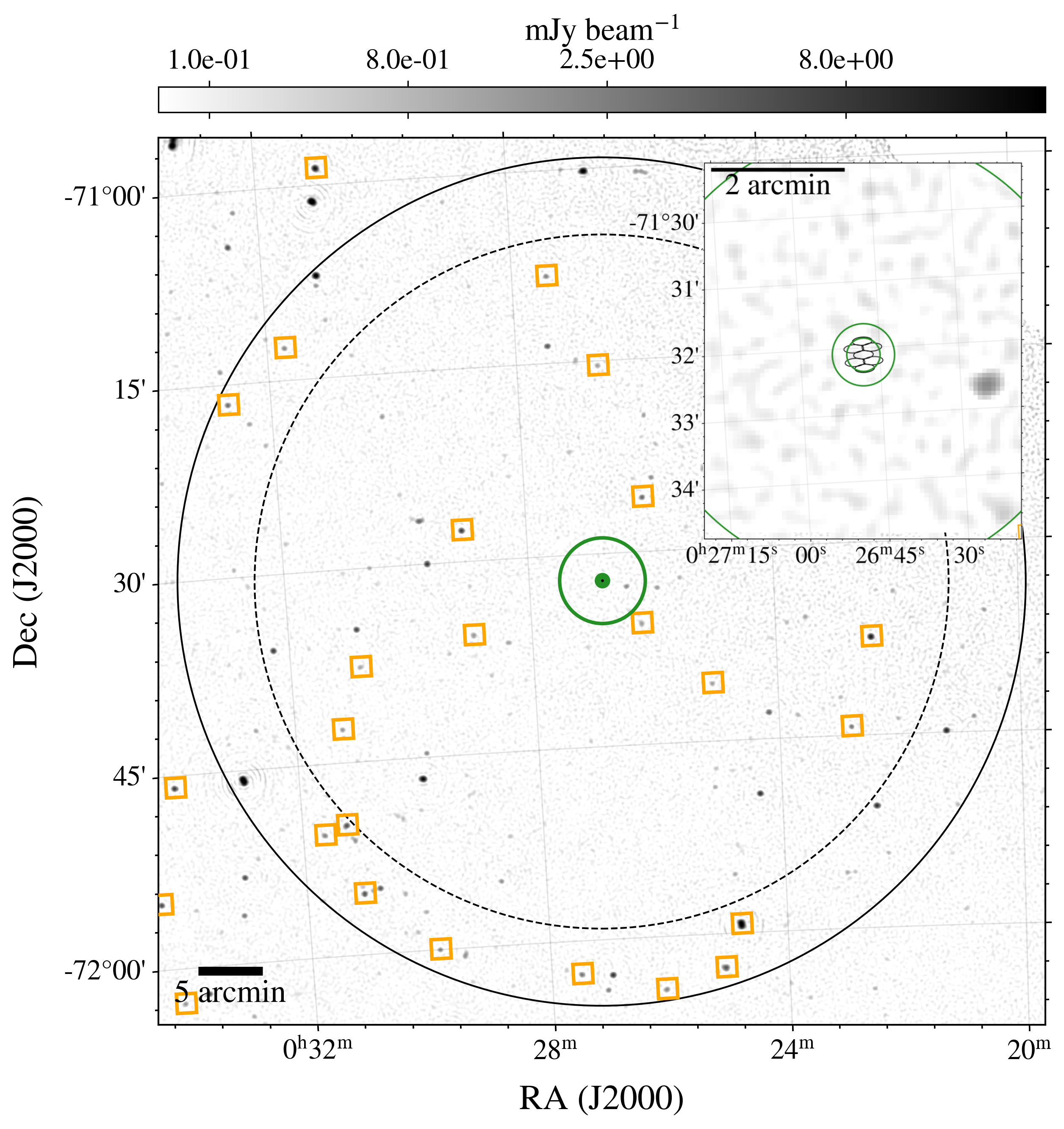}
    \caption{The second pass of NGC\,121.}
    \label{fig:NGC121_2ndpass}
\end{subfigure}

\label{fig:NGC121-both-passes}

\caption{(a): The first pass observation of the SMC Globular Cluster NGC\,121 as detailed in \autoref{NGC121}. The regions displayed are as detailed in \autoref{fig:pointing1}. Full array coherent beams are tiled to overlap at 75 per cent sensitivity. The three green circles are (from largest to smallest) the  total, half-light and core radii of the cluster respectively \protect\citep{Baume2008,Glatt2009}. 
\\
(b): The second pass observation of the SMC Globular Cluster NGC\,121 as detailed in \autoref{NGC121}. The regions displayed are as detailed in \autoref{fig:NGC121}. } 
\end{figure}

\end{landscape}

\bsp	
\label{lastpage}
\end{document}